\documentclass[10pt]{article}

\usepackage{amssymb}
\usepackage{graphicx}
\usepackage[numbers,compress]{natbib}
\begin{document}

\title{Operators of quantum theory of  Dirac's free field}

	\author{Ion I. Cot\u aescu\thanks{Corresponding author E-mail:~~i.cotaescu@e-uvt.ro}\\
	{\it West University of Timi\c soara,} \\{\it V. Parvan Ave. 4,
		RO-300223 Timi\c soara}}
	
	\maketitle

\begin{abstract}
	The Pryce (e) spin and position operators of the quantum theory of Dirac's free field were re-defined and 	studied recently with the help of a new spin symmetry and suitable spectral representations  [I. I. Cot\u aescu, Eur. Phys. J. C  (2022) 82:1073].   This approach is generalized here  associating a pair of integral operators acting directly on particle and
	antiparticle wave spinors in momentum representation to any integral operator in configuration representation, acting 	on mode spinors. This framework allows an effective quantization procedure giving a large set of  one-particle operators  with physical meaning as the spin and orbital parts of the  isometry generators, the  Pauli-Lubanski and position operators or other spin-type operators proposed so far.  A special attention is paid to the operators which mix the particle and antiparticle sectors whose off-diagonal associated operators  have oscillating terms producing Zitterbevegung.  The principal operators of this type including  the usual coordinate operator are derived here for the first time.  As an application, it is shown  that  an apparatus  measuring these new observables may prepare and detect one-particle wave-packets moving uniformly without Zitterbewegung or spin dynamics, spreading  in time normally as any other relativistic  even non-relativistic wave-packet.  

\end{abstract}


\newpage
\tableofcontents

\section{Introduction}

In the relativistic quantum mechanics (RQM) of Dirac's field one considers traditionally  the usual coordinate operator  which is affected by Zitterbewegung \cite{Th,Zit1,Zit2} and the Pauli-Dirac spin operator  whose components  generate the  rotations of the Dirac representation  of the $SL(2, \mathbb{C})$ group \cite{Th} but which  are not conserved.   For this reason  many authors struggled to  find a suitable {\em conserved} spin operator \cite{Fr,B,FW,FG,Ch,Cz}  giving rise to a rich literature (see for instance Refs. \cite{A0,A,A1,A2} and the literature indicated therein). As this problem remains of actual interest \cite{Ch1,Ch2,Ch3,Ch4}, we tried to do the next step to quantization \cite{Cot}. We observed thus that the spin operator we need  is known from long time being proposed  by Pryce in momentum representation (MR) according to his hypothesis (e) \cite{B}.   In fact Pryce studied the relativistic mass-center operator analyzing many possible definitions among them the versions (c), (d) and (e) are of interest in Dirac's theory.  Each version gives its own specific angular momentum  related to a convenient spin operator  assuring the conservation of the total angular momentum. The hypothesis Pryce(e) is the unique version with correct physical meaning giving a would-be mass-center vector-operator with commuting components related to a conserved spin one whose components  generate a $su(2)$ algebra.

Foldy and Wouthuysen  have shown later that their famous transformation   \cite{FW}   leads to the Newton-Wigner  representation \cite{NW}  in which the Dirac Hamiltonian is diagonal while  the Pryce(e) spin and position operators become the aforementioned usual ones.   Apart the Pauli-Dirac and Pryce(e) spin operators other versions  were proposed   by Frenkel \cite{Fr}, Pryce(c) and Czochor \cite{B,Cz}, Fradkin and Good \cite{FG} and  Chakrabarti \cite{Ch}.  Among them, only the components of Pauli-Dirac and Chakrabarti spin operators  generate $su(2)$ algebras but these operators are not conserved. In contrast, the operators proposed by  Frenkel,  Pryce(c)-Czochor and Fradkin-Good  are conserved but their components do not close $su(2)$ algebras. For this reason we say that these are spin-type operators.  

We understood the role of  the Pryce(e) spin operator studying the symmetry of Pauli polarization spinors which define the  fermion polarization. These spinors  enter in the structure of the plane wave solutions of the Dirac equation  that form the basis of  mode (or fundamental) spinors.  Technically speaking,  the fermion polarization depends on the direction of spin projection that can be chosen arbitrarily. When this direction  depends on momentum,  as in the case of the largely used momentum-helicity basis,  we say that the polarization is {\em peculiar}. Otherwise we have  a {\em common}  polarization, independent on momentum as, for example,  in the momentum-spin basis defined in Ref. \cite{BDR}. In both these cases the polarization spinors offer us the degrees of freedom of the new $SU(2)$ {\em spin} symmetry we need for constructing a spin operator conserved via Noether's theorem \cite{Cot}.  However, this symmetry  was neglected so far  because of the difficulties in finding suitable operators  in configuration (or coordinates) representation  (CR)  able to transform only the polarization spinors in MR without affecting other quantities. Fortunately, we have found a spectral representation of a class of integral operators allowing us to define the action of the little group $SU(2)$ upon the polarization spinors \cite{Cot} showing  that the generators of these transformations are  the components of the conserved spin operator whose Fourier transform  is just the operator proposed by the version Pryce(e)  (see the third of Eqs. 6.7 of Ref. \cite{B}).   In this new framework  we defined the operator of fermion polarization  and study how the principal operators of  Dirac's theory depend on polarization trough new  momentum-dependent Pauli-type matrices and  covariant momentum derivatives \cite{Cot}. 

This was a crucial step to quantization allowing us to derive the principal one-particle operators of  quantum field theory (QFT).  The would-be mass-center operator of the version Pryce(e)  becomes after quantization  the time-dependent {\em dipole} one-particle operator whose velocity is the conserved part of the Dirac current (unaffected by Zitterbewegung) called often the classical current  \cite{Z1,Z2} and referred here as the {\em conserved} current.  Quantifying, in addition, the spin, and polarization operators  as well  the isometry generators for any polarization we outlined a coherent version of Dirac's QFT.  \cite{Cot}. 

In this paper we would like to continue and complete this study improving the general formalism in order to eliminate  the difficulties that impeded one to approach to the aforementioned results for more than seven decades. In our opinion the principal impediment was the manner in which the action of the integral operators of  RQM was considered so far. The  Dirac free fields in CR can be expanded in terms of  particle and antiparticle Pauli wave spinors in MR in a basis of Dirac's mode spinors. The  matrix, differential or integral operators  act directly on the  mode spinors.  The difficulties arise because of some integral operators which have  complicated  actions that cannot be manipulated or interpreted, as happened in the case of all Pryce's operators. The solution is to  associate to each integral operator in CR a pair of integral operators acting in MR on the particle and respectively antiparticle wave spinors, without affecting the basis of mode spinors.  In this manner the kernels of the integral operators in CR can be related to those of the associated operators in MR through spectral representations generalized here to a large class of integral operators. We obtain thus  a friendly approach in which we may study and interpret the principal integral operators of RQM  taking a decisive option to quantization. 

In view of the above arguments we would like to present here an extended review of the operators of Dirac's theory following three major objectives. The first one is to improve the entire formalism focusing on the theory of integral operators acting on the wave spinors.   The second objective is to develop and complete the quantum theory outlined in Ref. \cite{Cot} studying the entire collection of operators with physical meaning  of Dirac's QFT derived from the  operators  of RQM proposed till now, including the  operators  having oscillating terms producing Zitterbewegung.   Finally, we would like to present for the first time an example of Dirac's wave-packet prepared and detected by an apparatus able to measure the new Pryce's spin and position operators, laying out the image of a natural smooth propagation without Zitterbewegung or spin dynamics.  

We start in the  next section with the  Dirac theory in CR and MR presenting our framework and defining explicitly the new spin and orbital symmetries in CR before considering the solutions in MR where the mode spinors are constructed according to Wigner's method allowing us to point out the role of the polarization spinors. In the next section we present the equal-time and Fourier integral operators acting  on the mode spinors through their kernels. We pay a special attention to the operators proposed by Pryce  but without neglecting the other historical proposals of spin or spin-type operators \cite{Fr,FW,FG,Ch,Cz}. 

Section 3 is devoted to our principal technical improvement of the operator theory, namely the method of associated operators, relating the operators acting on fields to pairs of operators acting  directly on the Pauli wave spinors  in MR we call associated operators.  The operators which  do not mix particle and antiparticle wave functions are said  reducible, otherwise these being irreducible. We show that the irreducible operators have associated operators whose off-diagonal kernels, between particle and antiparticle wave functions,  oscillate with high frequency.  Fortunately, the principal operators we need are  reducible, without oscillating terms.  We derive and study the associated operators to the spin, position,   polarization and Pauli-Lubanski ones, paying a special attention  to the   isometry generators of the covariant representation of the Dirac field in CR which is equivalent to a pair of associated Wigner's induced representations in MR \cite{Wig,Mc,W}. Remarkably, in our approach we can show that the spin part of the rotation generators of the covariant representation are just the components of Pryce's spin operator in CR  associated to the spin parts of the rotation generators of Wigner's representations defined in MR.  In addition, we study the conserved spin-type operators proposed by Frankel, Pryce(c)-Czogor and Fradkin-Good analyzing their algebraic properties.  In the last subsection we generalize the spectral representations defined in Ref. \cite{Cot} expressing the kernels of  the integral operators acting in CR in terms of kernels of associated operators defined in MR. This method allows us to focus especially on the  principal non-Fourier operators  whose kernels in MR are momentum derivatives of  Dirac's $\delta$-distributions of complicated arguments. 

The previous results prepare the quantization  presented in section 5 where we apply  the Bogolyubov quantization method \cite{Bog}, transforming the expectation values of RQM in operators of QFT. We find that the reducible operators of RQM become after quantization one-particle operators  we divide in even and odd operators according to the relative sign between the particle and antiparticle terms (i. e. the charge  parity).  We define the operators of unitary transformations under isometries giving the general calculation rules and  we study  the algebra of principal observables generated by the  reducible operators of RQM. The last subsection is devoted to the quantization of the irreducible operators having  oscillating terms. The  new results presented here are the operators of QFT corresponding to the traditional Pauli-Dirac spin and coordinate operators of RQM, that can be related to the vector or axial currents, and other interesting operators as the Chakrabarti spin operator and the generators of the Foldy-Wouthuysen transformations.   

Turning back to RQM but now as a one-particle restriction of QFT, we consider in section 6  wave-packets prepared and detected by two different observers. We present first the general theory assuming that the detector filters  momenta oriented along the direction source-detector such that this  measures a one-dimensional wave-packet governed by radial observables. An example of isotropic wave-packet  is worked out  showing that this has an inertial motion spreading in time just as other  scalar or even non-relativistic wave-packets, without Zitterbewegung or spin dynamics \cite{Pack}.

The concluding remarks are presented in section 7 while in four Appendices we present successively the Dirac representation of the $SL(2,\mathbb{C})$ group,  the commutation relations of the  algebra of associated operators in MR, the Pryce(c) and (d) position operators  and the examples of peculiar and common fermion polarizations we know. 

\section{Dirac's free field}

In special relativity the covariant free fields \cite{BDR,KH} are defined in  Minkowski's space-time $M$ having the  metric $\eta={\rm diag}(1,-1,-1,-1)$ and Cartesian coordinates $x^{\mu}$ labeled by Greek indices  ($\alpha,\,\beta,...\mu,\,\nu...=0,1,2,3$). These fields transform covariantly under  Poincar\' e  isometries, $(\Lambda,a): x\to x'=\Lambda x+a$, that form the  group  $P_{+}^{\uparrow} ={T}(4)\,\circledS\, {L}_{+}^{\uparrow}$ \cite{WKT} constituted by the transformations $\Lambda \in {L}_{+}^{\uparrow}$ of the  orthochronous proper Lorentz group,  preserving the metric $\eta$, and the four dimensional translations, $a\in  \mathbb{R}^4$ of the invariant  subgroup $T(4)$. For the fields with half integer spins one considers, in addition,  the universal covering group of the Poincar\' e one,  $ \bar{ P}^{\uparrow}_{+}={T}(4)\,\circledS\, SL(2,\mathbb{C})$, formed by the mentioned translations and transformations  $\lambda\in SL(2,\mathbb{C}) $ related to  those of the group  ${L}_{+}^{\uparrow}$ through the canonical homomorphism, $\lambda\to \Lambda(\lambda)\in {L}_{+}^{\uparrow}$ \cite{WKT} obeying the condition (\ref{canh}). In this framework the covariant fields with spin can be defined on $M$ with values in vector spaces carrying  reducible finite-dimensional representations of the   $SL(2,\mathbb{C}) $ group where invariant Hermitian  forms can be defined.

\subsection{Lagrangian theory and its symmetries}

The Dirac field $\psi:M\to {\cal V}_D$ takes values in the space of Dirac spinors ${\cal V}_D={\cal V}_P\oplus{\cal V}_P$ which is the orthogonal sum of two spaces of Pauli spinors, ${\cal V}_P$, carrying the irreducible representations $(\frac{1}{2},0)$ and $(0,\frac{1}{2})$ of the $SL(2,\mathbb{C})$ group. These form  the Dirac representation  $\rho_ D=(\frac{1}{2},0)\oplus(0,\frac{1}{2})$ where one may define the Dirac  $\gamma$-matrices and the invariant Hermitian form $\overline{\psi }\psi$ with the help of  the Dirac adjoint $\overline{\psi}=\psi^+\gamma^0$ of $\psi$ (see details in the Appendix A). The fields $\psi$  and   $\overline{\psi}$  are the canonical variables of the action 
\begin{equation}\label{actionD}
	{\cal S}[\psi,\overline{\psi}]=\int d^4 x {\cal L}_D(\psi, \overline{\psi})\,,
\end{equation}
defined by the Lagrangian density, 
\begin{equation}
	{\cal L}_D(\psi,\overline{\psi})=\frac{i}{2}\,[\overline{\psi}\gamma^{\alpha}\partial_{\alpha}\psi-
	(\overline{\partial_{\alpha}\psi})\gamma^{\alpha}\psi] - 	m\overline{\psi}\psi\,,
\end{equation}
depending on the mass $m\not=0$ of the Dirac field. This action gives rise to the Dirac equation $E_D\psi=(i\gamma^{\mu}\partial_{\mu}-m)\psi=0$ that can be put in Hamiltonian form,   
\begin{equation}\label{hamD}
	i\partial_t\psi(x)=H_D\psi(x)\,, \quad 	H_D=-i\gamma^0\gamma^i\partial_i+m\gamma^0\,.	
\end{equation}  	
In other respects,  the conservation of the  electric charge via Noether's theorem  \cite{BDR,KH} suggests the form of the Dirac relativistic scalar product 
\begin{equation}\label{sp}
	\langle\psi,\psi'\rangle_D=\int d^3x\overline{\psi}(x)\gamma^0\psi'(x)=\int d^3x \psi^ {+} (x)\psi'(x)	\,.
\end{equation}
related to. We denote  by ${\cal F}=\{\psi\,|\, E_D\psi=0\}$ the space of {\em free fields}  which can be organized  as a rigged Hilbert space by using the Dirac scalar product. 

The action (\ref{actionD}) is invariant under the transformations of the well-known symmetries, namely  the Poincar\' e  isometries and the  $U(1)_{\rm em}$ transformations of the electromagnetic gauge.  The Dirac field transforms   under isometries  according to the {\em covariant} representation  ${T}   \,:\,(\lambda, a)\to { T}_{\lambda,a}\in {\rm Aut}({\cal F})$ of the group $ \tilde{P}^{\uparrow}_{+}$ as \cite{WKT}
\begin{equation}\label{TAa}
	({T}_{\lambda,a}\psi  )(x)
	=\lambda \psi  \left(\Lambda(\lambda)^{-1}(x-a)\right)\,,
\end{equation}
generated by  the basis-generators of the  corresponding representation of  the algebra ${\rm Lie}({T})$ that read
\begin{eqnarray}\label{Pgen}
	P_{\mu}=-\left.i\frac{\partial {T}_{1,a}}{\partial a^{\mu}}\right|_{a=0}\,, \quad 
	J_{\mu\nu}=\left.i\frac{\partial {T}_{\lambda(\omega),0}}{\partial \omega^{\mu\nu}}\right|_{\omega=0}\,.
\end{eqnarray}
For pointing out the physical meaning of these generators one separates the momentum components,  $ P^i=-i\partial_i$, and the energy  operator,  $ H=P_0=i\partial_t$,  denoting  the $SL(2,{\Bbb C})$ generators as, 
\begin{eqnarray}
	J_i  &=&\frac{1}{2}\,\varepsilon_{ijk}	J  _{jk}=
	-i\varepsilon_{ijk}\underline{x}^j\partial_k+s_i  \,,\label{J}\\
	K_i  &=& J_{0i}  =i (\underline{x}^i
	\partial_t+t\partial_i)+s_{0i}  \,, \label{K}
\end{eqnarray}
where $\underline{x}^i$ are the components of  the  {\em coordinate} vector-operator $\underline{\vec x}$  acting as $(\underline{x}^i \psi)(x)=x^i\psi(x)$. The reducible matrices $s_i$ and $s_{0i}$ are given by Eqs. (\ref{si}) and respectively (\ref{s0i}). The operators  $\{ H,  P^i, J  _i, K  _i\}$ form the usual basis of the Lie algebra ${\rm Lie}({T})$ of  the representation (\ref{TAa}) \cite{WKT}.  

The scalar product (\ref{sp}) helps us to write simply the quantities conserved via Noether theorem as expectation values, $\langle \psi,X \psi\rangle_D$, of the generators of the symmetry transformations $\psi\to T\psi=\psi -i \xi X\psi+...$ which leave invariant the action (\ref{actionD}) and implicitly the scalar product,  
$\langle {T}\psi  ,{T}\psi'  \rangle_D=\langle\psi ,\psi' \rangle_D$. Hereby we deduce that   the generators  $X$  are self-adjoint obeying
\begin{eqnarray}
	\langle\psi, X^+\psi'  \rangle_D =\langle X\psi  ,\psi'  \rangle_D=\langle\psi, X\psi'  \rangle_D \,.	
\end{eqnarray}
Therefore, we may conclude that   the covariant representation (\ref{TAa}) is {\em unitary} with respect to the relativistic scalar product  (\ref{sp}).

The above operators may generate freely new ones as, for example, the  Pauli-Lubanski pseudo-vector \cite{WKT},
\begin{equation}\label{PaLu}
	W^{\mu}=-\frac{1}{2}\,\varepsilon^{\mu\nu\alpha\beta}P_{\nu}  J_{\alpha\beta}  \,,
\end{equation}
having the components
\begin{equation}\label{PaLu1}
	W^0={ J}  _i{ P}^i={s} _i{P}^i\,,\quad  W ^i= H\,{ J}_i  +
	\varepsilon_{ijk}{ P} ^j { K}_k\,,
\end{equation}
as we take  $\varepsilon^{0123}=-\varepsilon_{0123}=-1$. This  operator is  considered by many authors as the covariant four-dimensional spin operator  as long as $W_0$ is just the helicity operator \cite{C}. Moreover, this gives rise to the second Casimir operator of the pair \cite{Th}
\begin{eqnarray}
	{C}_1&=& P_{\mu} P^{\mu}\sim m^2\,,\label{CM1}\\
	{C}_2  &=& W^{\mu}	W_{\mu} \sim -m^2 s(s+1)\,,\quad s=\textstyle{ \frac{1}{2}}\,,\label{CM2}
\end{eqnarray}
whose eigenvalues depend on the invariants $(m,s)$ determining  the representation ${T}$. 

The subgroup $SU(2)\subset SL(2,\mathbb{C})$ will play here a special role in studying the spin operator. For this reason  we consider  the restriction of the covariant representation ${T}$ to this subgroup, ${ T^r}\equiv { T}|_{SU(2)}$, such that ${ T}_{r,0}={T}^r_{\hat r}$ for any $\hat r\in SU(2)$ or $r={\rm diag}(\hat r,\hat r)\in \rho_D$. The basis-generators of the representation ${T^r}$ are the components of the total angular momentum operator ${\vec J}=\underline{\vec x}\land{\vec P} +{\vec s}$, defined by Eq. (\ref{J}), which is formed by the orbital term $\underline{\vec x}\land{\vec P}$ and  the  Pauli-Dirac spin matrix ${\vec s}$. However,  as mentioned before, these operators are not conserved separately such that we must look for a new conserved spin operator ${\vec S}$ related to a suitable new position operator, ${\vec X}=\underline{\vec x}+\delta{\vec X}$, allowing the  new splitting
\begin{equation}\label{spli}
	{\vec J}=\underline{\vec x}\land{\vec P} +{\vec s}={\vec L}+{\vec S}	\,,\qquad {\vec L}={\vec X}\land {\vec P}\,,
\end{equation}
which impose  the correction $\delta{\vec X}$ to satisfy $\delta{\vec X}\land {\vec P}={\vec s}-{\vec S}$. This new splitting gives rise to a pair of  new $su(2)\sim so(3)$ symmetries, namely the {\em orbital} symmetry generated by $\{L_1,L_2,L_3\}$ and the {\em spin} one generated by $\{S_1,S_2,S_3\}$. Moreover, we have shown that the  Fourier transforms of the operators ${\vec S}$ and $\delta{\vec X}$ are just the Pryce(e) operators  \cite{Cot}.

It is known that  for writing down the plane wave solutions of the Dirac equation we must chose the same orthonormal  basis of polarization spinors $\xi=\{\xi_{\sigma}| \sigma=\pm\frac{1}{2}\}$ in both the spaces  ${\cal V}_P$ of Pauli spinors carrying the irreducible representations $(\frac{1}{2},0)$ and $(0,\frac{1}{2})$ of $\rho_D$. Because the polarization spinors are free parameters we may consider the Dirac field as $\psi: M\times {\cal V}_P\to {\cal V}_D$ denoting it explicitly   by $\psi_{\xi}$ instead of $\psi$. The basis of polarization spinors  can be changed at any time, $\xi\to \hat r\xi$, by  applying a rotation $\hat r \in SU(2)$ which change the form of the Dirac spinor  giving rise to the new representation ${T}^s:\hat r\to {T}^s_{\hat r}$ of the group $SU(2)$ which encapsulates the  spin symmetry.  The operators of this representation have the action     
\begin{equation}\label{Rs}
	\left(	{T}^s_{\hat r(\theta)}\psi_{\xi}\right)(x)=\psi_{\hat r(\theta)\xi}(x)\,,
\end{equation} 
where $\hat r(\theta)$ are the rotations (\ref{r}) with Cayley-Klein parameters. The components of the spin operator can be defined now as the generators of this representation, \cite{Cot}
\begin{equation}\label{Spipi}
	S_i=\left.i\frac{\partial T^s_{\hat r(\theta)}}{\partial \theta^i}\right|_{\theta^i=0} 	~~\Rightarrow~~ S_i\psi_{\xi}=\psi_{\hat s_i\xi}\,,
\end{equation}
whose action is obvious. Similarly, we define here for the first time  the {\em orbital} representation $T^o: \hat r\to T^o_{\hat r}$ as
\begin{equation}\label{Ro}
	\left(T^o_{\hat r(\theta)}\psi_{\xi}\right)(t,{\vec x})=r(\theta)\psi_{\hat r(\theta)^{-1}\xi}\left(t,R[\hat r(\theta)]^{-1}{\vec x} \right)\,,
\end{equation}
for accomplishing the factorization ${T}^r=T^o\otimes T^s$. The  basis-generators of the orbital representation,
\begin{equation}\label{Lpipi}
	L_i=\left.i\frac{\partial T^o_{\hat r(\theta)}}{\partial \theta^i}\right|_{\theta^i=0} \,,
\end{equation} 
are the components of the new conserved orbital angular momentum operator ${\vec L}$. In what follows we shall pay a special attention to the new operators ${\vec S}$,  ${\vec L}$ and ${\vec X}$.

\subsection{Momentum representation}

In MR all quantities are defined on orbits in momentum space, $\Omega_{\mathring p}=\{{\vec p}\,|\, {\vec p}=\Lambda \mathring{p}, \Lambda\in L_{+}^{\uparrow} \}$, that may be built by applying Lorentz transformations on a {\em representative} momentum $\mathring{p}$ \cite{Wig, BW1,Mc}.  In the case of massive particles, we discuss here,  the representative momentum is just the rest frame one,  $\mathring{p}=(m,0,0,0)$.  The rotations that leave $\mathring{p}$ invariant, $\Lambda(r)\mathring{p}=\mathring{p}$, form the {\em stable} group $SO(3)\subset L_{+}^{\uparrow}$ of $\mathring{p}$ whose  universal covering group $SU(2)$  is called the {\em little} group associated to the representative momentum $\mathring{p}$. 

The momenta ${\vec p}\in \Omega_{\mathring{p}}$ may be obtained as ${\vec p}=\Lambda_{{\vec p}}\,\mathring{p}$ by applying  transformations $\Lambda_{{\vec p}}=L_{\vec p}R(r({\vec p}))$ formed by genuine Lorentz boosts and arbitrary rotations $R(r({\vec p}))=\Lambda(r({\vec p}))$  that do not change the representative momentum.   The corresponding transformations  $\lambda_{{\vec p}}\in \rho_D$  which satisfy $\Lambda(\lambda_{{\vec p}})=\Lambda_{{\vec p}}$ and $\lambda_{{\vec p}=0}=1\in \rho_D$ have the form  
\begin{equation}\label{lambda}
	\lambda_{{\vec p}}= l_{\vec p}\, r({\vec p})\,,
\end{equation}
where  the transformations  $l_{\vec p}$ given by Eq. (\ref{Ap}) are related to the genuine Lorentz boosts  $L_{\vec p}=\Lambda(l_{\vec p})$ having the matrix elements (\ref{Lboost}). The invariant measure on the massive orbits, \cite{WKT}
\begin{equation}\label{measure}
	\mu({\vec p})=\mu(\Lambda{\vec p})=\frac{d^3p}{E(p)}\,, \quad \forall \Lambda\in L_{+}^{\uparrow}\,, 	
\end{equation}
is the last tool we need for relating CR and MR.

The general solutions of the free Dirac equation, $\psi\in {\cal F}$, may be expanded in terms of  mode spinors  spinors,  $U_{{\vec p},\sigma}$ and  $V_{{\vec p},\sigma}=C  U_{{\vec p},\sigma}^*$, of positive and respectively negative frequencies,   related through the charge conjugation defined by the matrix $C=C^{-1}=i\gamma^2$.  The mode spinors are particular solutions of the Dirac equation which satisfy  the eigenvalues problems,  
\begin{eqnarray}
	& H U_{{\vec p},\sigma}=E(p) U_{{\vec p},\sigma}\,,\quad ~~
	&	H V_{{\vec p},\sigma}=-E(p) V_{{\vec p},\sigma}\,,\\
	&{ P}^i U_{{\vec p},\sigma}={p}^i\, U_{{\vec p},\sigma}\,,\quad ~~~~~
	&	{ P}^i V_{{\vec p},\sigma}=-{p}^i\, V_{{\vec p},\sigma}\,,
\end{eqnarray}
depending explicitly on the polarization spinors which will be specified later. Therefore, the general solutions of the Dirac equation  are free fields  that can be expanded as \cite{KH,BDR}
\begin{eqnarray}\label{Psi}
	\psi  (x)&=&	\psi  ^+(x)+	\psi  ^-(x)\nonumber\\
	&=&\int d^3p \sum_{\sigma}\left[U_{{\vec p},\sigma}(x) \alpha_{\sigma}({\vec p}) +V_{{\vec p},\sigma}(x) \beta^ {*} _{ \sigma}({\vec p})\right]\,,~~~~
\end{eqnarray}
in terms of spinors-functions $\alpha: \Omega_{\mathring p}\to {\cal V}_P$ and   $\beta: \Omega_{\mathring p}\to {\cal V}_P$ representing  the  particle and respectively antiparticle {\em wave spinors}. The space of free fields ${\cal F}$  can be split  thus in two  subspaces of positive and respectively negative frequencies, ${\cal F}={\cal F}^+\oplus {\cal F}^-$, which are orthogonal with respect to the scalar product (\ref{sp}).

The mode spinors  prepared at the initial time $t_0=0$ by an observer staying at rest in origin  have the general form
\begin{eqnarray}
	U_{{\vec p},\sigma}(x)&=&u_{\sigma}({\vec p})\frac{1}{(2\pi)^{\frac{3}{2}}} \,e^{-iE(p)t+i{\vec p}\cdot{\vec x}}\,,\label{U} \\
	V_{{\vec p},\sigma}(x)&=&v_{\sigma}({\vec p})\frac{1}{(2\pi)^{\frac{3}{2}}}\, e^{iE(p)t-i{\vec p}\cdot{\vec x}}\,,\label{V}
\end{eqnarray} 
where   $v_{\sigma}({\vec p})=C  u^*_{s\sigma}({\vec p})$. According to Wigner's general method   \cite{Wig,BW1,Th} we use  the transformations (\ref{lambda}) and (\ref{Ap}) for writing down the spinors  
\begin{eqnarray}
	u_{\sigma}({\vec p})&=&n({p})\lambda_{{\vec p}}\,  \mathring{u}_{\sigma}=n(p)l_{{\vec p}}r({\vec p})\,  \mathring{u}_{\sigma}\nonumber\\
	&=&n(p)l_{{\vec p}}\,  \mathring{u}_{\sigma}({\vec p})\,,  \label{Ufin} \\  
	v_{\sigma}({\vec p})&=&C  u_{\sigma}^*({\vec p})=n(p)\lambda_{\vec p} \mathring{v}_{\sigma}=n(p)l_{\vec p}r({\vec p}) \mathring{v}_{\sigma}\nonumber\\
	&=&n(p)l_{\vec p} \mathring{v}_{\sigma}({\vec p})\,,   \label{Vfin}
\end{eqnarray}
depending on a normalization factor  satisfying $n(0)=1$.  The rest frame spinors  $\mathring{u}_{\sigma}=u_{\sigma}(0)$ and $\mathring{v}_{\sigma} =v_{\sigma}(0)=C\mathring{u}_{\sigma}^*$  are solutions of the Dirac equation in the rest frame obeying   $\gamma^0\mathring{u}_{\sigma}=\mathring{u}_{\sigma}$ and $\gamma^0\mathring{v}_{\sigma}=-\mathring{v}_{\sigma}$. If these equations are satisfied then  the spinors (\ref{U}) and (\ref{V}) are solutions of the Dirac equation in MR, 
\begin{equation}\label{mucu1}
	(\gamma p-m)u_{\sigma}({\vec p})=0\,, \quad (\gamma p+m)v_{\sigma}({\vec p})=0\,,
\end{equation}
since  $\gamma p=E(p)\gamma^0-{\gamma}^i p^i =m l_{\vec p}\gamma^0 l_{\vec p}^{-1}$. 

Taking into account that the rotations $r({\vec p})$ are arbitrary, we separate the quantities   
\begin{eqnarray}
	\mathring{u}_{\sigma}({\vec p})&=&r({\vec p})	\mathring{u}_{\sigma}=\frac{1}{\sqrt{2}}\left(
	\begin{array}{c}
		\xi_{\sigma}({\vec p})\\
		\xi_{\sigma}({\vec p})
	\end{array}\right)\,,\label{xy}\\
	\mathring{v}_{\sigma}({\vec p})&=&r({\vec p})	\mathring{v}_{\sigma}=\frac{1}{\sqrt{2}}\left(
	\begin{array}{c}
		\eta_{\sigma}({\vec p})\\
		-\eta_{\sigma}({\vec p})
	\end{array}\right)\,,\label{xy1}
\end{eqnarray}
which are eigenspinors of the matrix $\gamma^0$ corresponding to the eigenvalues $1$ and respectively $-1$ as $r({\vec p})$ commutes with $\gamma^0$. These Dirac spinors depend on the related Pauli spinors $\xi_{\sigma}({\vec p})$ and  $\eta_{\sigma}({\vec p})=i\sigma_2 \xi^*_{\sigma}({\vec p})$ we call here the polarization spinors observing that only the spinors $\xi_{\sigma}({\vec p})$ remain arbitrary. The orthogonality and completeness properties of these spinors (presented in the Appendix C)  ensure the normalization of the spinors  (\ref{xy}) and (\ref{xy1}) which give rise to  the complete orthogonal system of projection matrices   
\begin{eqnarray}
	\sum_{\sigma}\mathring{u}_{\sigma}({\vec p})\mathring{u}_{\sigma}^+({\vec p})&=& \sum_{\sigma}\mathring{u}_{\sigma}\mathring{u}_{\sigma}^+   =\frac{1+\gamma^0}{2}\,,\label{ortu}\\
	\sum_{\sigma}\mathring{v}_{\sigma}({\vec p})\mathring{v}_{\sigma}^+({\vec p})&=&	 	
	\sum_{\sigma}	\mathring{v}_{\sigma}\mathring{v}_{\sigma}^+=\frac{1- \gamma^0}{2}\,,\label{ortv}
\end{eqnarray}
on the proper subspaces of the matrix $\gamma^0$. 

Finally, by setting  the  normalization factor in accordance to Eq. (\ref{ullu}),
\begin{equation}\label{Nor}
	n(p)=\sqrt{\frac{m}{E(p)}}\,,
\end{equation}	
we obtain  the orthonormalization,  
\begin{eqnarray}
	\langle U_{{\vec p},\sigma}, U_{{{\vec p}\,}',\sigma'}\rangle_D &=&
	\langle V_{{\vec p},\sigma}, V_{{{\vec p}\,}',\sigma'}\rangle_=	\delta_{\sigma\sigma^{\prime}}\delta^{3}({\vec p}-{\vec p}\,^{\prime})\,,\label{ortU}\\
	\langle U_{{\vec p},\sigma}, V_{{{\vec p}\,}',\sigma'}\rangle_D &=&
	\langle V_{{\vec p},\sigma}, U_{{{\vec p}\,}',\sigma'}\rangle_D =0\,, \label{ortV}
\end{eqnarray}
and completeness,
\begin{eqnarray}
	&&	\int d^3p \sum_{\sigma}\left[ U_{{\vec p},\sigma}(t,{\vec x})U_{{\vec p},\sigma}^{+}(t,{\vec x}\,')+V_{{\vec p},\sigma}(t,{\vec x})V_{{\vec p},\sigma}^{+}(t,{\vec x}\,')\right]=\delta^3({\vec x}-{\vec x}\,')\,,\label{comp}
\end{eqnarray}
of the basis of mode spinors.

Eq. (\ref{Psi}) can be seen now as the expansion of the free field $\psi$ in the basis of mode spinors whose "coefficients" are just the wave spinors
\begin{equation}\label{alpha}
	\alpha=\left( 
	\begin{array}{l}
		\alpha_{\frac{1}{2}}\\
		\alpha_{-\frac{1}{2}} 
	\end{array}\right)	\in \tilde{\cal F}^+\,,
	\quad
	\beta=\left( 
	\begin{array}{llc}
		\beta_{\frac{1}{2}}\\
		\beta_{-\frac{1}{2}} 
	\end{array}\right) 	\in \tilde{\cal F}^-\,,
\end{equation}
which encapsulate the physical meaning of $\psi$. When the filed $\psi$ is known then the wave spinors can be derived applying the inversion formulas
\begin{equation}\label{inv}
	\alpha_{\sigma}({\vec p})=\langle U_{{\vec p},\sigma},\psi \rangle_D\,, \quad 
	\beta_{\sigma}({\vec p})=\langle \psi,  V_{{\vec p},\sigma} \rangle_D\,, 
\end{equation}
resulted from  Eqs. (\ref{ortU}) and (\ref{ortV}). We assume now that the spaces $\tilde{\cal F}^+\sim \tilde{\cal F}^-$ are rigged Hilbert spaces, including Hilbert spaces ${\cal L}^2(\Omega_{\mathring{p}}, d^3p,{\cal V}_P)$,  equipped with the same scalar product, 
\begin{eqnarray}\label{spa}
	\langle \alpha, \alpha'\rangle=\int d^3p \,\alpha^+({\vec p})\alpha'({\vec p})=\int d^3p \sum_{\sigma}\alpha_{\sigma}^*({\vec p})	\alpha_{\sigma}'({\vec p})\,,
\end{eqnarray}
and similarly for the spinors $\beta$. Then after using  Eqs.  (\ref{ortU}) and (\ref{ortV}) we obtain  the important identity 
\begin{equation}\label{spp}
	\langle \psi  , \psi  '\rangle_D =\langle \alpha,\alpha'\rangle + \langle \beta,\beta'\rangle	\,,
\end{equation}
expressing the Dirac scalar product in terms of wave spinors. We remind the reader that when $\langle \psi  , \psi  \rangle_D=1$ then the quantities $|{\alpha_{\sigma}({\vec p})}|^2$ and $|{\beta_{\sigma}({\vec p})}|^2$ are the densities of probability  in momentum space of a particle and respectively antiparticle of polarization $\sigma$.  

\section{Operators of Dirac's theory}

The observables  of Dirac's RQM are linear operators acting on the space of free fields, $A,\,B,...\in {\rm Aut}({\cal F})$,  that must be self-sdjoint with respect to the scalar product (\ref{sp}). Apart from the familiar multiplicative and differential operators there are integral operators that deserve a special attention.

\subsection{From differential to integral operators}

The differential operators are  $4\times 4$ matrices depending on derivatives, $f(i\partial_{\mu})\in \rho_D$, whose action on the mode spinors  
\begin{eqnarray}
	\left[f(i\partial_{\mu}) \psi\right](x)&=&\int d^3p \sum_{\sigma}\left[  f({p^{\mu}})U_{{\vec p},{\sigma}}( x)\alpha_{\sigma}({\vec p})+ f(-{p^{\mu}})V_{{\vec p},{\sigma}}(x)\beta^*_{\sigma}({\vec p})\right]\,,\label{F2}
\end{eqnarray}
is given by the momentum-dependent  matrices $f(p^{\mu})$.  The principal differential operators are the translation generators $P_{\mu}=i\partial_{\mu}$, the operator of Dirac equation and implicitly the Dirac Hamiltonian (\ref{hamD}). However, there are important operators as those proposed by Pryce which are integral operators that cannot be reduced to differential ones. 

In general, the integral operators, $Z: {\cal F}\to {\cal F}$, have the action  
\begin{eqnarray}
	(Z\psi)(x)=\int d^4 x' {\frak Z}(x,x')\psi(x')\,,
\end{eqnarray}
defined by their  kernels ${\frak Z}: M\times M\to \rho_D$ denoted here by the corresponding Fraktur symbol, e. g. $Z\to {\frak Z}$. These operators are linear forming an algebra in which the multiplication, $Z=Z_1Z_2$, is defined by the composition rule of the corresponding kernels, 
\begin{eqnarray}
	{\frak Z}(x,x')=\int d^4x"{\frak Z}_{1}(x,x"){\frak Z}_{2}(x",x')\,.
\end{eqnarray} 
The identity operator $I$ of this algebra  acting as $(I\psi)(x)=\psi(x)$  has the kernel ${\frak I}(x,x')=\delta^4(x-x')$.  For any integral operator $Z$ we may write the Dirac bracket at the given time $t$ as
\begin{eqnarray}
	\langle \psi,Z\psi'\rangle_D|_t=\int d^3x\, d^4x' \psi^+(t,{\vec x}){\frak Z}(t,{\vec x},x')\psi(x')\,, 
\end{eqnarray} 
integrating only on the space coordinates ${\vec x}$.  The multiplicative or differential operators are particular cases of integral ones. For example, the derivatives $\partial_{\mu}$ can be seen as  integral operators having the kernels $\partial_{\mu}\delta^4(x)$. In general, the operators having  kernels depending on $t$ and $t'$ or only on $t-t'$  play the role of  {\em propagators}.

For describing usual observables it is enough to consider {\em equal-time operators}, $A$,  whose kernels of the form 
\begin{eqnarray}
	{\frak A}(x,x')=\delta(t-t'){ \frak A}(t, {\vec x},{\vec x}^{\,\prime})\,,
\end{eqnarray}  
define the operator action
\begin{equation}\label{Y1}
	(A \psi)(t,{\vec x})=\int d^3x' {\frak A}(t,{\vec x},{\vec x}^{\,\prime})\psi(t,{\vec x}^{\,\prime})\,,
\end{equation}
preserving  the time. The operator multiplication takes over this property,  
\begin{eqnarray}
	A&=&A_1A_2	~\Rightarrow~{\frak A}(t,{\vec x},{\vec x}^{\,\prime})=\int d^3x'' {\frak A}_1(t,{\vec x},{{\vec x}\,}''){\frak A}_2(t,{{\vec x}\,}'',{\vec x}^{\,\prime})\,,~~~~
\end{eqnarray}
which means  that the set of equal-time operators forms an algebra, $E[t] \subset {\rm Aut}({\cal F})$, at any fixed time $t$. The expectation values  of these operators  at a given time $t$, 
\begin{equation}
	\left.	\langle \psi, A\psi'\rangle_D\right|_t=\int d^3x\,d^3x'\psi^+(t,{\vec x}){\frak A}(t,{\vec x},{\vec x}^{\,\prime})\psi'(t,{\vec x}^{\,\prime})\,,
\end{equation}
are  dynamic quantities evolving in time as  
\begin{eqnarray}\label{DA}
	\left.\partial_t 	\langle \psi, A\psi'\rangle_D\right|_t&=&\left.	\langle \psi,  dA\psi'\rangle_D\right|_t \nonumber\\ 
	dA&=&\partial A+i[H_D,A]	\,,
\end{eqnarray} 
where $dA$ plays the role of total time derivative assuming that the new operator $\partial A$ has the action
\begin{equation}\label{dY1}
	(\partial A \psi)(t,{\vec x})=\int d^3x' \partial_t{\frak A}(t,{\vec x},{\vec x}^{\,\prime})\psi(t,{\vec x}^{\,\prime})\,.
\end{equation}
As mentioned before, we say that an operator is conserved if its expectation value is independent on time. This means that an  equal-time operator $A$ is conserved if and only if this  satisfies $dA=0$.  We get thus a tool allowing us to identify the conserved operators without resorting to Noether's  theorem.

A special subalgebra, $F[t]\subset E[t]$, is formed by Fourier  operators with local kernels, ${\frak A}(t,{\vec x},{\vec x}')= {\frak A}(t,{\vec x}-{\vec x}')$, allowing  three-dimensional Fourier representations, 
\begin{equation}\label{KerY0}
	{\frak A}(t,{\vec x}) =\int d^3p\,\frac{e^{i {\vec p}\cdot{\vec x}}}{(2\pi)^3}  {\hat A}(t,{\vec p})\,,
\end{equation} 
depending on the matrices  $\hat A(t,{\vec p})\in{\rho}_D$  we call here the Fourier transforms of the operators $A$.  Then the action (\ref{Y1}) on a field  (\ref{Psi})  can be written as
\begin{eqnarray}
	(A \psi)(t, {\vec x})&=&	\int d^3x'\, {\frak A}(t,{\vec x}-{\vec x}^{\,\prime})\psi(t,{\vec x}^{\,\prime})\nonumber\\
	&=&\int d^3p \sum_{\sigma}\left[\hat A(t,{\vec  p})U_{{\vec p},{\sigma}}(t, {\vec x})\alpha_{\sigma}({\vec p})+\hat A(t,-{\vec  p})V_{{\vec p},{\sigma}}(t, {\vec x})\beta^*_{\sigma}({\vec p})\right]\,.\label{Y2}
\end{eqnarray}
One can verify that a Fourier operator $A$ is self-adjoint  with respect to the scalar product (\ref{sp}) if its Fourier transform is a Hermitian matrix, $\hat A(t,{\vec  p})=\hat A(t,{\vec  p})^+$. 

In the $F[t]$ algebra, the operator multiplication, $A=A_1A_2$,  is given by the convolution of the corresponding kernels, ${\frak A}={\frak A}_1*{\frak A}_2$, defined as
\begin{eqnarray}
	{\frak A}(t, {\vec x}-{\vec x}^{\,\prime})=\int d^3x''{\frak A}_1(t, {\vec x}-{{\vec x}\,}''){\frak A}_2(t, {{\vec x}\,}''-{\vec x}^{\,\prime})\,,	
\end{eqnarray} 
which leads to  the multiplication,  $\hat A(t,{\vec p})=\hat A_1(t,{\vec p})\hat 
A_2(t,{\vec p})$, of the  Fourier transforms. One obtains thus the new algebra $\hat F[t]$ in MR, formed by the Fourier transforms of the Fourier operators, in which the identity is the matrix $\hat I({\vec p})=1\in\rho _D$. Obviously, the operator $A\in F[t]$ is invertible if its Fourier transform is invertible in $\hat F[t]$. 

As there are many equal-time or Fourier operators whose kernels are independent on time we denote their algebras by  $F[0]\subset E[0]$ observing that the time-independent  Fourier transforms of the operators of the $F[0]$ algebra constitute the algebra $\hat F[0]$.  An  example is the Dirac Hamiltonian  (\ref{hamD}) whose Fourier transform 
\begin{equation}\label{HDp}
	\hat H_D({\vec p})=m\gamma^0+\gamma^0{\vec\gamma}\cdot {\vec p} \in \hat F[0]\,,
\end{equation}
acts as 
\begin{eqnarray}
	{\hat H}_D({\vec p})U_{{\vec p},\sigma}(x)&=&E(p) U_{{\vec p},\sigma}(x)\,, \\
	{\hat H}_D(-{\vec p})V_{{\vec p},\sigma}(x)&=&-E(p) V_{{\vec p},\sigma}(x)	\,.\label{HDU}
\end{eqnarray}
Other elementary examples are the momentum-independent matrices of $\rho_D$, $\gamma^{\mu}$, $s_{\mu\nu}$,...etc.  that can be seen as Fourier operators whose Fourier transforms are just the matrices themselves. 

During the last century many authors  preferred to work in $\hat F[0]$  algebra manipulating exclusively the time-independent  Fourier transforms of the operators under consideration.  In this manner Pryce proposed his versions (c), (d) and (e) of related spin and position operators and  a complete set of orthogonal projection operators, defining their Fourier transforms \cite{B}.  In the same paper Pryce proposed a transformation which differs only through a parity from the famous  Foldy-Wouthuysen transformation proposed two years later \cite{FW} whose action remain exclusively at the level of the $\hat F[0]$ algebra. 

\subsection{Diagonal and oscillating terms}

The Pryce projection operators,  $\Pi_{\pm}\in F[0]$,  are defined by their Fourier transforms from $\hat F[0]$ that read
\begin{eqnarray}
	\hat\Pi_+({\vec p})&=&\frac{m}{E({p})}l_{\vec p}	\frac{1+\gamma^0}{2} l_{\vec p}=\frac{1}{2}\left(1+\frac{\hat H_D({\vec p})}{E(p)}\right)\,,\label{Pip}\\ 
	\hat\Pi_-({\vec p})&=&\frac{m}{E({p})}l^{-1}_{\vec p}	\frac{1-\gamma^0}{2} l^{-1}_{\vec p}=\frac{1}{2}\left(1-\frac{\hat H_D({\vec p})}{E(p)}\right)\,,~~~~\label{Pim}
\end{eqnarray}
where $\hat H_D({\vec p})$, defined by Eq. (\ref{HDp}), can be put now in the form
\begin{equation}\label{HDp1}
	\hat H_D({\vec p})=E(p)\left[\hat \Pi_+({\vec p})-\hat\Pi_-({\vec p})\right]	\,.
\end{equation}
Moreover, according to Eqs. (\ref{HDU})  we verify that  
\begin{eqnarray}
	( \Pi_+ U_{{\vec p},\sigma})(x)&=&\hat\Pi_+({\vec p})U_{{\vec p},\sigma}(x)=U_{{\vec p},\sigma}(x)\,,\nonumber\\
	( \Pi_- U_{{\vec p},\sigma})(x)&=&\hat\Pi_-({\vec p})U_{{\vec p},\sigma}(x)=0\,,\nonumber	\\	
	( \Pi_+ V_{{\vec p},\sigma})(x)&=&\hat\Pi_+(-{\vec p})V_{{\vec p},\sigma}(x)=0\,,\nonumber\\
	( \Pi_- V_{{\vec p},\sigma})(x)&=&\hat\Pi_-(-{\vec p})V_{{\vec p},\sigma}(x)=V_{{\vec p},\sigma}(x)\,,\nonumber
\end{eqnarray}  
concluding that the operators $\Pi_+=\Pi_+^2$ and $\Pi_-= \Pi_-^2$  satisfy $ \Pi_+\Pi_-=\Pi_-\Pi_+=0$ and $\Pi_+ +\Pi_-=I$ forming thus a complete system of  orthogonal projection operators. With their help one may separate the subspaces  of positive and negative frequencies,  $\Pi_+{\cal F}={\cal F}^+$ and  $\Pi_-{\cal F}={\cal F}^-$ \cite{B}. These projection operators allow us to define the new operator $N\in F[0]$ giving its  Fourier transform, 
\begin{eqnarray}
	\hat	N({\vec p})&=&\hat\Pi_+({\vec p})-\hat\Pi_-({\vec p})=\frac{\hat H_D({\vec p})}{E(p)}~ 
	\Rightarrow~	\hat N^2({\vec p})=1\in \rho_D ~\Rightarrow~ N^2=I\,.,\label{Num}
\end{eqnarray}
but postponing its interpretation that has to be discussed later. 

The Pryce projection operators help us to study how an operator $A\in E[t]$ acts on the orthogonal subspaces of ${\cal F}={\cal F}^+\oplus{\cal F}^-$  resorting to the expansion   
\begin{eqnarray}\label{AAAA0}
	A&=&A^{(+)}+A^{(-)}+A^{(\pm)}+A^{(\mp)}	\nonumber\\
	&=&\Pi_+A\Pi_+ +\Pi_-A\Pi_- +\Pi_+A\Pi_- +\Pi_-A\Pi_+ \,,
\end{eqnarray}
suggested by Pryce \cite{B} and written here  in a self-explanatory notation. When $A$ is a Hermitian operator then we have 
\begin{equation}\label{HAHA}
	\left[ A^{(+)}\right]^+=A^{(+)}\,, \quad \left[ A^{(-)}\right]^+=A^{(-)}\,, \quad \left[ A^{(\pm)}\right]^+=A^{(\mp)}\,.	
\end{equation}
The first two terms form the {\em  diagonal} part of $A$, denoted by $A_{\rm diag}=A^{(+)}+A^{(-)}$,  which does not mix the subspaces ${\cal F}^+$ and ${\cal F}^-$ among themselves.  The off-diagonal terms, $A^{(\pm)}$ and $A^{(\mp)}$, are nilpotent operators changing the sign of frequency. Under such circumstances we adopt the following definition: an equal-time operator $A\in E[t]$ is said {\em reducible} if $A=A_{\rm diag}$ as  $A^{(\pm)}=A^{(\mp)}=0$.  Otherwise the operator is irreducible having off-diagonal terms. 

In the case of  time-dependent Fourier operators $A\in F[t]$  the expansion (\ref{AAAA0}) gives the equivalent expansion of  the Fourier transforms in $\hat F[t]$ algebra that reads
\begin{eqnarray}
	\hat A(t,{\vec p})&=&\hat A^{(+)}(t,{\vec p})+\hat A^{(-)}(t,{\vec p})+\hat A^{(\pm)}(t,{\vec p})+\hat A^{(\mp)}(t,{\vec p})	\nonumber\\
	&=&\hat\Pi_+({\vec p})\hat A(t,{\vec p})\hat\Pi_+({\vec p}) +\hat\Pi_-({\vec p})\hat A(t,{\vec p})\hat\Pi_-({\vec p})\nonumber\\
	&+&\hat\Pi_+({\vec p})\hat A(t,{\vec p})\hat\Pi_-({\vec p}) +\hat\Pi_-({\vec p})\hat A(t,{\vec p})\hat\Pi_+({\vec p}) \,.~~~\label{AAAA}
\end{eqnarray}
In addition, we observe that the total time derivative  (\ref{DA}) acts on the Fourier transforms of the operator $A$ as 
\begin{equation}
	d\hat A(t,{\vec p})=\partial_t \hat A(t,{\vec p}) +i\left[\hat H_D({\vec p}), \hat A(t,{\vec p})\right]\,.
\end{equation} 
Taking into account that the operator (\ref{HDp1}) depends on Pryce's projection operators we can calculate the following  commutators
\begin{eqnarray}
	\left[\hat H_D({\vec p}), \hat A^{(+)}(t,{\vec p})\right]&=&\left[\hat H_D({\vec p}), \hat A^{(-)}(t,{\vec p})\right]=0\,,	\\
	\left[\hat H_D({\vec p}),\hat  A^{(\pm)}(t,{\vec p})\right]&=&2E(p)\hat A^{(\pm)}(t,{\vec p})\,,\label{HDApm}\\
	\left[\hat H_D({\vec p}), \hat A^{(\mp)}(t,{\vec p})\right]&=&-2E(p)\hat A^{(\mp)}(t,{\vec p})\,,\label{HDAmp}
\end{eqnarray}
concluding that a Fourier operator $A$ is conserved (obeying $dA=0$) only if this is reducible  and independent on time, $A=A_{\rm diag}\in F[0]$. In fact, all the diagonal parts of the Fourier operators of the algebra $F[0]$  are conserved. In contrast, the off-diagonal terms are oscillating in time with the frequency $2E(p)$ as it results from Eqs. (\ref{HDApm}) and  (\ref{HDAmp}).  These terms  form the {\em oscillating} part   $A_{\rm osc}=A^{(\pm)}+A^{(\mp)}$ of the operator $A$. The well-known example is the operator of Dirac's current density whose oscillating terms give rise to Zitterbewegung \cite{Zit1,Zit2,Z1,Z2}.

We must stress that the criteria for selecting conserved Fourier operators cannot be extended to any equal-time operators even these satisfy a similar condition $A=A_{\rm diag}\in E[0]$. The example is the position operator which satisfies this condition but evolves linearly in time as we shall show in Sec. 4.2.

\subsection{Pryce(e) spin and related operators} 

The principal  Pryce's proposal is his version (e) defining the Fourier transforms of a conserved spin operator  ${\vec S}_{\rm Pr(e)}$ related to a suitable correction to the coordinate operator, $\delta{\vec X}_{\rm Pr(e)}$. These Fourier transforms 
\begin{eqnarray}
	\vec{\hat S}_{\rm Pr(e)}({\vec p})&=&\frac{m}{E(p)} {\vec s}+\frac{{\vec p}\, ({\vec s}\cdot{\vec p})}{E(p)(E(p)+m)}+\frac{i}{2E(p)}{\vec p}\land {\vec\gamma} \,,\label{PrS}\\
	\delta\vec{\hat X}_{\rm Pr(e)}({\vec p})&=&\frac{i{\vec\gamma}}{2E(p)}+\frac{{\vec p}\land {\vec s}}{E(p)(E(p)+m)}-\frac{i{\vec p}\, ({\vec \gamma}\cdot {\vec p})}{2E(p)^2(E(p)+m)}\,, \label{PrX}
\end{eqnarray} 
satisfy the identity $\delta\vec{\hat X}_{\rm Pr(e)}({\vec p})\land {\vec p}={\vec s}-\vec{\hat S}_{\rm Pr(e)}({\vec p})$ in order to ensure the conservation of the total angular momentum (\ref{spli}). The Pryce(e) spin operator was considered later by 
Foldy and Wouthuysen which showed that their operator  (\ref{FW}) transforms the Pryce(e) spin operator into the Pauli-Dirac one as in Eq. (\ref{FW2}). For this reason many authors consider the Pryce(e) spin operator as the Foldy-Wouthuysena one denoting it by ${\vec S}_{\rm FW}$ \cite{A,A1}.  In what follows we use the simpler notation of the spin operator 
${\vec S}\equiv {\vec S}_{\rm Pr(e)}\equiv{\vec S}_{\rm FW}\in F[0]$ and similarly for its Fourier transform, $\vec{\hat S}({\vec p})\equiv \vec{\hat S}_{\rm Pr(e)}({\vec p})\in\hat F[0]$, defined by Eq. (\ref{PrS}).

In Ref. \cite{Cot} we considered a spectral representation for showing that ${\vec S}$  is just the operator defined by Eq.  (\ref{Spipi}) whose components generate the spin symmetry. We found that its Fourier transform (\ref{PrS})  can be put in the form  \cite{Cot}, 
\begin{eqnarray}\label{Sip}
	\vec{\hat S}({\vec p})&=&\frac{m}{E(p)}\left[ l_{\vec p}\, {\vec s}\, \frac{1+\gamma^0}{2}l_{\vec p}+l^{-1}_{\vec p}{\vec s}\, \frac{1-\gamma^0}{2}l^{-1}_{\vec p}\right]\nonumber\\ 
	&=&{\vec s}({\vec p}) \, \hat\Pi_+({\vec p}) +  {\vec s} ( -{\vec p})\,\hat \Pi_-({\vec p})	\,,	
\end{eqnarray}
laying out the operator 
\begin{equation}\label{sCh}
	\vec{\hat S}_{\rm Ch}({\vec p})\equiv {\vec s}({\vec p})=l_{\vec p} {\vec s}\,  l_{\vec p}^{-1}  \in   \hat F[0]  \,,
\end{equation}
which  was proposed by  Chakrabarti \cite{Ch}  as the Fourier transform of an alternative spin operator, $\vec S_{\rm Ch}\in F[0]$. However, this operator is not  conserved,  having the same action as the Pryce(e) one but only in the particle sector while in the antiparticle sector there is a discrepancy generating  oscillating terms as we shall show in Sec. 5.3. Nevertheless, the properties of the Chakrabarti operator, 
\begin{equation}\label{sCh1}
	{\vec s}({\vec p})={\vec s}\,^+(-{\vec p})\,, \quad {\vec s}(\pm{\vec p})\hat\Pi_{\pm}({\vec p})=\hat\Pi_{\pm}({\vec p}) {\vec s}(\mp{\vec p})\,,
\end{equation}	 
guarantee that ${\vec S}$ is a conserved Hermitian operator, its Fourier transform obeying $\vec{\hat S}({\vec p})=\vec{\hat S}^+({\vec p})=\vec{\hat  S}_{\rm diag}({\vec p})\in \hat F[0]$.  In addition, the  components ${S}_i$ are translation invariant,  commuting with the momentum operator,  having similar  algebraic properties as the Pauli-Dirac operator, 
\begin{eqnarray}
	\left[\hat S_i({\vec p})  ,	\hat S_j({\vec p})\right]=i\epsilon_{ijk}	 \hat S_k({\vec p})~~&~\Rightarrow ~&~
	\left[ S_i  ,	 S_j\right]=i\epsilon_{ijk} S_k\,,\nonumber\\
	\left\{\hat S_i({\vec p})  ,\hat	S_j({\vec p})\right\}=\frac{1}{2}\delta_{ij} \cdot 1\in \rho_D  &~\Rightarrow ~&
	\left\{ S_i  ,	 S_j\right\}=\frac{1}{2}\delta_{ij} I\,,\nonumber\\
	\vec{\hat S}^{2}({\vec p})=\frac{3}{4}\cdot 1\in\rho_D ~&~\Rightarrow ~&~~~~~~~ \vec{ S}^{2}=\frac{3}{4} I\,,\nonumber
\end{eqnarray}
defining thus a spin half representation of the $SU(2)$ group.  Furthermore, for writing explicitly the action of this operator we re-denote $\psi\to \psi_{\xi}$, $U_{{\vec p},\sigma}\to U_{{\vec p},\xi_{\sigma}}$ and $V_{{\vec p},\sigma}\to V_{{\vec p},\eta_{\sigma}}$. By using then  the form of the spinors  (\ref{Ufin}) and (\ref{Vfin}) we may write the actions 
\begin{eqnarray}
	(S_i U_{{\vec p},\xi_{\sigma}})(x)&=&\hat S_i({\vec p})U_{{\vec p},\xi_{\sigma}}(x)=U_{{\vec p},\hat s_i\xi_{\sigma}}(x)\,,\\
	(S_i V_{{\vec p},\eta_{\sigma}})(x)&=&\hat S_i(-{\vec p})V_{{\vec p},\eta_{\sigma}}(x)=V_{{\vec p},\hat s_i\eta_{\sigma}}(x)\,,	
\end{eqnarray}
concluding that $\vec{\hat S}({\vec p})$ is just the Fourier transform of the spin operator ${\vec S}$ defined by Eq. (\ref{Spipi}). The integral representation helping us to derive the identity (\ref{Sip}) will be discussed and generalized in Sec. 4.4.  

In applications we may use the new auxiliary  operators ${\vec S}^{(+)}$ and ${\vec S}^{(-)}$ whose components have the Fourier transforms  
\begin{equation}\label{SteS}
	\hat S_i^{(+)}({\vec p})=\Theta_{ij}({\vec p})\hat S_j({\vec p})	\,, \quad \hat S_i^{(-)}({\vec p})=\Theta^{-1}_{ij}({\vec p})\hat S_j({\vec p})\,,
\end{equation}
where $\Theta({\vec p})$ is the $SO(3)$ tensor defined in Eq.  (\ref{tete}) as 
the space part of the Lorentz boost $L_{\vec p}$ given by Eq.  (\ref{Lboost}).  With these notations  the Fourier transform of the Pauli-Lubanski operator (\ref{PaLu1}) can be written now as
\begin{eqnarray}
	&&\hat W^{\mu}({\vec p })= m(L_{\vec p})^{\mu\,\cdot}_{\cdot\,i}\hat S_i({\vec p})\Rightarrow\nonumber\\
	&&	\hat W^0({\vec p})={\vec p}\cdot\vec{\hat S}({\vec p})={\vec p}\cdot{\vec s}\,, \quad \vec{\hat W}({\vec p})=m\, \vec{\hat S}\, ^{(+)}({\vec p})\,, \label{PaLu2}
\end{eqnarray}
satisfying $p^{\mu}\hat W_{\mu}({\vec p})=0$ and $\hat W^{\mu}({\vec p})\hat W_{\mu}({\vec p})=-m^2\frac {3}{4}\cdot 1\in \rho_D$.   

The form of Pryce(e) spin operator allowed us to define the operator of fermion polarization  for any  related polarization spinors, $\xi_{\sigma}({\vec p})$  and $\eta_{\sigma}({\vec p})$, satisfying the general eigenvalues problems
\begin{equation}\label{snpp}
	\hat s_i  {n}^i({\vec p})\xi_{\sigma}({\vec p})	=\sigma\, \xi_{\sigma}({\vec p}) \Rightarrow 
	\hat s_i  {n}^i({\vec p})\eta_{\sigma}({\vec p})	=-\sigma\, \eta_{\sigma}({\vec p}),
\end{equation}
where the unit vector ${\vec n}({\vec p})$ gives the peculiar direction with respect to which the polarization is measured.   The corresponding polarization operator may be defined as the Fourier operator $W_s\in F[0]$ whose Fourier transform reads \cite{Cot}
\begin{eqnarray}
	\hat W_s({\vec p})=w({\vec p})\hat\Pi_+({\vec p}) +  w(-{\vec p}) \hat\Pi_-({\vec p})
	\,,	\label{Pol}
\end{eqnarray}
where $w({\vec p})={\vec s}({\vec p})\cdot {\vec n}({\vec p})$. As in the case of the spin operator we find that the operator of fermion polarization acts as 
\begin{eqnarray}
	(W_s U_{{\vec p},\xi_{\sigma}({\vec p})})(x)&=&\hat W_s({\vec p})U_{{\vec p},\xi_{\sigma}({\vec p})}(x)\nonumber\\
	&=&U_{{\vec p},\hat s_i n^i({\vec p})\xi_{\sigma}({\vec p})}(x)=\sigma U_{{\vec p},\xi_{\sigma}({\vec p})}(x)\,,\label{WU}\\
	(W_s V_{{\vec p},\eta_{\sigma}({\vec p})})(x)&=&\hat W_s(-{\vec p})V_{{\vec p},\eta_{\sigma}({\vec p})}(x)\nonumber\\
	&=&V_{{\vec p},\hat s_i  n^i({\vec p}) \eta_{\sigma}({\vec p})}(x)= -\sigma V_{{\vec p},\eta_{\sigma}({\vec p})}(x)\,.~~~~~~\label{WV}	
\end{eqnarray}
These eigenvalues problems convince us that $W_s$ is just the operator we need for completing the system of commuting operators 	$\{H,P^1,P^2,P^3, W_s\}$  defining the momentum  bases of RQM.  

Finally, we remind the reader that the conserved spin operator (\ref{Sip}) is related to Pryce's position operator of version (e) whose  correction $\delta{\vec X}$ has the Fourier transform (\ref{PrX}) that can be put in the simpler form \cite{Cot}
\begin{eqnarray}\label{XFour}
	\delta\vec{\hat X}({\vec p})\equiv	\delta \vec{\hat X}_{\rm Pr(e)}({\vec p})=\delta {\vec x}_{+}({\vec p})\hat\Pi_+({\vec p}) +  \delta {\vec x}_{-}({\vec p})\hat \Pi_-({\vec p})\,,
\end{eqnarray}
where the components of  $ \delta {\vec x}_{\pm}( {\vec p})$ have the form
\begin{equation}\label{deltax}
	\delta x^i_{\pm}({\vec p})=-i\frac{1}{n(p)} \left(\partial_{p^i} n(p)l_{\pm{\vec p}}\right)l_{\mp{\vec p}}\,,
\end{equation}  
depending on the normalization factor (\ref{Nor}) and momentum derivatives. However, we cannot construct the whole position operator ${\vec X}=\underline{\vec x}+\delta {\vec X}$ with the tools we considered so far because of  the coordinate operator $\underline{\vec x}$ which  is no longer a Fourier one. For this reason we shall study this operator in Sec. 4.2 after constructing a convenient framework.

\subsection{Other spin-type and position operators}

There are other proposals of conserved spin-type Fourier operators that cannot be integrated naturally in Dirac's theory, as in the case of the Pryce(e) one,  because of their components which do not satisfy $su(2)$ commutation relations. Nevertheless, these operators deserve to be briefly examined as they represent  observables that could be measured in some dedicated experiments \cite{A,A1}. 

The oldest proposal is the Frankel (Fr) spin-type operator which is a Fourier operator, ${\vec S}_{\rm Fr}$,  having the Fourier transform  \cite{Fr}
\begin{eqnarray}
	\vec{\hat S}_{\rm Fr}({\vec p})&=&{\vec s}+\frac{i}{2m}{\vec p}\land{\vec \gamma} \nonumber\\
	&=& \frac{E(p)}{m}\left( \vec{\hat S}({\vec p})-\frac{{\vec p\,({\vec p}\cdot \vec{\hat S}({\vec p}))}}{E(p)(E(p)+m)}\right)=\frac{E(p)}{m}\vec{\hat S}\,^{(-)}({\vec p})\,, \label{SFr}
\end{eqnarray}
written with the notation (\ref{SteS}).  The components of this operator are conserved and translation invariant commuting with $H_D$ and $P^i$  but these do not satisfy the $su(2)$ algebra such that the squared norm, 
\begin{equation}
	\vec{\hat S}^2_{\rm Fr}({\vec p})=\frac{1}{4}\left( 1+2\frac{E(p)^2}{m^2}\right)\cdot 1\in\rho_D\,,	
\end{equation}
is larger than $\frac{3}{4}$.  The Frankel spin-type operator may be generated as
\begin{eqnarray}
	&&\left[ \hat S^{(+)}_i({\vec p}), \hat S^{(+)}_j({\vec p}) \right]=i \epsilon_{ijk}\hat S_{{\rm Fr}\,k}({\vec p})
	~~~\Rightarrow ~~~ \left[ S^{(+)}_i,  S^{(+)}_j \right]=i \epsilon_{ijk}S_{{\rm Fr}\,k}\,,\label{Com10}
\end{eqnarray}
having  specific commutation rules
\begin{eqnarray}
	&&	\left[ \hat S_{{\rm Fr}\, i}({\vec p}), \hat S_{{\rm Fr}\, j}({\vec p}) \right]=i \epsilon_{ijk}\hat C_{{\rm Fr}\,k}({\vec p})
 ~~~\Rightarrow~~~  \left[ S_{{\rm Fr}\, i},  S_{{\rm Fr}\, j} \right]=i \epsilon_{ijk}C_{{\rm Fr}\,k}\,,	\label{Com1}
\end{eqnarray}
which define the new Fourier operator ${\vec C} _{\rm Fr}$ whose Fourier transform reads
\begin{equation}\label{CFr}
	\vec{\hat C}_{\rm Fr}({\vec p})=\frac{E(p)}{m}\left( \vec{\hat S}({\vec p})+ \frac{{\vec p\,({\vec p}\cdot \vec{\hat S}({\vec p}))}}{m(E(p)+m)}\right)=\frac{E(p)}{m}\vec{\hat S}\,^{(+)}({\vec p})\,.	
\end{equation}

A similar spin-type operator was considered initially by Pryce  according to his hypothesis (c) \cite{B} and then re-defined and studied by Czochor \cite{Cz} such that this is called often the Czochor spin operator \cite{A,A1}.  Here we speak about the Pryce(c)-Czochor (PC)  operator defined as the diagonal part of the Pauli-Dirac one \cite{Cz},
\begin{equation}
	{\vec S}_{\rm PC}=\Pi_+{\vec s}\,\Pi_+ +\Pi_-{\vec s}\,\Pi_-\,.
\end{equation}
This has the Fourier transform   \cite{Cz,A,A1},
\begin{eqnarray}
	\vec{\hat S}_{\rm PC}({\vec p})&=&\hat\Pi_+({\vec p}){\vec s}\,\hat\Pi_+({\vec p})+\hat\Pi_-({\vec p}){\vec s}\,\hat\Pi_-({\vec p})\nonumber\\	
	&=&\frac{m^2}{E(p)^2}{\vec s}+\frac{{\vec p}\,({\vec p}\cdot {\vec s})}{E(p)^2}	+\frac{i m}{2 E(p)^2}{\vec p}\land {\vec \gamma}
	=\frac{m}{E(p)}\vec{\hat S}\,^{(+)}({\vec p})\,,\label{SCz1}
\end{eqnarray}
whose  squared norm  
\begin{equation}\label{SCz2}
	\vec{\hat S}^2_{\rm PC}({\vec p})=\frac{1}{4}\left( 1+2\frac{m^2}{E(p)^2}\right)\cdot 1\in\rho_D\,,	
\end{equation}
takes values in the domain $(\frac{1}{4}, \frac{3}{4}]$. The Pryce(c)-Czochor spin-type operator may be generated as
\begin{eqnarray}
	&&	\left[ \hat S^{(-)}_i({\vec p}), \hat S^{(-)}_j({\vec p}) \right]=i\epsilon_{ijk} \hat 
	S_{{\rm PC}\,k}({\vec p})
	~~~\Rightarrow ~~~	\left[ S^{(-)}_i,  S^{(-)}_j \right]=i\epsilon_{ijk} S_{{\rm PC}\,k} \,,
	\label{Com20}	
\end{eqnarray}
satisfying  the commutation relations 
\begin{eqnarray}
	&&	\left[ \hat S_{{\rm PC}\, i}({\vec p}), \hat S_{{\rm PC}\, j}({\vec p}) \right]=i\epsilon_{ijk} \hat C_{{\rm PC}\,k}({\vec p})
	~~~\Rightarrow ~~~	\left[ S_{{\rm PC}\, i},  S_{{\rm PC}\, j} \right]=i\epsilon_{ijk} C_{{\rm PC}\,k} \,,
	\label{Com2}	
\end{eqnarray}
where the Fourier transform of the new operator ${\vec C}_{{\rm PC}}$ reads
\begin{equation}\label{CCz}
	\vec{\hat C}_{\rm PC}({\vec p})=\frac{m}{E(p)}\vec{\hat S}\,^{(-)}({\vec p}) \,.	
\end{equation}
We conclude that  the Frankel and Pryce(c)-Czochor spin-type operators are elements of  a larger  algebraic structure  depending only on the pair of operators ${\vec S}^{(+)}$ and  ${\vec S}^{(-)}$. In other respects, all  the Fourier transforms of conserved spin and spin-type operators discussed so far have the same projection along the momentum direction such that 
\begin{equation}
	{\vec p}\cdot\vec{\hat S}_{\rm Fr}({\vec p})={\vec p}\cdot\vec{\hat S}_{\rm PC}({\vec p})={\vec p}\cdot \vec{\hat S}({\vec p})={\vec p}\cdot {\vec s}\in\hat F[0]	\,.
\end{equation}
This means that we can inverse Eqs. (\ref{SFr}) and (\ref{SCz1}) relating  at any time the operators $\vec{\hat S}_{\rm Fr}({\vec p})$ and $\vec{\hat S}_{\rm PC}({\vec p})$ and implicitly their commutator operators, $\vec{\hat C}_{\rm Fr}({\vec p})$ and $\vec{\hat C}_{\rm PC}({\vec p})$. 

Another conserved and translation invariant operator was proposed by Fradkin and Good \cite{FG} defining its Fourier transform, 
\begin{eqnarray}\label{SFG}
	\vec{\hat S}_{\rm FG}({\vec p})&=&\gamma^0{\vec s}+\frac{{\vec p\, ({\vec p}\cdot {\vec s)})}}{p^2}\left( \frac{\hat H_D({\vec p})}{E(p)}-\gamma^0 \right) 
	=\vec{\hat  S}({\vec p}) \hat N({\vec p}) ~~~\Rightarrow~~~ {\vec S}_{\rm FG}={\vec S} N\,,
\end{eqnarray}
where  the operator $N$ has the Fourier transform  (\ref{Num}).  As $N$ commutes with the spin operator ${\vec S}$ and $N^2=I$ we may write directly the commutators
\begin{equation}\label{SFCS}
	\left[ S_{{\rm FG}\, i},  S_{{\rm FG}\, j} \right]=i\epsilon_{ijk}NS_{{\rm FG}\, k}\,,
	~~\Rightarrow~~ 	\vec{S}^2_{\rm FG}=\vec{S}^2=\frac{3}{4} I\,,
\end{equation}
that guarantee a desired square norm but without defining a Lie algebra. The simple algebraic properties of the Fradkin-Good spin-type operator indicate that this is somewhat useless being equivalent with the Pryce(e) one.  Other operators proposed recently  \cite{Ch3,Ch4} could be related to the above spin and spin-type operators  in further investigations.  

The Pryce(c)-Czochor spin-type operator   was constructed from the beginning according to Pryce's hypothesis (c). Moreover, it is not difficult to verify that  the Frankel one  complies with the hypothesis (d) such that  both these operators are related to  specific position operators, ${\vec X}_{\rm Pr(c)}=\underline{\vec x}+\delta {\vec X}_{\rm Pr(c)}$ and respectively ${\vec X}_{\rm Pr(d)}=\underline{\vec x}+\delta {\vec X}_{\rm Pr(d)}$. Observing that the corrections are Fourier operators it is convenient to use the artifice 
\begin{equation}\label{Prcd1}
	{\vec X}_{\rm Pr(c)}={\vec X}+\delta {\vec X}_{\rm Pr(c)}-\delta{\vec X}\,, \quad 	{\vec X}_{\rm Pr(d)}={\vec X}+\delta {\vec X}_{\rm Pr(d )}-\delta{\vec X}\,,
\end{equation}
providing us with simple Fourier transforms,
\begin{eqnarray}
	\delta \vec{\hat X}_{\rm Pr(c)}({\vec p})-\delta\vec{\hat X}({\vec p})	&=&\frac{{\vec p}\land\vec{\hat S}({\vec p})}{E(p)(E(p)+m)}\,,\label{Prcd2}\\
	\delta \vec{\hat X}_{\rm Pr(d)}({\vec p})-\delta\vec{\hat X}({\vec p})	&=&-\frac{{\vec p}\land\vec{\hat S}({\vec p})}{m(E(p)+m)}\,,\label{Prcd3}
\end{eqnarray}
resulted from the formulas of Ref. \cite{B}. These position operators give alternative splittings of the total angular momentum,
\begin{eqnarray}
	{\vec J}= {\vec X}_{\rm Pr(c)}\land{\vec P}+{\vec S}_{\rm PC}= {\vec X}_{\rm Pr(d)}\land{\vec P}+{\vec S}_{\rm Fr}\,,\nonumber
\end{eqnarray}
but which are formal, without a precised physical meaning,  as the components of  the position operators  do not commute among themselves  while those of the spin-type operators do not satisfy a $su(2)$ algebra. The only attribute of the above spin-type and related orbital angular momentum operators is that they are conserved.

We conclude that  the study of  various  position operators reduces to the Pryce(e) one which has to be derived after  passing beyond the technical difficulties constructing another suitable effective framework.  

\section{Method of associated operators}  

The difficulties arising in Dirac's theory  come from the fact that there are  many  equal-time integral operators having bi-local kernels which do not have Fourier transforms. For  studying such operators  we must resort to  integral representations but which  can be defined properly only by relating  the operators acting on the free fields to pairs of operators acting on the wave spinors (\ref{alpha}), we call here {\em associated} operators.  In other worlds we transfer the action of a given operator from mode spinors to the wave spinors obtaining thus a tool for deriving systematically expectation values in terms of wave spinors we need for preparing the quantization.

\subsection{Associated operators}

We start associating to each operator $A:{\cal F}\to {\cal F}$ in CR the pair of operators, $\tilde A: \tilde{\cal F}^+\to \tilde{\cal F}$ and $\tilde A^c: \tilde{\cal F}^-\to \tilde{\cal F}$, obeying
\begin{eqnarray}
	(A\psi)(x)&=&\int d^3p \sum_{\sigma}\left[(AU_{{\vec p},\sigma})(x) \alpha)_{\sigma}({\vec p})+(AV_{{\vec p},\sigma})(x) \beta^ {*}_{ \sigma}({\vec p})\right]\nonumber\\	
	&\equiv&\int d^3p \sum_{\sigma}\left[U_{{\vec p},\sigma}(x) (\tilde A\alpha)_{\sigma}({\vec p})+V_{{\vec p},\sigma}(x) (\tilde A^c\beta)^ {*}_{ \sigma}({\vec p})\right]\,,\label{AAAc}
\end{eqnarray}
such that the brackets  of $A$ for two different fields, $\psi$ and $\psi'$,  can be calculated as
\begin{equation}\label{expA}
	\langle \psi, A\psi'\rangle_D=\langle \alpha, \tilde A \alpha'\rangle+\langle \beta, \tilde A^{c\, +} \beta' \rangle\,.
\end{equation}
Hereby we deduce that if $A=A^+$ is Hermitian with respect to the Dirac scalar product (\ref{sp}) then the associated operators are Hermitian with respect to the scalar product (\ref{spa}),  $\tilde A={\tilde A}^+$ and $\tilde A^c={\tilde A^c\,}^+$. For simplicity we denote the Hermitian conjugation of the operators acting on the spaces  ${\cal F}$ and $\tilde{\cal F}$ with the same symbol but bearing in mind that the scalar products of these spaces are different. 

In general,  the operators $A\in E[t]$ and their associated operators $(\tilde A,\tilde A^c)$ may depend on time such that we must be careful considering the entire algebra we manipulate as frozen at a fixed time $t$. The new operators $\tilde A$ and $\tilde A^c$ are well-defined at any time as their action can be derived by applying the inversion formulas (\ref{inv}) to Eq. (\ref{AAAc}) at a given instant $t$. We find thus  that $\tilde A$ and $\tilde A^c$ are integral operators that may depend on time acting as
\begin{eqnarray}
		\left.(\tilde A\alpha)_{\sigma}({\vec p})\right|_t&=&\int d^3p'\sum_{\sigma'}\left.\langle U_{{\vec p},\sigma},AU_{{\vec p}\,',\sigma'}\rangle_D\right|_t \alpha_{\sigma'}({\vec p}\,')\nonumber\\
	&+&\int d^3p'\sum_{\sigma'}\left.\langle U_{{\vec p},\sigma},AV_{{\vec p}\,',\sigma'}\rangle_D\right|_t  \beta^*_{\sigma'}({\vec p}\,')	\,,
	\label{Aab}\\
		\left.(\tilde A^c\beta)_{\sigma}({\vec p})\right|_t&=&\int d^3p'\sum_{\sigma'}\left.\langle U_{{\vec p}\,',\sigma'},AV_{{\vec p},\sigma}\rangle_D\right|_t  \alpha^*_{\sigma'}({\vec p}\,')\nonumber\\
	&+&\int d^3p'\sum_{\sigma'}\left.\langle V_{{\vec p}\,',\sigma'},AV_{{\vec p},\sigma}\rangle_D\right|_t  \beta_{\sigma'}({\vec p}\,')	\,,\label{Acab}
\end{eqnarray}
through kernels which are the matrix elements of the operator $A$ in the basis of mode spinors.  We obtain thus the association $A\Leftrightarrow (\tilde A,\tilde A^c)$ defined through Eq. (\ref{AAAc}) which is a bijective mapping between two isomorphic operator algebras, ${E}[t]\subset {\rm Aut}({\cal F})$ and $\tilde{E}[t]\oplus \tilde E^c[t] \subset {\rm Aut}(\tilde {\cal F})$, preserving the linear and multiplication properties. Obviously, the identity operator of the algebras $\tilde E[t]$ and $\tilde E[t]^c$ is the matrix $1_{2\times 2}$.  For analyzing the actions of these operators we rewrite Eqs. (\ref{Aab}) and (\ref{Acab}) as,
\begin{eqnarray}
	&&`	\left.(\tilde A\alpha)_{\sigma}({\vec p})\right|_t=\left.(\tilde A^{(+)}\alpha)_{\sigma}({\vec p})\right|_t +\left.(\tilde A^{(\pm)}\beta^*)_{\sigma}({\vec p})\right|_t \,,\label{splitA}\\	
	&&	\left.(\tilde A^c\beta)_{\sigma}({\vec p})\right|_t=\left.(\tilde A^{(\mp)}\alpha^*)_{\sigma}({\vec p})\right|_t +\left.(\tilde A^{(-)}\beta)_{\sigma}({\vec p})\right|_t \,.~~~~~~\label{splitAc} 
\end{eqnarray}
in terms of the  new associated operators, 
\begin{eqnarray}
	\tilde A^{(+)}\in& {\rm Aut}(\tilde{\cal F}^+)\,,~~~~~~~~~~&	\tilde A^{(-)}\in {\rm Aut}(\tilde{\cal F}^-)\nonumber\\
	\tilde A^{(\pm)}\in &{\rm Lin}(\tilde{\cal F}^+,\tilde{\cal F}^{-\,*})\,,\quad&
	\tilde A^{(\mp)}\in {\rm Lin}(\tilde{\cal F}^-,\tilde{\cal F}^{+\,*})\,,\nonumber
\end{eqnarray} 
which  are integral operators in MR whose  kernels are the matrix elements of the operators $A^{(+)}, \,A^{(-)}, \, A^{(\pm)}$ and $A^{(\mp)}$ defined by the expansion (\ref{AAAA0}). Therefore, if $A\in E[t]$ is reducible then we have 
\begin{equation}\label{redcon}
	A^{(\pm)}= A^{(\mp)}=0 ~\Rightarrow~	\tilde  A^{(\pm)}=\tilde  A^{(\mp)}=0~\Rightarrow~\left\{ 
	\begin{array}{l}
		~\tilde A=\tilde A^{(+)}\,,\\
		\tilde A^c=\tilde A^{(-)}\,.
	\end{array}\right.
\end{equation}
Anticipating, we specify that all the Hermitian reducible operators $A\in E[t]$  we study here have associated operators related through {\em charge parity},  $\tilde A^c=\pm \tilde A$.

In the particular case of  Fourier operators, $A\in\, F[t]$, having time-dependent Fourier transforms $\hat A(t,{\vec p})$, the matrix elements can be calculated easier as 
\begin{eqnarray}
		\left.\langle U_{{\vec p},\sigma},AU_{{\vec p}\,',\sigma'}\rangle_D\right|_t& =&\left.\langle U_{{\vec p},\sigma},\hat A(t,{\vec p}\,')U_{{\vec p}\,',\sigma'}\rangle_D\right|_t\nonumber\\
	&=&\delta^3({\vec p}-{\vec p}\,')\frac{m}{E(p)}\mathring{u}_{\sigma}^+({\vec p})l_{\vec p}\hat A(t,{\vec p})l_{\vec p}\,\mathring{u}_{\sigma'}({\vec p})	\,,\label{mat1}\\
		\left.\langle U_{{\vec p},\sigma}, AV_{{\vec p}\,',\sigma'}\rangle_D\right|_t& =&\left.\langle U_{{\vec p},\sigma},\hat A(t,-{\vec p}\,')V_{{\vec p}\,',\sigma'}\rangle_D\right|_t\nonumber\\
	&=&\delta^3({\vec p}+{\vec p}\,')\frac{m}{E(p)}\mathring{u}_{\sigma}^+({\vec p})l_{\vec p}\hat A(t,{\vec p})l_{-{\vec p}}\,\mathring{v}_{\sigma'}(-{\vec p})e^{2iE(p)t}	\,,~~~~~~~
\end{eqnarray}
\begin{eqnarray}
		\left.\langle V_{{\vec p}\,',\sigma'}, AU_{{\vec p},\sigma}\rangle_D\right|_t& =&\left.\langle V_{{\vec p}\,',\sigma'},\hat A(t,{\vec p})U_{{\vec p},\sigma}\rangle_D\right|_t\nonumber\\
	&=&\delta^3({\vec p}+{\vec p}\,')\frac{m}{E(p)}\mathring{v}_{\sigma'}^+(-{\vec p})l_{-{\vec p}}\hat A(t,{\vec p})l_{{\vec p}}\,\mathring{u}_{\sigma}({\vec p})e^{-2iE(p)t}	\,,~~~~~~~\\
	\left.\langle V_{{\vec p},\sigma}, AV_{{\vec p}\,',\sigma'}\rangle_D\right|_t &=&\left.\langle V_{{\vec p},\sigma},\hat A(t,-{\vec p}\,')V_{{\vec p}\,',\sigma'}\rangle_D\right|_t\nonumber\\
	&=&\delta^3({\vec p}-{\vec p}\,')\frac{m}{E(p)}\mathring{v}_{\sigma}^+({\vec p})l_{\vec p}\hat A(t,-{\vec p})l_{\vec p}\,\mathring{v}_{\sigma'}({\vec p})	\,,\label{mat2}
\end{eqnarray}
observing that in this case the associated operators are simple $2\times 2$ matrix-operators acting on the spaces $\tilde{\cal F}^+$ and  $\tilde{\cal F}^-$. Hereby we deduce the matrix elements of the associated diagonal operators
\begin{eqnarray}
	&&	\tilde A^{(+)}_{\sigma\sigma'}(t,{\vec p})=\frac{m}{E(p)}\,\mathring{u}^+_{\sigma}({\vec p})l_{\vec p}\hat A(t,{\vec p})l_{\vec p}\,\mathring{u}_{\sigma'}({\vec p})\,,\label{Aa1}\\
	&&	\tilde A^{(-)}_{\sigma\sigma'}(t,{\vec p})=\frac{m}{E(p)}\,\mathring{u}^+_{\sigma}({\vec p})l_{{\vec p}}\,C\hat A(t,-{\vec p})^T Cl_{\vec p}\,\mathring{u}_{\sigma'}({\vec p})\,,~~~~~~~~\label{Aa2}
\end{eqnarray}
and those of the off-diagonal ones
\begin{eqnarray}
	&&	\tilde A^{(\pm)}_{\sigma\sigma'}(t,{\vec p})=\frac{m}{E(p)}\,\mathring{u}^+_{\sigma}({\vec p})l_{\vec p}\hat A(t,{\vec p})l_{-{\vec p}}\,\mathring{v}_{\sigma'}(-{\vec p})e^{2iE(p)t}\,,\label{Aa1p}\\
	&&	\tilde A^{(		\mp)}_{\sigma\sigma'}(t,{\vec p})=\frac{m}{E(p)}\,\mathring{v}^+_{\sigma'}(-{\vec p})l_{-{\vec p}}\hat A(t,{\vec p})l_{\vec p}\,\mathring{u}_{\sigma}({\vec p})e^{-2iE(p)t}\,,~~~~~~~
	\label{Aa2p}
\end{eqnarray}
which oscillate with the frequency $2E(p)$. 

\subsection{Associated spin, polarization and position  operators}

The simplest examples of reducible Fourier operators are the projection operators related to the operators $ I,\,N\in F[0]$  for which  we have to substitute the expressions (\ref{Pip}) and (\ref{Pim}) in Eqs. (\ref{Aa1}) and (\ref{Aa2}) using then the identities  (\ref{idll}) for obtaining  the associated operators, 
\begin{eqnarray}
	\Pi_+~~&\Rightarrow&~~\tilde \Pi_+ =1_{2\times2}\,, \quad \tilde \Pi^c_+ =0\,,\nonumber\\
	\Pi_-~~&\Rightarrow&~~\tilde \Pi_- =0 \,, \quad\quad~~ \tilde \Pi^c_- =1_{2\times2}\,,\nonumber\\
	I=\Pi_++\Pi_-~~&\Rightarrow&~~ \tilde I=\tilde I^c=1_{2\times2}\,,\nonumber	\\
	N=\Pi_+-\Pi_-~~&\Rightarrow&~~ \tilde N=-\tilde N^c=1_{2\times2}\,,\nonumber
\end{eqnarray}
depending on  the identity operator $1_{2\times2}$ of  $\tilde F[0]\simeq \tilde F^c[0]$ algebras. More interesting are the associated operators to the new observables of our approach, namely the spin, fermion polarization and position operators we have to study in this section.  

For deriving the associated operators to the Pryce(e) spin ${\vec S}$ we substitute  its Fourier transform  (\ref{Sip}) in Eqs. (\ref{Aa1}) and (\ref{Aa2}) taking into account that these operators are reducible, ${\vec S}={\vec S}_{\rm diag}$.  By using again the identity (\ref{idll})  we find that  the associated operators of ${\vec S}$ have the components \cite{Cot}
\begin{eqnarray}
	{S}_i~~\Rightarrow~ \tilde S_i=-\tilde S_i^c=\frac{1}{2}\Sigma_i({\vec p}) \label{tilS}	\,,
\end{eqnarray}
where the $2\times 2$ matrices $\Sigma_i({\vec p})$ have the matrix elements 
\begin{equation}\label{Dxx}
	\Sigma_{i\,\sigma\sigma'}({\vec p})=2\mathring{u}_{\sigma}^+({\vec p})s_i \mathring{u}_{\sigma'}({\vec p})=\xi^+_{\sigma}({\vec p})\sigma_i\,\xi_{\sigma'}({\vec p})\,,	
\end{equation}
depending on the polarization spinors and having the same algebraic properties as the Pauli matrices. Similar procedures give the associated operators 
\begin{eqnarray}
	{S}^{(+)}_i&\Rightarrow&\tilde S^{(+)}_i=-\tilde S_i^{(+)\,c}=\frac{1}{2}\,\Theta_{ij}({\vec p})\Sigma_j({\vec p}) \label{tilS+}	\,,\label{Splus}	\\
	{S}^{(-)}_i&\Rightarrow&\tilde S^{(-)}_i=-\tilde S_i^{(-)\,c}=\frac{1}{2}\,\Theta^{-1}_{ij}({\vec p})\Sigma_j({\vec p}) \label{tilS-}	\,,	\label{Sminus}
\end{eqnarray}
to those defined by Eqs. (\ref{SteS}) and  the simple associated operators of the polarization operator (\ref{Pol}), 
\begin{equation}
	W_s~\Rightarrow~~\tilde W_s=-\tilde W_s^c=\frac{1}{2}\sigma_3\,,\label{tilW}	
\end{equation}
according to the definition (\ref{snpp}) of the polarization spinors. 

The position operator,  ${\vec X}$, is  reducible  but is no longer a Fourier operator even though the correction $\delta{\vec X}$ of the Pryce(e) version is of this type having the Fourier transform given by Eqs. (\ref{XFour})  and (\ref{deltax}). For extracting the action of this operator we apply the Green theorem after deriving the  identities \cite{Cot} 
\begin{eqnarray}
		\left(\delta X^i U_{{\vec p},\xi_{\sigma}}\right)(t,{\vec x})&=&\delta \tilde X^i({\vec p})U_{{\vec p},\xi_{\sigma}}(t,{\vec x})	\nonumber\\
	&=&-i\partial_{p^i}U_{{\vec p},\xi_{\sigma}}(t,{\vec x})-x^i U_{{\vec p},\xi_{\sigma}}(t,{\vec x})+\frac{t p^i}{E(p)}U_{{\vec p},\xi_{\sigma}}(t,{\vec x})\nonumber\\
	&&\hspace*{12mm}+\sum_{\sigma'}U_{{\vec p},\xi_{\sigma'}}(t,{\vec x})\Omega_{i\,\sigma' \sigma}({\vec p})\,,\label{X1}\\
		\left(\delta X^i V_{{\vec p},\eta_{\sigma}}\right)(t,{\vec x})&=&\delta \tilde X^i(-{\vec p})V_{{\vec p},\eta_{\sigma}}(t,{\vec x})	\nonumber\\
	&=&i\partial_{p^i}V_{{\vec p},\eta_{\sigma}}(t,{\vec x})-x^i V_{{\vec p},\eta_{\sigma}}(t,{\vec x})+\frac{t p^i}{E(p)}V_{{\vec p},\eta_{\sigma}}(t,{\vec x})\nonumber\\
	&&\hspace*{12mm}-\sum_{\sigma'}V_{{\vec p},\eta_{\sigma'}}(t,{\vec x})\Omega^*_{i\,\sigma' \sigma}({\vec p})\,.\label{X2}
\end{eqnarray} 
We find that this operator depends linearly on time, ${\vec X}(t)={\vec X}+t {\vec V}$,  its components having  simple and intuitive associated operators  \cite{Cot},
\begin{eqnarray}
	X^i~~&\Rightarrow&~~ \tilde X^i=\tilde X^{c\,i}=i\tilde \partial_i \,, \label{tilX}\\
	V^i~~&\Rightarrow&~~\tilde V^i=\tilde V^{c\,i}=\frac{p^i}{E(p)}\,,\label{tilV}
\end{eqnarray}
where the {\em covariant} derivatives \cite{Cot},
\begin{equation}\label{covD}
	\tilde\partial_i=\partial_{p^i} 1_{2\times 2}+\Omega_i({\vec p})\,,
\end{equation}
are defined such that $\tilde\partial_i [\xi_{\sigma}({\vec p})\alpha_{\sigma}({\vec p})] =\xi_{\sigma}({\vec p})\tilde\partial_i \alpha_{\sigma}({\vec p}$. Therefore, the connections, 
\begin{eqnarray}
	\Omega_{i\,\sigma\sigma'}({\vec p})=\xi^+_{\sigma}({\vec p})\left[\partial_{p^i}\xi_{\sigma'}({\vec p})\right]=\left\{\eta^+_{\sigma}({\vec p})\left[\partial_{p^i}\eta_{\sigma'}({\vec p})\right]\right\}^*\,,\label{Omega}
\end{eqnarray}
are anti-Hermitian,   $\Omega_{i\,\sigma\sigma'}({\vec p})=-\Omega_{i\,\sigma'\sigma}^*({\vec p})$, which means that the operators $i\tilde \partial_i$ are Hermitian.  We must stress that the principal property of the covariant derivatives is of commuting with the spin components, $[\tilde\partial_i, \tilde S_j]=0$.  In the case of peculiar polarization the connections $\Omega_i({\vec p})$ guarantee this property which becomes trivial in the case of common polarization when $\Omega_i=0$ and $\tilde S_i$ are independent on ${\vec p}$. 

Initially, Pryce proposed the operator ${\vec X}$ as the relativistic mass-center operator of RQM. However,   we have shown in Ref. \cite{Cot} that after quantization this becomes in fact the operator of center of charges, or simpler the dipole operator, while the velocity operator ${\vec V}$ becomes just the corresponding conserved vector current. For this reason we defined another mass-center operator changing by hand the sign of the antiparticle term. Now we have the opportunity of using the operator $N$ for defining the mass-center operator from the beginning, at the level of RQM. We assume that this has the form  ${\vec X}_{MC}(t)={\vec X}_{MC}+t{\vec V}_{MC}$ where
\begin{equation}\label{XMC}
	{\vec X}_{MC}(t)=N{\vec X}(t)~~~\Rightarrow~~~ X_{MC}^i=NX^i\,, ~~V_{MC}^i=NV^i\,,
\end{equation} 
such that the associated operators $\tilde X^i_{MC}=-\tilde X^{c\, i}_{MC}=\tilde X^i$ and $\tilde V^i_{MC}=-\tilde V^{\c\,i}_{MC}=\tilde V^i$ guarantee the desired sign of the antiparticle term after quantization. 

Other position operators are the Pryce (c) and (d) ones depending on the principal position operator (e) as in Eqs. (\ref{Prcd1}-\ref{Prcd3}). As these operators  are of a marginal interest we restrict ourselves to present briefly  in the Appendix C   their  associated operators and some algebraic properties.    

\subsection{Associated isometry generators}

Let us see now how the Pryce(e) spin operator is related to the generators of the Poincar\' e isometries.   In our approach  we may establish explicitly the equivalence between the covariant representation and a pair of Wigner's induced ones transforming the Pauli wave spinors.  The covariant representation ${T}$ defined by Eq. (\ref{TAa})   may be associated to a  pair of Wigner's representations whose operators $\tilde { T}\in {\rm Aut}(\tilde{\cal F}^+)$ and $\tilde {T}^c\in {\rm Aut}(\tilde{\cal F}^-)$  satisfy \cite{WKT,W,Th},
\begin{eqnarray}
	({T}_{\lambda,a}\psi)(x)&=&\int d^3p \sum_{\sigma}\left[U_{{\vec p},\sigma}(x)(\tilde { T}_{\lambda,a}\, \alpha)_{\sigma}({\vec p})+V_{{\vec p},\sigma}(x) (\tilde { T}^c_{\lambda,a}\,\beta)^ {*} _{ \sigma}({\vec p})\right]\,.\label{basic}
\end{eqnarray}
In other respects, by using  the identity $(\Lambda x)\cdot p=x\cdot (\Lambda^{-1}p)$ and the invariant measure (\ref{measure}) we expand Eq.  (\ref{TAa}) changing the integration variable  as  
\begin{eqnarray}
	({T}_{\lambda,a}\psi  )(x)&=&\lambda \psi  \left(\Lambda(\lambda)^{-1}(x-a)\right)\nonumber\\
	&=&\int d^3p \frac{E(p_{\lambda})}{E(p)}\sum_{\sigma} \left[ \lambda U'_{{\vec p},\sigma}(x)\alpha_{\sigma}({\vec p}_{\lambda})e^{i {a}\cdot{p}} + \lambda V'_{{\vec p},\sigma}(x)\beta^*_{\sigma}({\vec p}_{\lambda})e^{-i {a}\cdot{p}}\right]\,,\label{subT}
\end{eqnarray}
where we denote $a\cdot p=a_{\mu}p^{\mu}=E(p)a^0-{\vec p}\cdot {\vec a}$ while the new mode spinors
\begin{eqnarray}
	U'_{{\vec p},\sigma}(x)&=&u_{\sigma}({\vec p}_{\lambda})\frac{1}{(2\pi)^{\frac{3}{2}}} \,e^{-iE(p)t+i{\vec p}\cdot{\vec x}}\,,\label{Upr} \\
	V'_{{\vec p},\sigma}(x)&=&v_{\sigma}({\vec p}_{\lambda})\frac{1}{(2\pi)^{\frac{3}{2}}}\, e^{iE(p)t-i{\vec p}\cdot{\vec x}}\,,\label{Vpr}
\end{eqnarray} 
depend on the transformed momentum of components
\begin{equation}\label{pLp}
	{p}^{\mu}_{\lambda}=\left<\Lambda(\lambda)^{-1}\right>^{\mu\,\cdot} _{\cdot\,\nu} p^{\nu}\,,
\end{equation}
through the spinors (\ref{Ufin}) and (\ref{Vfin}). Hereby, we deduce that   $\tilde {T}_{\lambda,a}\simeq \tilde {T}^c_{\lambda,a}$ acting alike on the spaces  $\tilde{\cal F}^+$ and  $\tilde{\cal F}^-$ as \cite{Wig,WKT,Th},
\begin{eqnarray}
	(\tilde {T}_{\lambda,a}\, \alpha)_{\sigma}({\vec p})=\sqrt{\frac{E(p_{\lambda})}{E( p)}}e^{i {a}\cdot{p}}\sum_{\sigma'}{D}_{\sigma\sigma'}(\lambda,{\vec p}) \alpha_{\sigma'}({\vec p}_{\lambda}) \,,
	\label{Wig}
\end{eqnarray}
and similarly for $\beta$,  because of their related matrices,  
\begin{eqnarray}
	{D}_{\sigma\sigma'}(\lambda,{\vec p})={\mathring{u}}_{\sigma}^+({\vec p})w(\lambda,{\vec p})\mathring{u}_{\sigma'}({\vec p}_{\lambda})
	=\left[{\mathring{v}}^+_{\sigma}({\vec p})w(\lambda,{\vec p})\mathring{v}_{\sigma'}({\vec p}_{\lambda})\right]^*\,.\label{preD}
\end{eqnarray}
These depend on the well-known Wigner transformations 
\begin{equation}\label{wwig}
	w(\lambda,{\vec p})=l^{-1}_{{\vec p}} \lambda\, l_{{\vec p}_{\lambda}}\in \rho_D	\,,
\end{equation}
whose  corresponding  Lorentz transformations   leave invariant the representative momentum,
\begin{equation}
	\Lambda[w(\lambda,{\vec p})]\mathring{p}=L^{-1}_{{\vec p}}\Lambda(\lambda){ p}_{\lambda}=L^{-1}_{{\vec p}}{p}=\mathring{p}\,,\nonumber 	
\end{equation}
which means that    $\Lambda[w(\lambda,{\vec p})]\in SO(3)$ is a rotation and consequently   $w(\lambda,{\vec p})\in \rho_D[SU(2)]$.  Furthermore, bearing in mind that the $SU(2)$ rotations of $\rho_D$ have  the form  (\ref{r}) we obtain the definitive expression of the matrix elements (\ref{preD}) as, 
\begin{equation}
	{D}_{\sigma\sigma'}(\lambda,{\vec p})=\xi^+_{\sigma}({\vec p}) \hat l^{-1}_{{\vec p}} \hat\lambda\, \hat l_{{\vec p}_{\lambda}}\xi_{\sigma'}({\vec p}_{\lambda}) \,,
	\label{Wrot}
\end{equation}
observing that these depend explicitly on the polarization spinors. As these matrices form the representation of spin $s=\frac{1}{2}$ of the little group $SU(2)$ one says that the equivalent Wigner representation $\tilde {T}\simeq \tilde{T}^c$ are  {\em  induced} by the subgroup $T(4)\,\circledS\,SU(2)$ \cite{Wig,WKT,Th}.  Note that for rotations, $\lambda=r \in\rho _D[SU(2)]$, we obtain the usual $SU(2)$ linear representation as $E(p_{\lambda})=E(p)$ and $\hat r \hat l_{\vec{p}_{\lambda}}\hat r^{-1}=\hat l_{\vec p} \Rightarrow \hat l^{-1}_{{\vec p}} \hat r\, \hat l_{{\vec p}_{\lambda}}=\hat r \Rightarrow D(r,{\vec p}) = D(\hat r)$ where
\begin{equation}\label{Wigr}
	D_{\sigma\sigma'}(\hat r)=\xi_{\sigma'}^+\hat r \xi_{\sigma}=\left(\eta_{\sigma'}^+\hat r \eta_{\sigma} \right)^*\,.
\end{equation}
We understand thus  that the specific mechanism of  the induced representations acts  only for the Lorentz boosts,   
$\lambda\in \rho_D[SL(2,\mathbb{C})/SU(2)]$.  

The Wigner induced representations  are unitary with respect to the scalar product (\ref{spa}) \cite{Wig,Mc}, 
\begin{eqnarray}
	\langle \tilde {T}_{\lambda,a} \alpha, \tilde {T}_{\lambda,a} \alpha'\rangle =\langle  \alpha,  \alpha'\rangle	\,,
\end{eqnarray}
and similarly for  $\beta$. Bearing in mind  that the covariant representations are unitary with respect to the scalar product (\ref{sp}) which can be decomposed as in Eq. (\ref{spp}) we conclude that the expansion (\ref{Psi}) establishes the unitary equivalence, ${T}  =\tilde {T}  \oplus \tilde {T} $, of the covariant representation with the {orthogonal} sum  of Wigner's unitary irreducible ones  \cite{Mc}.  Under such circumstances, the self-adjoint generators $\tilde X\in {\rm Lie}(\tilde {T})$   defined as
\begin{eqnarray}
	\tilde	P_{\mu}=-\left.i\frac{\partial \tilde {T}_{1,a}}{\partial a^{\mu}}\right|_{a=0}\,, \quad 
	\tilde	J_{\mu\nu}=\left.i\frac{\partial \tilde {T}_{\lambda(\omega),0}}{\partial \omega^{\mu\nu}}\right|_{\omega=0}\,,
\end{eqnarray}
are just the associated operators of the generators  $X\in {\rm Lie}({T})$ such that
\begin{eqnarray}
	(X\psi)(x)&=&\int d^3p \sum_{\sigma}\left[U_{{\vec p},\sigma}(x)(\tilde X\, \alpha)_{\sigma}({\vec p})- \,V_{{\vec p},\sigma}(x) (\tilde X\,\beta)^ {*} _{ \sigma}({\vec p})\right]\,,\label{basicx}
\end{eqnarray}
as we deduce deriving Eq. (\ref{basic}) with respect to the corresponding group parameter $\zeta\in (\omega, a)$ in $\zeta=0$. We find thus that the isometry generators are reducible  whose associated operators  obeying  $\tilde X^c=-\tilde X$ as a consequence of the fact that $\tilde {T}^c\simeq\tilde {T}$ \cite{Cot}. 

The associated Abelian generators are trivial being diagonal in momentum basis,
\begin{equation}\label{tilHP}
	\tilde H=-\tilde H^c=E(p)\,,\qquad \tilde P^i=-\tilde P^{c\,i}=p^i\,.
\end{equation} 
For rotations we use the Cayley-Klein parameters as in Eq. (\ref{r}) recovering the natural splitting (\ref{spli}),   
\begin{eqnarray}\label{tilJ}
	J_i=L_i+S_i~~&\Rightarrow&~~ \tilde J_i=-\tilde J^c_i=\tilde L_i+\tilde S_i\,,
\end{eqnarray}
laying out the components of the Pryce(e) spin operator (\ref{tilS})  and intuitive components of the orbital angular momentum operator,
\begin{equation}\label{tilL}
	L_i~~\Rightarrow~~\tilde L_i =-\tilde L_i^c=-i\epsilon_{ijk}p^j\tilde \partial_k\,.
\end{equation}
The sets of conserved operators $\{\tilde L_1,\tilde L_2,\tilde L_3\}$ and $\{\tilde S_1,\tilde S_2,\tilde S_3\}$ satisfying Eqs. (\ref{ssuu2}) generate the representations $\tilde T^o$ and $\tilde T^s$ of the associated factorization 
\begin{equation}\label{factor}
	{T}^r=	T^o\otimes T^s~~ \Rightarrow~~\tilde{T}^r=	\tilde T^o\otimes \tilde T^s\,,
\end{equation}	
of the $SU(2)$ restriction $\tilde{T}^r\equiv\left.\tilde{T}\right|_{SU(2)}$ of the representation $\tilde{T}$. For the Lorentz boosts we perform a similar calculation with $ \lambda=l(\tau)$ as in Eq. (\ref{l}) obtaining    a similar splitting, 
\begin{eqnarray}\label{tilK}
	K_i~~&\Rightarrow&~~ \tilde K_i=-\tilde K^c_i=\tilde K_i^o+\tilde K_i^s\,,
\end{eqnarray}
where the orbital and spin components,  
\begin{eqnarray}
	\tilde K_i^o=-{\tilde K_i^o\,}^c&=&iE(p)\tilde \partial_i+i\frac{p^i}{2 E(p)}=\frac{1}{2}\left\{\tilde X^i,E(p)\right\}\,,\label{kaka0}\\
	\tilde K_i^s=-{\tilde K_i^s\,}^c&=&\frac{1}{E(p)+m}\epsilon_{ijk}p^j \tilde S_{k}\,,\label{kakas}
\end{eqnarray}
are no longer commuting among themselves as we see from Eq. (\ref{KiKjs}). This means that  the factorization (\ref{factor}) cannot be extended to the entire $SL(2,\mathbb{C})$ group. Note that the form  (\ref{kaka0}) guarantees that the operators $K_i^o$ are Hermitian with respect to the scalar product (\ref{spa}). \footnote{The second term of Eq. \ref{kaka0}) was omitted in In Eq. (124) of Ref. \cite{Cot} but without affecting other results.} The algebraic properties of these operators are presented in the Appendix B where we show how an algebra of orbital operators in MR can be selected. This is formed by the orbital subalgebra  Lie$(\tilde{T^o})$  generated by the set $\{E(p), p^i,\tilde L_i,\tilde K_i^o\}$  and the kinetic operators $\tilde X^i$ and $\tilde V^i$ which do not have spin parts. 

Finally, let us turn back to the Pauli-Lubanski  operator whose components are formed by products of isometry generators as in  Eq. (\ref{PaLu2}). After a few manipulation we find the associated operators 
\begin{eqnarray}
	W^0&\Rightarrow&\tilde W^0=\tilde W^{c\,0}=p^i\tilde S_i\,,\label{tilW0}\\
	W^i&\Rightarrow&\tilde W^i=\tilde W^{c\,i}=E(p)\tilde J_i+\epsilon_{ijk}p^j\tilde K^s_k=m\tilde S_i^{(+)}\,,\label{tilWi}
\end{eqnarray}
expressed  in terms of  operators (\ref{tilS}) and (\ref{tilS+}) . Hereby we recover the identity $P^{\mu}W_{\mu}=E(p)\tilde W_0-p^i\tilde W^i=0$  and the well-known  invariant $\tilde W^{\mu} \tilde W_{\mu}=-\frac{3}{4}m^2 1_{2\times 2}$. In the Appendix B we give the commutation relations of the components $\tilde W^{\mu}$  with our new  operators  $\tilde S_i$ and $\tilde X^i$ that complete the  algebraic properties we already  know  \cite{Th,WKT}.  

\subsection{Spectral representations}

The correspondence $A \Leftrightarrow (\tilde A,\tilde A^c)$ defined by Eq. (\ref{AAAc}) is bijective. We have seen how $A$ generates the operators $\tilde A$ and $\tilde A^c$ so we have to face now with the inverse problem we try to  solve resorting to spectral representations as those defined in Ref. \cite{Cot} in the particular case when $\tilde A$ and $\tilde A^c$ are matrix-operators. In what follows we  generalize these spectral representations  to any equal-time associated operators whose action on the wave spinors is given by arbitrary kernels. 

Let us start with  the equal-time integral operator (\ref{Y1}) whose action in CR is given by the time-dependent bi-local kernel ${\frak A}$. In addition, we assume that $A$ is reducible  its associated operators acting as
\begin{eqnarray}
	&&	(\tilde A\alpha)_{\sigma}(t,{\vec p})=\int d^3p' \sum_{\sigma'}\tilde{\frak A}_{\sigma \sigma\,' }(t,{\vec p},{\vec p}\,' )\alpha_{\sigma'}({\vec p}\,' )\,,\\
	&&	(\tilde A^c\beta)_{\sigma}(t,{\vec p})=\int d^3p' \sum_{\sigma'}\tilde{\frak A}_{\sigma \sigma'}^c(t,{\vec p},{\vec p}\,' )\beta_{\sigma\,' }({\vec p}')\,.~~~~~~
\end{eqnarray} 
In this case  we may exploit the orthonormalization and the completeness properties of the mode spinors, given by Eqs. (\ref{ortU}), (\ref{ortV}) and (\ref{comp}), for relating the kernels of the associated operators through the spectral representation,  
\begin{eqnarray}
	{ \frak A}(t,{\vec x},{\vec x}\,' )
	&=&\int d^3p\,d^3p'\sum_{\sigma\sigma'}\left[ U_{{\vec p},\sigma}(t,{\vec x})\tilde{\frak A}_{\sigma \sigma'}(t,{\vec p},{\vec p}')U^+_{{\vec p}\,' ,\sigma'}(t,{\vec x}\,' )\right.\nonumber\\
	&&\left.\hspace*{12mm} +V_{{\vec p},\sigma}(t,{\vec x})\tilde{\frak A}_{\sigma \sigma'}^{c\,*}(t,{\vec p},{\vec p}\,' )V^+_{{\vec p}\,' ,\sigma'}(t,{\vec x}\,' )	\right] \,,  \label{intr1}
\end{eqnarray}
giving the action of the operator $A$ in CR when we know the actions of the associated operators $\tilde A$ and $\tilde A^c$.

This mechanism is useful for taking over in CR  the principal properties of our operators we defined in  MR where we studied the induced Wigner representations and their generators.  In spite of their manifest covariance, the operators $T_{\lambda,a}$ can be seen as equal-time operators after the transformation (\ref{subT}). Their kernels in CR, ${\frak T}_{\lambda,a}(t,{\vec x},{\vec x}\,' )$, may be derived  according to the spectral representation (\ref{intr1}) where we have to substitute the kernels in MR that are time-independent having the form
\begin{eqnarray}
	\tilde{\frak T}_{\lambda,a}({\vec p},{\vec p}\,' )=\tilde{\frak T}^c_{\lambda,a}({\vec p},{\vec p}\,')
	=\delta^3\left( {\vec p}_{\lambda}-{\vec p}\,' \right) e^{ia\cdot p}\sqrt{\frac{E(p')}{E(p)}}D(\lambda, {\vec p})\,,~~~~\label{KerT}
\end{eqnarray}
depending on the momentum (\ref{pLp}) and matrix (\ref{Wrot}). In a similar manner we may write the spectral representations of the kernels of the basis-generators for which we separated the orbital parts, $\tilde L_i$, $\tilde K^o_i$ and $\tilde X^i$,  depending on momentum derivatives. According to the results of  Sec. 4.2 we may write the kernels of these operators in MR,   
\begin{eqnarray}
	\tilde{\frak L}_i({\vec p}, {\vec p}\,' )&=&-\tilde{\frak L}^c_{i}({\vec p}, {\vec p}\,' )=-i\epsilon_{ijk}p^j\tilde \partial_k \delta^3({\vec p}-{\vec p}\,') 1_{2\times 2}	\,,\label{Lulu}\\
	\tilde{\frak K}^o_{i}({\vec p}, {\vec p}\,' )&=&-{\tilde{\frak K}_{i}^o\,}^c({\vec p}, {\vec p}\,')	=\left[\delta^3({\vec p}-{\vec p}\,' )\frac{ip^i}{2E(p)}+i E(p)\tilde \partial_i \delta^3({\vec p}-{\vec p}')\right] 1_{2\times 2}	\,,\\
	\tilde{\frak X}^i({\vec p}, {\vec p}\,' )&=&{\tilde{\frak X}^i\,}^c({\vec p}, {\vec p}\,' )=i\tilde \partial_i \delta^3({\vec p}-{\vec p}\,' ) 1_{2\times 2}	\,,
\end{eqnarray} 
that substituted in Eq. (\ref{intr1}) will give the kernels of the operators $L_i$, $K^o_i$ and $X^i$ acting in CR as integral operators that may depend on time.

In the particular case when $A\in F[t]$ is a Fourier operator then the associated operators have the kernels
\begin{eqnarray}
	\tilde{\frak A}(t,{\vec p},{\vec p}\,' )&=&\delta^3({\vec p}-{\vec p}\,' )\tilde A(t,{\vec p})  \,,\\
	\tilde{\frak A}^c(t,{\vec p},{\vec p}\,' )&=&\delta^3({\vec p}-{\vec p}\,' )\tilde A^c(t,{\vec p})  \,,	
\end{eqnarray} 
which solve one of the integrals of the spectral representation (\ref{intr1}) remaining with the simpler form
\begin{eqnarray}
	{ \frak A}(t,{\vec x}-{\vec x}\,' )=\int d^3p\sum_{\sigma\sigma'}\left[ U_{{\vec p},\sigma}(t,{\vec x})\tilde A_{\sigma \sigma'}(t,{\vec p})U^+_{{\vec p},\sigma'}(t,{\vec x}\,' ) +V_{{\vec p},\sigma}(t,{\vec x})\tilde A^c_{\sigma \sigma'}(t,{\vec p})^*V^+_{{\vec p},\sigma'}(t,{\vec x}\,' )\right]  \,,\label{intr2}
\end{eqnarray}
that can be applied to all the spin parts of our operators. 

In ref. \cite{Cot} we used this type of spectral representation for studying the transformations  (\ref{Rs}) of the spin symmetry starting with the identities
\begin{eqnarray}
	\hat r \xi_{\sigma}&=&\sum_{\sigma'}\xi_{\sigma'}D_{\sigma'\sigma}(\hat 
	r) 
~~~	\Rightarrow~~~ 	U_{{\vec p},\hat r \xi_{\sigma}}(x)=\sum_{\sigma'} U_{{\vec p},\xi_{\sigma'}}(x)D_{\sigma'\sigma}(\hat r)\,,\label{Uxi}\\
	\hat r \eta_{\sigma}&=&\sum_{\sigma'}\eta_{\sigma'}D^*_{\sigma'\sigma}(\hat 
	r) 
~~~	\Rightarrow~~~ V_{{\vec p},\hat r \eta_{\sigma}}(x)=\sum_{\sigma'} V_{{\vec p},\eta_{\sigma'}}(x)D^*_{\sigma'\sigma}(\hat r)\,,\label{Veta}
\end{eqnarray}
where $r$ are the rotations (\ref{r0})  of $\rho_D$ while the matrices $D(\hat r)$ are defined by Eq. (\ref{Wigr}).  Under such circumstances, the operator $T^s_{\hat r}$ can be defined as the integral Fourier operator having the local time-independent  kernel
\begin{equation}\label{KerY}
	{\frak T}^s_{\hat r}({\vec x}-{\vec x}\,' ) =\int d^3p\,\frac{e^{i ({\vec p}-{\vec p}\,' )\cdot{\vec x}}}{(2\pi)^3}  {T}^s_{\hat r}({\vec p})\,,
\end{equation} 
given by Eq. (\ref{intr2}) where we substitute
\begin{equation}
	\tilde A_{\sigma\sigma'}(t,{\vec p})	=\tilde A^c_{\sigma\sigma'}(t,{\vec p})=D_{\sigma\sigma'}(\hat r)\,.
\end{equation}
The Fourier transform of $T^s_{\hat r}({\vec p})$ can be derived now considering the form of   the mode spinors (\ref{Ufin}) and (\ref{Vfin}) and using the identities (\ref{Uxi}), (\ref{Veta}) and (\ref{idll}). After a little calculation we obtain
\begin{eqnarray}
	{T}^s_{\hat r} ({\vec p})&=&\frac{m}{E(p)}\left[ l_{\vec p}\,  r\, \frac{1+\gamma^0}{2}l_{\vec p}+l^{-1}_{\vec p} r\, \frac{1-\gamma^0}{2}l^{-1}_{\vec p}\right] \nonumber\\
	&=& l_{\vec p} \, r\,  l_{\vec p}^{-1}\tilde \Pi_+({\vec p}) + l_{\vec p}^{-1} r\,  l_{\vec p}\tilde \Pi_-({\vec p})	\,.\label{FR}
\end{eqnarray}  
This spectral representation  was crucial for showing that the spin components defined by Eq. (\ref{Spipi}) have just the Fourier transforms (\ref{PrS}) proposed by Pryce as his version (e). In Ref.  \cite{Cot} we started with the  Fourier transform (\ref{FR}) where we substituted $\hat r=\hat r(\theta)$ given by Eq. (\ref{r}). Applying then the definition (\ref{Spipi}) we found the Fourier transforms (\ref{Sip}) which, fortunately, coincide with those proposed by Pryce as we deduced after using a suitable code on computer.

We have now all the elements we  need for writing down for the first time the kernels of the operators $T^o_{\hat r}$ of the orbital representation of the $SO(3)$ group which are no longer Fourier operators. These operators are  defined by Eq. (\ref{Ro}) which combines the actions of $T_{r,0}$ and $T^s_{\hat r}$ such that, according to Eqs. (\ref{KerT}) and (\ref{KerY}), we may write the associated kernels in MR,
\begin{eqnarray}
	\tilde{\frak T}^o_{\hat r} ({\vec p},{\vec p}\,' )&=&\tilde{\frak T}^{o\,c}_{\hat r} ({\vec p},{\vec p}\,' )\nonumber\\
	&=&\delta^3\left( {\vec p}_{\hat r}-{\vec p}\,' \right)D^{-1}(\hat r)D(\hat r, {\vec p})\,,\nonumber\\
	&=&\delta^3\left( {\vec p}_{\hat r}-{\vec p}\,' \right)1_{2\times 2}\,, \quad p_{\hat r}=R(\hat r)^{-1}{p}\,,~~~~~~
	\label{KerT1}
\end{eqnarray}
as it results from Eq. (\ref{Wigr}).  Substituting these kernels in Eq. (\ref{intr1}) we obtain the kernels of the operators $T^o_{\hat r}$ of the orbital representation acting on the free fields in CR.   Finally, substituting  again $\hat r \to \hat r(\theta)$ in $T^o_{\hat r}$ and applying the definition (\ref{Lpipi}) we obtain the kernels (\ref{Lulu}) giving  the action of the operators (\ref{tilL})  in MR directly, without resorting to Wigner's theory as in Sec. 3.3.

We conclude that the action of the operators of the spin and orbital symmetries can be properly defined thanks to our spectral representations outlined in Ref. \cite{Cot} for Fourier operators and generalized here to any equal-time integral operators.

\section{Quantum theory}

The quantization reveals the physical meaning of the quantum observables of RQM transforming them in  operators of QFT. The principal benefit of our approach  is the association between the operator actions  in  CR  and MR allowing us to derive at any time the expectation values of  the operators defined in MR according to the general rule (\ref{expA}).  We get thus the opportunity of applying the Bogolyubov method for quantizing the operators of RQM.   

\subsection{Quantization}

In special relativistic QFT  each observer has its own measure apparatus formed by the set of observables defined in its proper frame at a fixed initial time. As we adopted already the point of view of an observer staying at rest in origin preparing  the free fields at the initial time $t=0$, we assume that this observer keeps the same initial condition for quantization.  

Applying the Bogolyubov method of quantization \cite{Bog} we replace first  the wave spinors of MR with field operators, $(\alpha, \alpha^*)\to ({\frak a},{\frak a}^{\dag})$ and $(\beta, \beta^*)\to ({\frak b},{\frak b}^{\dag})$,  satisfying  canonical   anti-commutation relations among them  the non-vanishing ones are, 
\begin{eqnarray}
	\left\{{\frak a}_{\sigma}({\vec p}),{\frak a}_{\sigma'}^{\dag}({\vec p}^{\,\prime})\right\}=	\left\{{\frak b}_{\sigma}({\vec p}),{\frak b}_{\sigma'}^{\dag}({\vec p}^{\,\prime})\right\}
	=\delta_{\sigma\sigma'}\delta^3({\vec p}-{\vec p}^{\,\prime})\,.
\end{eqnarray}
The Dirac free field becomes thus the field operator  
\begin{eqnarray}\label{Psiq}
	\psi(x)=\int d^3p \sum_{\sigma}\left[U_{{\vec p},\sigma}(x) {\frak a}_{\sigma}({\vec p}) +V_{{\vec p},\sigma}(x) {\frak b}^ {\dag} _{ \sigma}({\vec p})\right]\,,
\end{eqnarray}  
denoted with the same symbol but acting on  the Fock state space equipped with the scalar product $\langle~~|~~\rangle$ and a normalized vacuum state $|0\rangle$ accomplishing
\begin{eqnarray}
	{\frak a}_{\sigma}({\vec p})|0\rangle={\frak b}_{\sigma}({\vec p})|0\rangle=0\,,\quad \langle 0|{\frak a}_{\sigma}^{\dagger}({\vec
		p})=\langle 0|  {\frak b}_{\sigma}^{\dagger}({\vec p})=0\,.
\end{eqnarray}
The sectors with different number of particles have to be constructed applying the standard method for constructing generalized momentum bases of various polarizations.

Through quantization the  expectation value of any  time-dependent operator $A(t)$  of RQM becomes  an operator, 
\begin{equation}\label{qA} 
	A(t)~\to ~ \mathsf{A}=\left.:\langle\psi , A(t)\psi\rangle_D :\right| _{t=0}\,,
\end{equation}
calculated  respecting the normal ordering of the operator products \cite{BDR} at the initial time $t=0$.  This procedure allows us to write down any operator $\mathsf{A}$ directly in terms of  the operators  associated to the operator $A=A(t)|_{t=0}$. We consider first the reducible operators complying with the condition (\ref{redcon}) for which we  obtain the  general formula
\begin{equation}\label{Aq}
	\mathsf{A}=\int d^3{p} \left[ {\frak a}^{\dag}({\vec p})(\tilde A {\frak a})({\vec p}) -  {\frak b}^{\dag}({\vec p})(\tilde A^{c\,+} {\frak b})({\vec p})\right]\,,
\end{equation}
written with the compact notation
\begin{equation}\label{compnot}
	{\frak a}^{\dag}({\vec p})(\tilde A {\frak a})({\vec p})\equiv\sum_{\sigma} {\frak a}^{\dag}_{\sigma}({\vec p})(\tilde A {\frak a})_{\sigma}({\vec p})\,,
\end{equation}
and similarly for the second term. For shortening the terminology we say here that the associated operators $A\Leftrightarrow (\tilde A, \tilde A^c)$ are the {\em parents} operators of $\mathsf{A}$. We specify that the bracket (\ref{qA}) is calculated according to Eq. (\ref{expA}) in which the last term changes its sign after introducing the normal ordering of the operator products. When $\tilde A^c=- \tilde A$ we say that the one-particle operator (\ref{Aq}) is even (of positive charge parity) describing an additive property which is similar for particles and antiparticles as, for example,  the energy, momentum, spin, etc. The odd operators (with negative charge parity), for which   $\tilde A^c= \tilde A$, describe electrical properties depending on the opposite charges of particles and antiparticles. We introduce thus the operator signature which behaves in commutation relations as the usual algebraic signs in multiplication, e. g. $[A_{\rm odd}, B_{\rm odd}]=C_{\rm even},\, [A_{\rm odd}, B_{\rm even}]=C_{\rm odd},...$ etc.

Given an arbitrary operator $A\in {\rm Aut}({\cal F})$ and its Hermitian conjugated $A^+$ we define the adjoint operator of $\mathsf{A}$, 
\begin{eqnarray}
	A^+(t)\Rightarrow \mathsf{A}^{\dagger}=\left.:\langle \psi , A(t)^+\psi\rangle_D :\right| _{t=0}=\left.:\langle A(t) \psi , \psi\rangle_D :\right| _{t=0}\,,
\end{eqnarray}
complying with the standard definition  $\langle \alpha |\mathsf{A}^{\dagger}\beta\rangle =\langle\mathsf{A} \alpha |\beta\rangle$ on the Fock space.  In what follows we shall meet only  self-adjoint one-particle operators  as all their parent operators of RQM are reducible and Hermitian with respect to the scalar products of the spaces in which they act. We obtain thus an  operator algebra  formed by fields  and self-adjoint one-particle operators which have the obvious properties
\begin{eqnarray}
	\left[\mathsf{A}, \psi(x)\right]&=&-(A\psi)(x)\,, \label{algXX}\\
	\left[\mathsf{A}, \mathsf{B}\right]&=&:\left<\psi, [A,B]\psi\right>_D: \,,\label{algXX1}
\end{eqnarray} 
preserving the structures of Lie algebras but without taking over other algebraic properties of their parent operators from  RQM as the product of two one-particle operators is no longer an operator of the same type. Therefore, we must restrict ourselves to the Lie algebras of symmetry generators  and unitary transformations  whose actions reduce to sums of successive commutations according to the well-known rule
\begin{equation}\label{eXY}
	e^{\mathsf{X}}\mathsf{Y}e^{-\mathsf{X}}=\mathsf{Y}+[\mathsf{X},\mathsf{Y}]+\frac{1}{2} [\mathsf{X}, [\mathsf{X},\mathsf{Y}]]+\frac{1}{3!}[\mathsf{X}, [\mathsf{X}, [\mathsf{X},\mathsf{Y}]]]... 
\end{equation}
we have to use in what follows. 

The Poincar\' e generators (\ref{Pgen}) give rise to the self-adjoint one-particle operators calculated at the initial time $t=0$,
\begin{eqnarray}
	\mathsf{P}_{\mu}=:\langle \psi, P_{\mu}\psi\rangle_D:\,, \quad 	\mathsf{J}_{\mu\nu}=\left.:\langle \psi, J_{\mu\nu}\psi\rangle_D:\right|_{t=0}\,.
\end{eqnarray}
The brackets corresponding to  the operators $\mathsf{P}^{\mu}$ and $\mathsf{S}_{ij}$ are independent on time but for the operators $\mathsf{S}_{0i}$ we must impose the initial condition following to see later how these operators evolve in time. With these generators we may construct unitary transformations with various parametrizations among them we choose here those of the first kind defining the unitary  operators of translations and $SL(2,\mathbb{C})$ transformations as,
\begin{eqnarray}
	\mathsf{U}(a)&=&\exp\left(-i a^{\mu}\mathsf{P}_{\mu}\right)\,, \quad\quad\quad~~	a\in T(4)\,,\\
	\mathsf{U}(\omega)&=&\exp\left(\frac{i}{2}\, \omega^{\mu\nu}\mathsf{J}_{\mu\nu}\right)\,, ~~	\lambda(\omega)\in \rho_D[SL(2,\mathbb{C})]\,,~~~~~~
\end{eqnarray} 
in accordance with our definition (\ref{Pgen}) of the isometry generators and the rule (\ref{algXX}).  This construction guarantees the expected isometry transformations of the field operators, 
\begin{eqnarray}
		\mathsf{U}(a){\frak a}_{\sigma}({\vec p})\mathsf{U}^{\dagger}(a)&=&\left( \tilde T_{1,a}\,{\frak a}\right)_{\sigma}({\vec p}) =e^{ia\cdot p}{\frak a}_{\sigma}({\vec p})\,,\label{Tra}\\
		\mathsf{U}(\omega){\frak a}_{\sigma}({\vec p})\mathsf{U}^{\dagger}(\omega)&=&\left( \tilde T_{\lambda(\omega),0}\,{\frak a}\right)_{\sigma}({\vec p})	=\sqrt{\frac{E(p_{\lambda} )}{E( p)}}\sum_{\sigma'}{D}_{\sigma\sigma'}\left(\lambda(\omega),{\vec p}\right) {\frak a}_{\sigma'}({\vec p}_{\lambda}) \,,	\label{Lala}
\end{eqnarray}
where the matrix $D$ is given by Eq. (\ref{Wrot}) and ${\vec p}_{\lambda} $ by Eq. (\ref{pLp}).
As the operators ${\frak a}_{\sigma}$ and ${\frak b}_{\sigma}$ transform alike under isometries we obtain from Eq. (\ref{subT}) the transformations of the quantum field,  
\begin{eqnarray}
	\mathsf{U}(a)\psi(x)\mathsf{U}^{\dagger}(a)&=&\left( T_{1,a}\,\psi\right)(x)=\psi(x-a)\,,\\
	\mathsf{U}(\omega)\psi(x)\mathsf{U}^{\dagger}(\omega)&=&\left( T_{\lambda(\omega),0}\,\psi\right)(x)=\lambda(\omega)\psi\left(\Lambda^{-1}(\lambda(\omega)) x\right)\,.
\end{eqnarray}
Moreover, the isometry generators transform usually according to the adjoint representation of the Poincar\' e group \cite{WKT} assuring thus the relativistic covariance. In the case of Lorentz transformations   $\lambda(\omega)\in \rho_D[SL(2,\mathbb{C})]$ we have 
\begin{eqnarray}
	\mathsf{U}(\omega)P_{\mu}\mathsf{U}^{\dagger}(\omega)&=&\Lambda_{\mu\,\cdot}^{\cdot\,\alpha}(\omega) P_{\alpha}\,,\\
	\mathsf{U}(\omega)J_{\mu\nu}\mathsf{U}^{\dagger}(\omega)&=&\Lambda_{\mu\,\cdot}^{\cdot\,\alpha}(\omega) \Lambda_{\nu\,\cdot}^{\cdot\,\beta}(\omega)J_{\alpha\beta}\,,	
\end{eqnarray}
where $\Lambda(\omega)$ is defined in the Appendix A. We may say  thus that the unitary operators $\mathsf{U}(a)$ and $\mathsf{U}(\omega)$ encapsulate the entire theory of the relativistic covariance under Poincar\' e isometries. More specific, the transformations 
\begin{eqnarray}
	\mathsf{U}(\omega,a)=\mathsf{U}(\omega) \mathsf{U}(a) \,:\,  \mathsf{A}~\to~ \mathsf{A}'=\mathsf{U}(\omega,a)\mathsf{A}\mathsf{U}^{\dag}(\omega, a)\,,
\end{eqnarray}
of an operator expressed in terms of particle and antiparticle operators can be derived by using Eqs.  (\ref{Tra}) and  (\ref{Lala}). In general,  these transformations are not manifest covariant because of their momentum-dependent transformation matrices remaining under the integral over momenta.

We have seen that the quantization is performed at the initial time $t=0$ when one obtains a set of one-particle operators among them we may find conserved operators which commute with the energy one  $\mathsf{H}=\mathsf{P}_0$ or dynamical operators whose time evolution is governed by the translation operator generated by  $\mathsf{H}$, 
\begin{equation}\label{evolve}
	\mathsf{U}(t)=\exp\left(-i t \mathsf{H}\right)\,: \, \mathsf{A}~\to~\mathsf{A}(t)=\mathsf{U}^{\dagger}(t)\mathsf{A}\mathsf{U}(t) \,.	
\end{equation} 
Thus the observer staying at rest in origin recovers the time evolution of the observables obtained through quantization at the initial time $t=0$.  

\subsection{Reducible operators}

The reducible operators of RQM give rise to the one-particle operators of  QFT.  There are two such operators  commuting with the entire algebra of observables namely, the charge operator $\mathsf{Q}=\mathsf{N}_+-\mathsf{N}_-$ and that of the total  number  of particles $\mathsf{N}=\mathsf{N}_++\mathsf{N}_-$, formed by the particle and antiparticle number operators
\begin{eqnarray}
	\mathsf{N_+}&=&:\langle\psi,\Pi_+ \psi\rangle_D:=\int d^3p\,{\frak a}^{\dag}({\vec p}){\frak a}({\vec p}) \,,	\label{Nplus}\\
	\mathsf{N}_-&=&:-\langle\psi, \Pi_- \psi\rangle_D:=\int d^3p\, {\frak b}^{\dag}({\vec p}){\frak b}({\vec p})\,,\label{Nminus}
\end{eqnarray}
coming from the parent operators $\pm\Pi_{\pm}$ of RQM.  Other diagonal operators in momentum basis are the translations generators, energy and momentum,  
\begin{eqnarray}
	\mathsf{H}&=&:\langle\psi, H\psi\rangle_D:=\int d^3p\,E(p)\left[{\frak a}^{\dag}({\vec p}){\frak a}({\vec p}) +{\frak  b}^{\dag}({\vec p}){\frak b}({\vec p})\right]\,,\label{Hom}\\ 
	\mathsf{P}^i&=&:\langle\psi, P^i\psi\rangle_D:=\int d^3p\,p^i\left[{\frak a}^{\dag}({\vec p}){\frak a}({\vec p}) +{\frak b}^{\dag}({\vec p}){\frak b}({\vec p})\right]\,, \label{Pom}	
\end{eqnarray}
as well as our new operator of fermion polarization  \footnote{In Eqs. (115) and (149) of Ref. \cite{Cot} the factor $\frac{1}{2}$ must be ignored.},	
\begin{eqnarray}
	\mathsf{W}_s& =&:\langle \psi, W_s\psi\rangle_D:=\frac{1}{2}\int d^3p\left[{\frak a}^{\dag}({\vec p}){\sigma}_3
	{\frak a}({\vec p}) +{\frak b}^{\dag}({\vec p}){\sigma}_3{\frak b}({\vec p})\right]\,, \label{Polq}
\end{eqnarray}
which completes the set 
$\{\mathsf{H},\mathsf{P}^1,\mathsf{P}^2, \mathsf{P}^3,\mathsf{W}_s,\mathsf{Q}\}$	
of commuting operators determining the momentum bases of the Fock state space. 

Applying the general rule (\ref{Aq})  to the associated rotation generators (\ref{tilJ}) we find the splitting of the total angular momentum 
\begin{eqnarray}\label{split}
	\mathsf{J}_i=:\langle\psi, J_i\psi\rangle_D:&=&:\langle\psi, L_i\psi\rangle_D:+:\langle\psi, S_i\psi\rangle_D: =\mathsf{L}_i+\mathsf{S}_i\,,
\end{eqnarray} 
where  the = of the orbital angular momentum, $\mathsf{L}_i$, and  spin operator, $\mathsf{S}_i$, can be written as  
\begin{eqnarray}
	&&	\mathsf{L}_i =-\frac{i}{2}\int d^3p\, \epsilon_{ijk} p^j \left[{\frak a}^{\dag}({\vec p}){\stackrel{\leftrightarrow}{\tilde\partial_{i}}}{\frak a}({\vec p})+{\frak b}^{\dag}({\vec p}){\stackrel{\leftrightarrow}{\tilde\partial_{i}}}{\frak b}({\vec p})\right]\,,\label{Lang}\\
	&&	\mathsf{S}_i =\frac{1}{2}\int d^3p\left[{\frak a}^{\dag}({\vec p})\Sigma_{i}({\vec p}){\frak a}({\vec p})+{\frak b}^{\dag}({\vec p})\Sigma_{i}({\vec p}){\frak b}({\vec p})\right]\,,~~~~~~~~~
	\label{Spin}
\end{eqnarray}
according to Eqs. (\ref{tilL}) and (\ref{tilS}). Here we  use the special notation
\begin{equation}\label{stak}
	\alpha^+ \stackrel{\leftrightarrow}{\tilde\partial_{i}} \beta =\alpha^+ (\partial_{p^i}\beta)-(\partial_{p^i}\alpha^+)\beta +2\alpha^+\Omega_i({\vec p}) \beta\,,
\end{equation}  
inspired by Green's theorem, which points out explicitly that $\mathsf{L}_i$ are self-adjoint operators. The components $\mathsf{L}_i $ and  $\mathsf{S}_i$  form the bases of two {\em independent} unitary representations of the  $su(2)\sim so(3)$ algebra,  $\left[\mathsf{L}_i,\mathsf{S}_j\right]=0$, generating the orbital and respectively spin symmetries. These operators are {\em conserved} as they commute with $\mathsf{H}$ while the commutation relations 
\begin{equation}
	\left[\mathsf{L}_i,\mathsf{P}^j\right]=i\epsilon_{ijk}\mathsf{P}^k\,, \qquad 	\left[\mathsf{S}_i,\mathsf{P}^j\right]=0\,,
\end{equation}
show that only the spin operator is   invariant under space translations.  Moreover,  using Eqs. (\ref{Lala}) and (\ref{rsigr}) and changing then the integration variable, $\vec{p}_{\lambda}\to \vec{p}$, we obtain the transformation of the spin operator under arbitrary transformations $\lambda(\omega)\in SL(2,\mathbb{C})$ as
\begin{eqnarray}
		\Lambda(\omega)\,:\, \mathsf{S}_i\to \mathsf{S}_i'&=&\mathsf{U}(\omega)\mathsf{S}_i\mathsf{U}^{\dag}(\omega)\nonumber\\
	&=&\frac{1}{2}\int d^3p\,\left[{\frak a}^{\dag}({\vec p})\Sigma'_{i}({\vec p}){\frak a}({\vec p})+{\frak b}^{\dag}({\vec p})\Sigma'_{i}({\vec p}){\frak b}({\vec p})\right]\,,~~~~~~	\label{Spintr}
\end{eqnarray}
where $\Sigma_i'({\vec p})= R_{ij}(\omega,\vec{p}) \Sigma_j({\vec p})$ are the transformed $\Sigma$-matrices under the Wigner rotations 
\begin{equation}
	R(\omega,\vec{p})=\Lambda\left( w[\lambda(\omega), \Lambda(\omega)\vec{p}]\right)=L_{\Lambda(\omega)\vec{p}}^{-1}\Lambda(\omega) L_{\vec{p}}\,.
\end{equation} 
For genuine rotations, $\lambda(\omega)=r\in \rho_D[SU(2)]$, the matrix $R(r)$ is independent on momentum such that the spin operator transforms as a $SO(3)$ vector-operator,  $\mathsf{S}_i\to R_{ij}(r)\mathsf{S}_j$. We may conclude that  the quantum version of the Pryce(e) spin operator $\vec{\mathsf{S}}$ transforms covariantly only under rotations. 

The generators of the Lorentz boosts have the general form (\ref{Aq}) depending on the operators (\ref{tilK}) which have orbital and  spin terms suggesting the splitting,  
\begin{equation}\label{splitK}
	\mathsf{K}_i=:\langle\psi, K_i\psi\rangle_D:=\mathsf{K}^o_i+\mathsf{K}^s_i	\,,
\end{equation}
in orbital and spin parts that read 
\begin{eqnarray}
	&&	\mathsf{K}^o_i 	=
	\frac{i}{2}\int d^3p\, E(p) \left[{\frak a}^{\dag}({\vec p}){\stackrel{\leftrightarrow}{\tilde\partial_{i}}}{\frak a}({\vec p})+{\frak b}^{\dag}({\vec p}){\stackrel{\leftrightarrow}{\tilde\partial_{i}}}{\frak b}({\vec p})\right] \,, ~~~~~~\label{Kt0}\\
	&&	\mathsf{K}^s_i 	=\int d^3p  \left[{\frak a}^{\dag}({\vec p})\tilde K^s_{i}{\frak a}({\vec p}) +{\frak b}^{\dag}({\vec p})\tilde K^s_{i}{\frak b}({\vec p})\right]\,, \label{Kt0s}
\end{eqnarray} 
as it results from Eqs. (\ref{kaka0}) and (\ref{kakas}).  The  commutation relations 
\begin{eqnarray}
	&&\left[\mathsf{H},\mathsf{K}^o_i\right]=-i\mathsf{P}^i\,, \quad  \left[\mathsf{P}^i,\mathsf{K}^o_j\right]=-i \delta^i_j\mathsf{H}\,,\\
	&&\left[\mathsf{H},\mathsf{K}^s_i\right]=0\,, \quad\quad~~  \left[\mathsf{P}^i,\mathsf{K}^s_j\right]=0\,,
\end{eqnarray}
show that only the operators $\mathsf{K}^s_i$ are conserved and invariant under translations while $\mathsf{K}^o_i$ satisfy usual orbital commutation relations  evolving as 
\begin{equation}
	\mathsf{K}^o_i(t)=\mathsf{U}^{\dagger}(t)\mathsf{K}^o_i \mathsf{U}(t) =\mathsf{K}^o_i+ \mathsf{P}^i\,t\,,\label{coordK}
\end{equation} 
which means that the generators (\ref{splitK}) are time-dependent,
\begin{eqnarray}
	\mathsf{K}_i(t)=\mathsf{U}^{\dagger}(t)\mathsf{K}_i \mathsf{U}(t) =\mathsf{K}^o_i(t)+\mathsf{K}^s_i=\mathsf{K}_i+ \mathsf{P}^i\,t\,,
\end{eqnarray}
evolving linearly in time.

The operators discussed above satisfy commutation relations similar to those given in the Appendix B for their associated  parent operators of RQM. The set  $\{\mathsf{H}, \mathsf{P}^i,\mathsf{J}_i,\mathsf{K}_i\}$ generates the representation of the Lie$(\tilde P_{+}^{\uparrow})$   algebra with values in one-particle operators which includes the orbital subalgebra generated by $\{\mathsf{H}, \mathsf{P}^i,\mathsf{L}_i,\mathsf{K}^o_i\}$. In contrast, the operators $\mathsf{S}_i$ and $\mathsf{K}^s_i$ do not close an algebra, each commutator giving rise to a new operator  generating thus an infinite Lie algebra.

The operators  (\ref{tilW0}) and (\ref{tilWi}) associated to the components of the Pauli-Lubanski operator give rise to the odd one-particle operators
\begin{eqnarray}
	\mathsf{W}^0&=&\frac{1}{2}\int d^3p\, p^i \left[{\frak a}^{\dag}({\vec p})\Sigma_{i}({\vec p}){\frak a}({\vec p})-{\frak b}^{\dag}({\vec p})\Sigma_{i}({\vec p}){\frak b}({\vec p})   \right]
	\,,\\
	\mathsf{W}^i&=&m\frac{1}{2}\int d^3p \,\Theta_{ij}({\vec p}) 	\left[{\frak a}^{\dag}({\vec p})\Sigma_{j}({\vec p}){\frak a}({\vec p}) -{\frak b}^{\dag}({\vec p})\Sigma_{j}({\vec p}){\frak b}({\vec p})   \right],
\end{eqnarray}
where the tensor $\Theta$ is defined in Eq. (\ref{tete}). The operator $\mathsf{W}^0$ is known as the {\em helicity} operator as in the momentum-helicity basis (presented in the Appendix D) this takes the form    
\begin{equation}\label{W0}
	\mathsf{W}^0=\frac{1}{2}\int d^3p\,p\left[{\frak a}^{\dag}({\vec p}){\sigma_3}{\frak a}({\vec p}) -{\frak b}^{\dag}({\vec p}){\sigma_3}{\frak b}({\vec p})\right]	\,,
\end{equation}
resulted from the identity (\ref{pSig}). A dimensionless version of this operator called the helical operator  was defined recently  for any peculiar polarization  as \cite{V1,V2} 
\begin{equation}\label{Wh}
	\mathsf{W}_h=\frac{1}{2}\int d^3p\, \frac{p^i}{p} \left[{\frak a}^{\dag}({\vec p})\Sigma_{i}({\vec p}){\frak a}({\vec p})-{\frak b}^{\dag}({\vec p})\Sigma_{i}({\vec p}){\frak b}({\vec p})   \right]\,,
\end{equation} 
becoming in momentum-helicity basis  the odd replica  of our polarization operator (\ref{Polq}) which is even by definition.

A special set of operators whose quantization  deserves to briefly examined is formed by the operators  (\ref{SteS}) related to the historical Frankel and Pryce(c)-Czochor proposals.  The associated operators (\ref{tilS+}) and (\ref{tilS-})  give the corresponding even one-particle operators 
\begin{eqnarray}
	\mathsf{S}^{(+)}_i &=&\frac{1}{2}\int d^3p\,\Theta_{ij}({\vec p})\left[{\frak a}^{\dag}({\vec p})\Sigma_{i}({\vec p}){\frak a}({\vec p})+{\frak b}^{\dag}({\vec p})\Sigma_{i}({\vec p}){\frak b}({\vec p})\right]\,,\\
	\mathsf{S}^{(-)}_i &=&\frac{1}{2}\int d^3p\,\Theta^{-1}_{ij}({\vec p})\left[{\frak a}^{\dag}({\vec p})\Sigma_{i}({\vec p}){\frak a}({\vec p})+{\frak b}^{\dag}({\vec p})\Sigma_{i}({\vec p}){\frak b}({\vec p})\right]\,.	
\end{eqnarray}
Similarly,  the parent operators  (\ref{SFr}),  (\ref{CFr}), (\ref{SCz1}) and (\ref{CCz}) give rise to the one-particle operators
\begin{eqnarray}
	&&	{\mathsf S}_{{\rm Fr}\, i}= \frac{1}{2}\int d^3p \frac{E(p)}{m} \,\Theta^{-1}_{ij}({\vec p})
	\left[{\frak a}^{\dag}({\vec p})\Sigma_{j}({\vec p}){\frak a}({\vec p}+{\frak b}^{\dag}({\vec p})\Sigma_{j}({\vec p}){\frak b}({\vec p})   \right]\,,  \label{SFrr}\\
	&&	{\mathsf C}_{{\rm Fr}\,i}= \frac{1}{2}\int d^3p \frac{E(p)}{m} \,\Theta_{ij}({\vec p})	\left[{\frak a}^{\dag}({\vec p})\Sigma_{j}({\vec p}){\frak a}({\vec p})+{\frak b}^{\dag}({\vec p})\Sigma_{j}({\vec p}){\frak b}({\vec p})   \right]\,,  
\end{eqnarray}
\begin{eqnarray}		
	&&	{\mathsf S}_{{\rm PC}\, i}= \frac{1}{2}\int d^3p \frac{m}{E(p)} \,\Theta_{ij}({\vec p}) 	\left[{\frak a}^{\dag}({\vec p})\Sigma_{j}({\vec p}){\frak a}({\vec p})+{\frak b}^{\dag}({\vec p})\Sigma_{j}({\vec p}){\frak b}({\vec p})   \right]\,, \label{SCzz} \\	
	&&	{\mathsf C}_{{\rm PC}\,i}= \frac{1}{2}\int d^3p \frac{m}{E(p)} \,\Theta^{-1}_{ij}({\vec p}) 	\left[{\frak a}^{\dag}({\vec p})\Sigma_{j}({\vec p}){\frak a}({\vec p})+{\frak b}^{\dag}({\vec p})\Sigma_{j}({\vec p}){\frak b}({\vec p})   \right]\,,
\end{eqnarray}
which are conserved and translation invariant, behaving as $SO(3)$ vectors. They satisfy similar commutation relations as in Eqs. (\ref{Com10}),  (\ref{Com1}),  (\ref{Com20}) and (\ref{Com2}) but cannot close an algebra as each new commutator defines a new operator.
Note that after quantization the Fradkin-Good operator (\ref{SFG}) becomes the odd version of the Pryce(e) one such that this brings nothing new.

An important set of kinetic observables is formed by the  components of position and velocities operators. In Ref. \cite{Cot} we have shown that the original Pryce(e) operator proposed as a mass-center one becomes after quantization  the dipole operator which can be  transformed in the mass-center one changing by hand the sign of the antiparticle term. For improving this apparently arbitrary procedure  we defined here the mass-center operator  (\ref{XMC})  in RQM before quantization. Bearing in mind all these results we define now the particle and antiparticle center operators at the initial time $t_0=0$ and the corresponding velocities  as
\begin{eqnarray}
	\mathsf{X}^i_+&=&:\langle\psi,\Pi_+ X^i\psi\rangle_D:=\frac{i}{2}\int d^3p{\frak a}^{\dag}({\vec p})\stackrel{\leftrightarrow}{\tilde\partial_{i}} {\frak a}({\vec p})\,,	\label{Xplus}\\
	\mathsf{V}^i_+&=&:\langle\psi,\Pi_+ V^i\psi\rangle_D:=\int d^3p\frac{p^i}{E(p)}{\frak a}^{\dag}({\vec p}) {\frak a}({\vec p})\,,\label{Vplus}	\\
	\mathsf{X}^i_-&=&-:\langle\psi,\Pi_- X^i\psi\rangle_D:=\frac{i}{2}\int d^3p{\frak b}^{\dag}({\vec p})\stackrel{\leftrightarrow}{\tilde\partial_{i}} {\frak b}({\vec p})\,,\label{Xmin}	\\
	\mathsf{V}^i_-&=&-:\langle\psi,\Pi_- V^i\psi\rangle_D:=\int d^3p\frac{p^i}{E(p)}{\frak b}^{\dag}({\vec p}) {\frak b}({\vec p})\,,	~~~~~~~
\end{eqnarray} 
by using the derivative (\ref{stak}).  These operators satisfy 
\begin{eqnarray}
	[\mathsf{H}, \mathsf{X^i}_{\pm}]=-i\mathsf{V}_{\pm}^i\,, \quad 	[\mathsf{H}, \mathsf{V}_{\pm}^i]=0\,,
\end{eqnarray}
showing that the velocity components $\mathsf{V}^i_{\pm}$ are conserved operators while the position  ones evolve as
\begin{equation}\label{Xplmi}
	\mathsf{X}^i_{\pm}(t)=\mathsf{U}^{\dag}(t)\mathsf{X}^i_{\pm}\mathsf{U}(t)= \mathsf{X}^i_{\pm}+t\, \mathsf{V}^i_{\pm}\,.
\end{equation}
Moreover, we can verify that $\mathsf{X}_{\pm}^i(t)$  satisfy canonical relations coordinate-momentum,
\begin{equation}\label{Xcan}
	\left[ 	\mathsf{X}_{\pm}^i(t), \mathsf{X}_{\pm}^j(t) \right]=0\,,\quad	\left[ 	\mathsf{X}_{\pm}^i(t), \mathsf{P}^j \right]=i\delta_{ij}\mathsf{N}_{\pm}\,,
\end{equation}
as was expected according to the  hypothesis Pryce(e), but with $\mathsf{N}_{\pm}$ instead of the identity operator. These position operators transform under rotations  as  $SO(3)$ vector-operators satisfying 
\begin{eqnarray}
	\left[\mathsf{L}_i , \mathsf{X}_{\pm}^j(t) \right]=i\epsilon_{ijk} \mathsf{X}_{\pm}^k(t) \,, \quad 	\left[\mathsf{S}_i , \mathsf{X}_{\pm}^j(t) \right]=0\,.
\end{eqnarray} 
The transformations under Lorentz boosts are quite complicated because of the transformation matrices which depend on momentum remaining under integral as in Eq. (\ref{Spintr}).  For this reason the relativistic covariance of the position and other orbital operators will be studied elsewhere.

The above results allow us to bring into  intuitive forms the components of the {\em  dipole}  and  {\em mass-center} operators 
\begin{equation}\label{XVmc}
	\mathsf{X}^i(t)=	\mathsf{X}^i_+(t)-\mathsf{X}^i_-(t) \,,\quad 		\mathsf{X}^i_{MC}(t)=\mathsf{X}^i_+(t)+\mathsf{X}^i_-(t)\,,
\end{equation}
whose  velocities 
\begin{equation}\label{vel}
	\mathsf{V}^i= \mathsf{V}^i_{+}- \mathsf{V}^i_{-}\,,\quad 	 \mathsf{V}^i_{MC}= \mathsf{V}^i_{+}+ \mathsf{V}^i_{-}\,,
\end{equation}
have conserved components. The dipole velocity of components    $\mathsf{V}^i$,  known as the classical current  \cite{Z1}, is referred here as the  conserved current.  Note that the position operators at different instants $t'\not=t$ do not commute,
\begin{eqnarray}
	\left[	\mathsf{X}^i(t), \mathsf{X}^j(t')\right]=\left[	\mathsf{X}_{MC}^i(t), \mathsf{X}_{MC}^j(t')\right] =i(t'-t) \mathsf{G}_{ij}\,,
\end{eqnarray} 
giving rise to  the new even  one-particle operator 
\begin{eqnarray}
	\mathsf{G}_{ij}&=&\int \frac{d^3p}{E(p)}\left( \delta_{ij}-\frac{p^i p^j}{E(p)^2}\right)	\left[{\frak a}^{\dag}({\vec p}){\frak a}({\vec p})+{\frak  b}^{\dag}({\vec p}){\frak b}({\vec p})\right]\,,
\end{eqnarray}
derived according to Eq. (\ref{comXV}).

The principal observables of QFT we studied above  are Hermitian one-particle operators, whose parents operators are reducible.  These observables  are either conserved, commuting with $\mathsf{H}$, or evolving linearly in time as the boost generators and position operators. The conserved spin operator  of components (\ref{Spin}) associated to  position operators (\ref{XVmc}) whose velocities (\ref{vel}) are conserved may describe a smooth inertial motions without Zitterbewegung. However, it is not forbidden to measure the traditional observables  $\underline{\vec x}$ and ${\vec s}$ whose components are no longer reducible operators, generating after quantization oscillating terms.

\subsection{Irreducible operators}

For analyzing the behaviour of the irreducible operators it is convenient to split each Hermitian operator $A=A_{\rm diag}+A_{\rm osc}$ in its diagonal and oscillating parts as defined in Sec. 3.2.  After quantization we obtain the operator
$\mathsf{A}=\mathsf{A}_{\rm diag}+\mathsf{A}_{\rm osc}$ 
whose diagonal part is a one-particle operator expressed in terms of associated operators as 
\begin{eqnarray}
	\mathsf{A}_{\rm diag}=\int d^3p \left[ {\frak a}^{\dag}({\vec p})\left(\tilde A^{(+)}{\frak a}\right)({\vec p})- {\frak b}^{\dag}({\vec p})\left(\tilde A^{(-)}{\frak b}\right)({\vec p})\right] \,,\label{Adiag}
\end{eqnarray}
while the oscillating term, 
\begin{eqnarray}
	\mathsf{A}_{\rm osc}&=&\int d^3p \left[ {\frak a}^{\dag}({\vec p})\left[\left(\tilde A^{z}{\frak b}\right)^{\dag}(-{\vec p})\right]^T+ [{\frak b}({-\vec p})]^T\left(\tilde A^{z\,+} {\frak a}\right)({\vec p})\right] \,,\label{Azitt}	
\end{eqnarray}
depends only on the operator  $\tilde A^z=\tilde A^{(\pm)}=[\tilde A^{(\mp)}]^+$. This may be written either in compact notation,  
\begin{eqnarray}
	{\frak a}^{\dag}({\vec p})\left[\left(\tilde A^{z}{\frak b}\right)^{\dag}(-{\vec p})\right]^T=
	\sum_{\sigma\sigma'}	
	{\frak a}^{\dag}_{\sigma}({\vec p})\tilde A^{z}_{\sigma\sigma'}({\vec p}){\frak b}^{\dag}_{\sigma'}(-{\vec p})\,,
\end{eqnarray}
or by using explicitly the matrix elements (\ref{Aa1p}).  

We focus here on the operators of QFT whose parents are either Fourier operators or simple momentum-indepen\-dent matrix-operators of $\rho_D$ that can be seen as particular Fourier operators for which the Fourier transform is the operator itself. Therefore, we may derive the matrix elements of the associated operators according to Eqs,  (\ref{Aa1}-\ref{Aa2p}) where we have to substitute the operators under consideration. We obtain thus the diagonal terms which are one-particle operators and oscillating parts having the specific form (\ref{Azitt}). All these operators form an open algebra with obvious commutation rules, $[A_{\rm diag}, B_{\rm diag}]=C_{\rm diag}$,  $[A_{\rm osc}, B_{\rm osc}]=C_{\rm diag}$ and $[A_{\rm osc}, B_{\rm diag}]=C_{\rm osc}$, showing that only the diagonal terms may form a sub-algebra.

Let us consider first the  quantization of the  coordinate  operator $\underline {\vec x}={\vec X}-\delta{\vec X}$ which can be done  as we derived already the Pryce(e) dipole operator with components  (\ref{XVmc}) and we know that $\delta{\vec X}$ is a Fourier operator. Applying the canonical quantization procedure at the initial time $t=0$ and translating then the result at an arbitrary instant $t$ we obtain  the operators 
\begin{eqnarray}
	\delta\mathsf{X}^i(t)=	\delta\mathsf{X}_{\rm diag}^i+	\delta\mathsf{X}_{\rm osc}^i(t) \,,
\end{eqnarray}
having conserved odd diagonal parts,
\begin{eqnarray}\label{dXdiag}
	\delta\mathsf{X}_{\rm diag}^i&=&-\frac{1}{2}\int d^3p\,\frac{\epsilon_{ijk}p^j}{E(p)(E(p)+m)}\left[{\frak a}^{\dag}({\vec p})\Sigma_k({\vec p}){\frak a}({\vec p}) -	{\frak b}^{\dag}({\vec p})\Sigma_k({\vec p}){\frak b}({\vec p})\right]\,,
\end{eqnarray}
and oscillating terms of the form
\begin{eqnarray}
	\delta\mathsf{X}_{\rm osc}^i (t)=\int d^3p \sum_{\sigma,\sigma'}\left[ \delta\tilde X^{z\, i}_{\sigma\sigma'}(t,{\vec p}){\frak a}^{\dag}_{\sigma}({\vec p}){\frak b}_{\sigma'}^{\dag} (-{\vec p}) +{\rm H.c.}\right]	\,,
\end{eqnarray}
where, according to Eq. (\ref{tete}), we have
\begin{eqnarray}
	&&\delta\tilde X^{z\, i}_{\sigma\sigma'}(t,{\vec p})=-\frac{i e^{2iE(p)t}}{2E(p)}\,\Theta^{-1}_{ij}({\vec p}) \xi^+_{\sigma}({\vec p})\sigma_j\eta_{\sigma'}(-{\vec p})\,.~~~~~~~
\end{eqnarray} 
Hereby we obtain the components of  the coordinate operator of QFT,
\begin{equation}\label{coord}
	\underline{\mathsf{x}}^i(t)=\mathsf{X}^i(t)-\delta\mathsf{X}^i(t)=\underline{\mathsf{x}}_{\, 0}^i+t\mathsf{V}^i-\delta\mathsf{X}_{\rm osc}^i (t)	\,,  
\end{equation}
having the static  terms     
\begin{equation}\label{incoord}
	\underline{\mathsf{x}}_{\,0}^i= \mathsf{X}^i-	\delta\mathsf{X}_{\rm diag}^i \equiv	 \mathsf{X}_{\rm Pr(c)}^i\,,
\end{equation} 
we interpret as the components of  the {\em  initial coordinate} operator as this is just the diagonal part  of the coordinate operator (\ref{coord})  at the instant $t=0$. This one-particle operator corresponding to the hypothesis Pryce(c) \cite{B} has    components which  satisfy canonical coordinate-momentum commutation relations  but do not commute among themselves as we verify  in the Appendix C.

The oscillating term of  Eq. (\ref{coord}) produces the Zitterbewegung which was discovered studying the vector current  \cite{Zit1,Zit2}   produced by  the Dirac current density, $j^{\mu}(x)=:\bar{\psi}(x)\gamma^{\mu} \psi(x):$. Its time-like component gives rise to the conserved charge operator, 
\begin{eqnarray}
	\mathsf{Q}=\int d^3x  :\bar\psi (t,{\vec x})\gamma^0\psi(t,{\vec x}):=:\langle \psi,\psi\rangle_D:=\mathsf{N}_+-\mathsf{N}_-\,,
\end{eqnarray}
expressed in terms of the operators (\ref{Nplus}) and (\ref{Nminus}) while its space part produces  the vector current having the components 
\begin{eqnarray}
	\mathsf{I}_V^i(t)&=&\int d^3x :\bar\psi (t,{\vec x})\gamma^i\psi(t,{\vec x}):	=\left.:\langle \psi, \gamma^0\gamma^i\psi\rangle_D:\right|_t \nonumber\\
	&=&2 i	\left.:\langle \psi, s_{0i}\psi\rangle_D:\right|_t =2 i \mathsf{s}_{0i}(t)\,,\label{zitts}
\end{eqnarray}
proportional with the generators (\ref{s0i}) we split as 
\begin{eqnarray}
	\mathsf{I}_V^i(t)=		\mathsf{I}_{V\,\rm diag}^i +	\mathsf{I}_{V\,\rm osc}^i (t) ~~~\Rightarrow~~~ \mathsf{s}_{0i}(t)=		\mathsf{s}_{{\rm diag}\,0i} +	\mathsf{s}_{{\rm osc}\,0i} (t)\,.
\end{eqnarray}
Calculating these components we recover the well-known result 
\begin{eqnarray}
	\mathsf{I}_V^i(t)&=&\frac{d}{dt}\underline{\mathsf{x}}^i(t)
~~~\Rightarrow~~~	\mathsf{I}_{V\,\rm diag}^i=\mathsf{V}^i\,,\quad 	\mathsf{I}_{V\,\rm osc}^i (t)=-\frac{d}{dt}\delta\mathsf{X}_{\rm osc}^i (t)\,,
\end{eqnarray}
which  was discussed in Refs. \cite{Z1,Z2} but using particular polarization spinors. 

Apart from the conserved Dirac current density one uses the axial current density $j_A^{\mu}(x)=-:\bar{\psi}(x)\gamma^5\gamma^{\mu} \psi(x):$ which is conserved only in the massless case. This gives rise to  the axial charge,
\begin{eqnarray}
	\mathsf{Q}_A(t)=\int d^3x j_A^0=\left.:\langle \psi,\gamma^5\psi\rangle_D:\right|_t =\mathsf{Q}_{A\, \rm diag}+\mathsf{Q}_{A\, \rm osc}(t)\,,
\end{eqnarray}
having a conserved  diagonal part
\begin{eqnarray}
	\mathsf{Q}_{A\, \rm diag}=\int d^3p \frac{p^i}{E(p)}\left[ {\frak a}^{\dag}({\vec p}) \Sigma_i ({\vec p}){\frak a}({\vec p})+{\frak b}^{\dag}({\vec p}) \Sigma_i ({\vec p}){\frak b}({\vec p})\right]\,,
\end{eqnarray}
which is an even one-particle operator in contrast with the charge operator which is odd.  In addition, this  has the oscillating term
\begin{eqnarray}
	\mathsf{Q}_{A\, \rm osc}(t)&=&-\int d^3p\frac{m}{E(p)}\left[ e^{2iE(p) t}{\frak a}^{\dag}({\vec p})\left( {\frak b}^{\dag}(-{\vec p})   \right)^T +\,{\rm H.c.}\right]\,.
\end{eqnarray}
The corresponding components of axial current,
\begin{eqnarray}
	\mathsf{I}_A^i(t)&=&-\int d^3x :\bar\psi (t,{\vec x})\gamma^5\gamma^i\psi(t,{\vec x}):\nonumber\\
	&=&-\left.:\langle \psi, \gamma^0\gamma^5\gamma^i\psi\rangle_D:\right|_t 
	=2	\left.:\langle \psi, s_i\psi\rangle_D:\right|_t = 2 \mathsf{s}_i(t)\,,\label{zitspin}
\end{eqnarray}
are proportional with the generators (\ref{si}) we split again as 
\begin{eqnarray}
	\mathsf{I}_A^i(t)=\mathsf{I}_{A\,\rm diag}^i +	\mathsf{I}_{A\,\rm osc}^i (t) ~\Rightarrow~ \mathsf{s}_i(t)=		\mathsf{s}_{{\rm diag}\,i} +	\mathsf{s}_{{\rm osc}\,i} (t)\,.	
\end{eqnarray}
pointing out the conserved diagonal terms $\mathsf{I}_{A\,\rm diag}^i=2\, {\mathsf S}_{{\rm PC}\, i}$ depending on the components  (\ref{SCzz}) of the Pryce(c)-Czochor operator which is by definition the diagonal part of Pauli's one.   
The oscillating parts read
\begin{eqnarray}
	\mathsf{I}_{A\,\rm osc}^i (t)=\int d^3p \sum_{\sigma,\sigma'}\left[ \tilde I^{z\, i}_{A\,\sigma\sigma'}(t,{\vec p}){\frak a}^{\dag}_{\sigma}({\vec p}){\frak b}_{\sigma'}^{\dag} (-{\vec p}) +{\rm H.c.}\right]	\,,
\end{eqnarray}
where
\begin{eqnarray}
	\tilde I^{z\, i}_{A\,\sigma\sigma'}(t,{\vec p})=i  e^{2iE(p) t}\epsilon_{ijk}\frac{p^j}{E(p)}\xi_{\sigma}^+({\vec p})\sigma_k\eta_{\sigma'}(-{\vec p})\,.
\end{eqnarray}
We have thus  a complete image about the time evolution of the principal currents of Dirac's theory related to the operators $\mathsf{s}_{\mu\nu}(t)$ defined by Eqs. (\ref{zitspin}) and (\ref{zitts}) that represent the generators of the operator-valued representation of QFT equivalent to $\rho_D[SL(2,\mathbb{C})]$.   

Other matrix-operators  of  RQM,  irreducible on $\tilde{\cal F}$, are the generators of various transformations that can be defined in  $\rho_D$.  For example, the Foldy-Wouthuysen transformation (\ref{FW}), which relates the Pauli-Dirac and Pryce spin operators as in Eq. (\ref{FW2}), are  generated by the Hermitian matrices $-i\gamma^i$ which are the parents of the operators
\begin{eqnarray}
	\mathsf{F}^i(t)=-i\left.:\langle\psi, \gamma^i\psi\rangle_D:	\right|_t=\mathsf{F}^i_{\rm diag}+\mathsf{F}_{\rm osc}^i(t)\,, 
\end{eqnarray}
having the diagonal parts 
\begin{eqnarray}
	\mathsf{F}^i_{\rm diag}&=&\int d^3 p\, \epsilon_{ijk}p^j \left[  {\frak a}^{\dag}({\vec p}) \Sigma_k ({\vec p}){\frak a}({\vec p})-{\frak b}^{\dag}({\vec p}) \Sigma_k ({\vec p}){\frak b}({\vec p}) \right]\,,	
\end{eqnarray}
which are now odd one-particle operators. The oscillating terms read
\begin{eqnarray}
	\mathsf{F}_{\rm osc}^i (t)=\int d^3p \sum_{\sigma,\sigma'}\left[ \tilde F^{z\, i}_{\sigma\sigma'}(t,{\vec p}){\frak a}^{\dag}_{\sigma}({\vec p}){\frak b}_{\sigma'}^{\dag} (-{\vec p}) +{\rm H.c.}\right]	\,,
\end{eqnarray}
where by using again the tensor (\ref{tete}) we may write
\begin{eqnarray}
	\tilde F^{z\, i}_{\sigma\sigma'}(t,{\vec p})= i e^{2iE(p) t}\frac{m}{E(p)}\, \Theta_{ij}({\vec p})\xi_{\sigma}^+({\vec p})\sigma_j\eta_{\sigma'}(-{\vec p})\,.
\end{eqnarray}
This behaviour explains why the  particular Foldy-Wouthuysen transformation (\ref{FW}) can relate the conserved Pryce(e) spin operator  to the non conserved Pauli-Dirac one as in Eq. (\ref{FW2}).

The Chakrabarti spin operator ${\vec S}_{\rm Ch}$  can be quantized starting with its Fourier transform  (\ref{sCh}), deriving the associated operators and applying the quantization procedure.  We find thus that the components of this operator, 
\begin{eqnarray}
	\mathsf{S}_{\rm Ch\,i}(t)=	 \mathsf{S}_i + 	 \mathsf{S}_{\rm osc\,i}(t)\,,
\end{eqnarray}
are formed by  those of the Pryce(e) spin operator with supplemental oscillating  terms  of the form
\begin{eqnarray}
	\mathsf{S}_{\rm osc\,i} (t)=\int d^3p \sum_{\sigma,\sigma'}\left[ \tilde S^{z\, i}_{\sigma\sigma'}(t,{\vec p}){\frak a}^{\dag}_{\sigma}({\vec p}){\frak b}_{\sigma'}^{\dag} (-{\vec p}) +{\rm H.c.}\right]	\,,
\end{eqnarray}
where
\begin{eqnarray}
	\tilde S^{z\, i}_{\sigma\sigma'}(t,{\vec p})=	\frac{ i e^{2iE(p) t}}{ m}\, \epsilon_{ijk}p^j\xi_{\sigma}^+({\vec p})\sigma_k\eta_{\sigma'}(-{\vec p})\,.
\end{eqnarray}
This result was expected as we know that the parent operator (\ref{sCh}) is not conserved. 

Finally let us focus on the scalar and pseudo-scalar charges starting with the scalar one we may split as  
\begin{eqnarray}
	\mathsf{Q}^{\rm sc}(t)=\int d^3x :\bar\psi(t,{\vec x})\psi(t,{\vec x}): =:\langle\psi, \gamma^0\psi\rangle_D:=	\mathsf{Q}^{\rm sc}_{\rm diag}+\mathsf{Q}^{\rm sc}_{\rm osc}(t)\,,	
\end{eqnarray}
where the conserved diagonal term
\begin{eqnarray}
	\mathsf{Q}_{\rm diag}^{\rm sc}&=&m\int \, \frac{d^3 p}{E(p)} \left[  {\frak a}^{\dag}({\vec p}){\frak a}({\vec p})+{\frak b}^{\dag}({\vec p}) {\frak b}({\vec p}) \right]\,,	
\end{eqnarray}	
is is an even one-particle operator while the oscillating part can be written as 	
\begin{eqnarray}
	\mathsf{Q}_{\rm osc}^{\rm sc} (t)&=&\int \frac{d^3p}{E(p)} \sum_{\sigma,\sigma'}\left[ \tilde Q^{{\rm sc}\,z}_{\sigma\sigma'}(t,{\vec p}){\frak a}^{\dag}_{\sigma}({\vec p}){\frak b}_{\sigma'}^{\dag} (-{\vec p}) +{\rm H.c.}\right]	\,,\nonumber\\
	\tilde Q^{{\rm sc}\,z}_{\sigma\sigma'}(t,{\vec p})&=&-	 e^{2iE(p) t}\, p^j\xi_{\sigma}^+({\vec p})\sigma_j\eta_{\sigma'}(-{\vec p})\,.
\end{eqnarray}
It is interesting that the pseudoscalar charge  does not have diagonal terms reducing to the oscillating form
\begin{eqnarray}
	&&	\mathsf{Q}^{\rm ps}(t)=\int d^3x :\bar\psi(t,{\vec x})\gamma^5\psi(t,{\vec x}): =:\langle\psi, \gamma^0\gamma^5\psi\rangle_D:\nonumber\\
	&&=- \int d^3p\, \sum_{\sigma,\sigma'}\left[e^{2iE(p) t}\xi_{\sigma}^+({\vec p})\eta_{\sigma'}(-{\vec p}) {\frak a}^{\dag}_{\sigma}({\vec p}){\frak b}_{\sigma'}^{\dag} (-{\vec p})+{\rm H.c.}\right]\,,
\end{eqnarray}
that could be of some interest in QFT.

Concluding we may say that our method of associated operators  allows us to quantize all the operators we need in QFT including the irreducible ones.  The oscillating terms of these  operators give vanishing expectation values and real-valued contributions to dispersion  in pure states but  they may lay out significant observable effects when  are measured in mixed states.    

\section{Propagation}

In applications we may turn back to RQM but considered now as the one-particle restriction of QFT. We have thus the advantage of the mathematical rigor and correct physical  interpretations offered by QFT. We assume that the quantum states are prepared or measured by an ideal apparatus represented by a set of one-particle operators without oscillating parts, including the Pryce(e) spin and position operators.

\subsection{Preparing and detecting wave-packets}

In what follows we study the propagation of the plane-wave packets generated by the one-particle physical states, 
\begin{equation}\label{packO}
	|\alpha\rangle=\int d^3 p \sum_{\sigma} \alpha_{\sigma}({\vec  p}){\frak a}^{\dagger}_{\sigma}({\vec p})|0\rangle \,, 
\end{equation}
defined by normalized wave spinors, $\alpha\in \tilde{\cal F}^+$, which  satisfy the normalization condition, 
\begin{equation}\label{normal}
	\langle\alpha|\alpha\rangle= \langle\alpha,\alpha\rangle=\int d^3 p\, \alpha^+({\vec p})\alpha({\vec p})=1\,.
\end{equation} 
The corresponding  wave spinors in CR, 
\begin{equation}\label{wp}
	\Psi_{\alpha}(x)=\langle 0|\psi(x)|\alpha\rangle=\int d^3 p \sum_{\sigma} U_{\vec p,\sigma}(x)\alpha_{\sigma}({\vec p})\,,
\end{equation} 
are normalized, $\left<\Psi_{\alpha},\Psi_{\alpha}\right>_D=1$, with respect to the scalar product (\ref{sp}).  This is a particular case of local relativistic wave function that can be obtained from the one-particle restriction of QFT. In general, one can construct directly such functions as Fourier transforms of momentum dependent wave functions obtained  in the  generalized Bargmann-Wigner approach  \cite{BW1} proposed  recently (see for instance Ref. \cite{BW2} and references therein). In this framework wave functions for massive and massless particles of different discrete or even continuous spins may be constructed and studied without resorting explicitly to the field operators of QFT.

The wave functions are not  measurable quantities but  are studied often  by using numerical and graphical methods for extracting intuitive information about propagation in the presence of Zitterbewegung and spin dynamics produced by the traditional observables of Dirac's RQM. Such methods were used for the first time in  Ref. \cite{Th1}.

In our approach we  avoid  these effects assuming that our  apparatus measures only the reducible observables as the energy, momentum, position, velocity, spin and polarization which are one-particle operators.  The physical meaning is given then only by the statistical quantities generated by these operators that can be derived easily  by using our previous results.  More specific, for any one-particle operator  $\mathsf{A}$ the expectation value and dispersion in the state $|\alpha\rangle$   denoted as 
\begin{eqnarray}
	\langle \mathsf{A}\rangle&\equiv&\langle \alpha|\mathsf{A}|\alpha\rangle =\langle \alpha , \tilde A \alpha\rangle\,, \label{expect}	\\
	{\rm disp}(\mathsf{A})&\equiv &\langle \mathsf{A}^2\rangle-\langle \mathsf{A}\rangle^2=\langle \tilde A\alpha,\tilde A\alpha\rangle-\langle \alpha , \tilde A \alpha\rangle^2\,,~~~~~\label{disp}
\end{eqnarray} 
may be written in terms of associated operators acting in MR of RQM. Once we have the dispersion we may write the uncertainty $\Delta  \mathsf{A}=\sqrt{{\rm disp}(\mathsf{A})}$.

For exploiting these formulas we need to specify the structure of the functions $\alpha_{\sigma}$.
We observe first that it is important to know where the state $\left|\alpha\right>$ is prepared translating the state in that point. If the state was prepared initially in origin then for a state prepared by the same apparatus  in the point of  position vector ${\vec x}_0$ we must perform the back translation  $\left|\alpha\right>\to \mathsf{U}(0,-{\vec x}_0)\left|\alpha\right>=e^{-i{\vec x}_0\cdot{\vec p}}\left|\alpha\right>$ as defined by Eq. (\ref{Tra}). On the other hand, we know  the position operator defined with the help of the covariant derivatives (\ref{covD}) that can be quite complicated in the case of peculiar polarization. Therefore, for a rapid inspection of a relevant example it is convenient to chose the simplest polarization spinors (\ref{etapm}) of the standard momentum-spin basis where $\Sigma_i=\sigma_i$ and $\Omega_i=0$.  

Starting  with these arguments we assume that the wave-packet  with the mentioned polarization is prepared at the initial time $t=0$ by an observer $O$  in the initial point ${\vec x}_0$.  Therefore, we may consider the wave spinor
\begin{eqnarray}
	\alpha({\vec p})=\left( 
	\begin{array}{c}
		\alpha_{\frac{1}{2}}({\vec p})\\
		\alpha_{-\frac{1}{2}}({\vec p})
	\end{array}\right)
	=\phi({\vec p})	e^{-i{\vec x}_0\cdot{\vec p}}\left( 
	\begin{array}{c}
		\cos\frac{\theta_s}{2}\\
		\sin\frac{\theta_s}{2}	
	\end{array}\right)\,,\label{alphapm}
\end{eqnarray}    
where $\theta_s$ is the polarization angle while $\phi: \mathbb{R}^3_{\vec p}\to \mathbb{R}$ is a real-valued scalar function which is normalized as
\begin{equation}\label{normphi}
	\langle\alpha|\alpha\rangle=1 ~~ \Rightarrow~~\int d^3p\, \phi({\vec p})^2=1\,.	
\end{equation}
With this function we may calculate the expectation values and dispersions of  the operators without spin terms as in  the scalar theory. For example,  in the case of the energy operator (\ref{Hom}) we may write
\begin{eqnarray}
	\langle \mathsf{H}\rangle&=&\int d^3p\, E(p)\phi({\vec p})^2\,,\label{bau1}	\\
	{\rm disp}(\mathsf{H})&=&\int d^3p\, E(p)^2\phi({\vec p})^2-\langle \mathsf{H}\rangle^2\,,
\end{eqnarray}
and similarly for the momentum components (\ref{Pom}). 

The polarization angle helps us to find rapidly the measurable quantities related to the spin components (\ref{Spin}) and polarization $\mathsf{W}_s=\mathsf{S}_3$.  Taking into account that now $\tilde S_i=\frac{1}{2}\sigma_i$ we obtain from Eqs. (\ref{expect}) and (\ref{disp}) the quantities, 
\begin{eqnarray}
	\langle \mathsf{S}_1\rangle&=&\frac{1}{2}\sin\theta_s \,,\quad {\rm disp}(\mathsf{S}_1) =\frac{1}{4}\cos^2\theta_s \,,\nonumber	\\
	\langle \mathsf{S}_2\rangle&= &0\,,\qquad\quad ~\,{\rm disp}(\mathsf{S}_2)=\frac{1}{4}\,,\nonumber\\
	\langle \mathsf{S}_3\rangle&=&\frac{1}{2}\cos\theta_s \,,\quad  {\rm disp}(\mathsf{S}_3)=\frac{1}{4}\sin^2\theta_s\,,\nonumber
\end{eqnarray}
with an obvious physical meaning  as the polarization angle is defined on the interval $[0,\pi]$ such that for $\theta_s=0$ the polarization is $\sigma=\frac{1}{2}$ ($\uparrow$) while for  $\theta_s=\pi$ this is $\sigma=-\frac{1}{2}$ ($\downarrow$). In both these cases of {\em total} polarization the measurements are exact with  ${\rm disp}(\mathsf{W}_s)={\rm disp}(\mathsf{S}_3)=0$.

The propagation of the wave packet is described  by the position operator  of components $ {X}^i_+(t)={X}^i_++t {V}^i_+$  defined by Eqs. (\ref{Xplus}) and (\ref{Vplus}).  In momentum-spin basis we use here we have the advantage of  $\Omega=0$  which means that  the covariant derivatives (\ref{covD}) become now the usual ones, $\tilde \partial_i\to \partial_{p^i}$. We find thus the quantities
\begin{eqnarray}
	\langle \mathsf{X}_+^i\rangle &=&\frac{i}{2}\int d^3p\,{\alpha}^{+}({\vec p})\stackrel{\leftrightarrow}{\partial_{p_i}} {\alpha}({\vec p})=x_0^i \int d^3p\, \phi({\vec p})^2=x_0^i\,,\label{evXplus}\\
	{\rm disp}(\mathsf{X}_+^i) &=&\int d^3p\, \partial_{p^i}\alpha^+({\vec p})\partial_{p^i}\alpha({\vec p}) \left({\rm no~ sum} \right) -(x_0^i)^2= \int d^3p\,\left( \partial_{p^i}\phi({\vec p}) \right)^2\,,\label{disXplus}\\
	\langle \mathsf{V}_+^i\rangle &=& \int d^3p\,\frac{p^i}{E(p)} \phi({\vec p})^2\,,\\
	{\rm disp}(\mathsf{V}_+^i) &=& \int d^3p\,\left(\frac{p^i}{E(p)}\right)^2 \phi({\vec p})^2-\langle \mathsf{V}_+^i\rangle ^2\,,
\end{eqnarray}
that depend only on the scalar function $\phi$. Finally we obtain the remarkable but expected result 
\begin{equation}\label{dispxt}
	{\rm disp}(\mathsf{X}_+^i(t))=	{\rm disp}(\mathsf{X}_+^i) +t^2 {\rm disp}(\mathsf{V}_+^i)\,,
\end{equation}
which lays out the dispersive character of this type of wave-packets that spread as other scalar or  non-relativistic wave-packets \cite{Pack}. A similar calculation can be done for the angular momentum which is conserved in our approach but less relevant in analyzing the inertial motion. 

Let us imagine now that another observer, $O'$, detects the above prepared wave-packet  performing measurement with a similar apparatus in the point ${\vec x}_0'$.  We denote by ${\vec x_0}-{{\vec x}'_0} ={\vec n} d$ the relative position vector assuming that the observers $O$ and $O'$ use the same Cartesian coordinates and, therefore, same observables.     The wave-packet evolves causally until the detector measures some of its parameters selecting (or filtering) only the fermions coming from the source $O$ whose momenta are in a narrow domain $\Delta\subset\mathbb{R}^3_{\vec p}$ along the direction ${\vec  n}$. Therefore, the measured state $|\alpha'\rangle$,  is  given now by the corresponding projection operator $\Lambda_{\Delta}$ as
\begin{equation}\label{alphaprim}
	|\alpha'\rangle = \Lambda_{\Delta}|\alpha\rangle= \int_{\Delta} d^3 p\, \alpha({\vec p})\, {\frak a}^{\dagger}({\vec p}) |0\rangle\,.
\end{equation}
This state is strongly  dependent on the domain $\Delta$ of measured momenta. Here we assume that this is a cone of axis ${\vec  n}$ and a very small solid angle $\Delta\Omega$ such that we may apply   the mean value theorem, 
\begin{eqnarray}
	\int_{\Delta}d^3p F({\vec p})\simeq\Delta\Omega \int_{0}^{\infty} dp p^2 F({\vec n}p)\,,\label{evlam}
\end{eqnarray} 
in spherical coordinates  ${\vec p}=(p,\vartheta,\varphi)$ to all the integrals over $\Delta$. We evaluate first  the quantity
\begin{eqnarray}
	\langle \alpha| \Lambda_{\Delta}|\alpha\rangle&=&  \int_{\Delta} d^3 p\, \alpha^+({\vec p}) \alpha({\vec p})
	=\int_{\Delta} d^3 p\, \phi({\vec p})^2\nonumber\\
	&\simeq&\Delta\Omega \int_{0}^{\infty} dp\,p^2 \phi({\vec n}p)^2=\Delta\Omega \kappa  \label{kkapa}	\,,
\end{eqnarray}
giving the probability $P_{\Delta}=| \langle \alpha| \Lambda_{\Delta}|\alpha\rangle |^2$ of measuring any momentum ${\vec p}\in \Delta$. Obviously, when one measures the whole continuous spectrum, $\Delta={\Bbb R}^3_k$, then $\Lambda_{\Delta}$ become the identity operator and  $P_{\Delta}=1$.  

Under such circumstances, the observer $O'$ measures new expectation values 
\begin{equation}\label{expv}
	\langle\mathsf{A}\rangle'=\langle \alpha'| \mathsf{A}|\alpha'\rangle  =
	\frac{\langle \alpha| \Lambda_{\Delta} \mathsf{A}|\alpha\rangle}{\langle \alpha| \Lambda_{\Delta}|\alpha\rangle}\,,
\end{equation}
for all the common observables of  $O$ and $O'$ which depend on momentum. Applying the above calculation rules we obtain the expectation values 
\begin{eqnarray}
	\langle\mathsf{H}\rangle'	&=& \frac{1}{\kappa} \int_{0}^{\infty} p^2dp\,E(p) \phi({\vec n}p)^2\,,\\
	\langle\mathsf{P^i}\rangle'	&=&n^i \frac{1}{\kappa} \int_{0}^{\infty} p^2dp\,p \phi({\vec n}p)^2=n^i \langle \mathsf{P}\rangle'\,,\label{Pund}\\
	\langle\mathsf{V_+^i}\rangle'	&=&n^i \frac{1}{\kappa} \int_{0}^{\infty}p^2 dp\,\frac{p}{E(p)} \phi({\vec n}p)^2=n^i \langle \mathsf{V_+}\rangle'\,,	\label{Vund}
\end{eqnarray}   
which show that $O'$ observes in fact a one-dimensional motion along the direction ${\vec n}$ measuring the new observables  
\begin{eqnarray}
	\mathsf{P}&=&\int d^3p\,p\left[{\frak a}^{\dag}({\vec p}){\frak a}({\vec p}) +{\frak b}^{\dag}({\vec p}){\frak b}({\vec p})\right]\,,\label{Prad}\\
	\mathsf{V}_+&=&\int d^3p\frac{p}{E(p)}{\frak a}^{\dag}({\vec p}) {\frak a}({\vec p})\,,	\label{Vrad}	
\end{eqnarray}
whose expectation values result from Eqs.  (\ref{Pund}) and (\ref{Vund}). We say that these operators and $\mathsf{V}_-$, defined similarly for antiparticles, are the {\em radial} observable of the common list of observables of $O$ and $O'$.  

Therefore, $O'$ measures a one dimensional wave-packet $\left|\alpha'\right>$ whose wave spinors  depend now on the new normalized scalar function
\begin{equation}\label{scfunct}
	\phi'(p)=\frac{1}{\sqrt{\kappa}}\,p\, \phi({\vec n} p)\,,
\end{equation}
allowing us to write the statistical quantities of the radial operators measured by $O'$ as    
\begin{eqnarray}
	\langle \mathsf{H}\rangle'&=&\int_{0}^{\infty} dp\,E(p) \phi'(p)^2\,,\\
	{\rm disp}(\mathsf{H})'&=&\int_{0}^{\infty} dp\,E(p)^2 \phi'(p)^2 -		{\langle\mathsf{P}\rangle'\,}^2\,, \label{list1}\\
	\langle \mathsf{P}\rangle'&=&\int_{0}^{\infty} dp\,{p} \phi'(p)^2\,,\\
	{\rm disp}(\mathsf{P})'&=&\int_{0}^{\infty} dp\,{p}^2 \phi'(p)^2 -		{\langle\mathsf{P}\rangle'\,}^2\,, \\
	\langle \mathsf{V}_+\rangle'&=&\int_{0}^{\infty} dp\,\frac{p}{E(p)} \phi'(p)^2	\,,\\
	{\rm disp}(\mathsf{V_+})'&=& \int_{0}^{\infty} dp\,\left(\frac{p}{E(p)}\right)^2 \phi'(p)^2 -		{\langle\mathsf{V_+}\rangle'\,}^2\,.~~~~~
\end{eqnarray} 
The expectation values of the operators $\mathsf{X}_+^i$  are not affected by the projection on the domain $\Delta$,  $\langle\mathsf{X}_+^i\rangle'=	\langle\mathsf{X}_+^i\rangle=x_0^i$, but the dispersions may be different as $O'$ measures
\begin{equation}\label{list2}
	{\rm disp}(\mathsf{X}_+^i)' =\frac{1}{\kappa}\int _0^{\infty}dp\,p^2\left.\left( \partial_{p^i}\phi({\vec p}) \right)^2\right|_{{\vec p}={\vec n}p}\,.
\end{equation}  
The only operators whose measurement is independent on the  momentum filtering  are the spin components for which we have $\langle\mathsf{S}_i\rangle'=	\langle\mathsf{S}_i\rangle$ and ${\rm disp}(S_i)'={\rm disp}(S_i)$. 

In this manner we derived all the statistical quantities of  prepared  or detected wave-packets  using only analytical methods  without resorting to a visual study of  the packet profile  in CR which might be intuitive but is sterile from the point of view of QFT.  

 \begin{figure}
	\centering
	\includegraphics[scale=0.28]{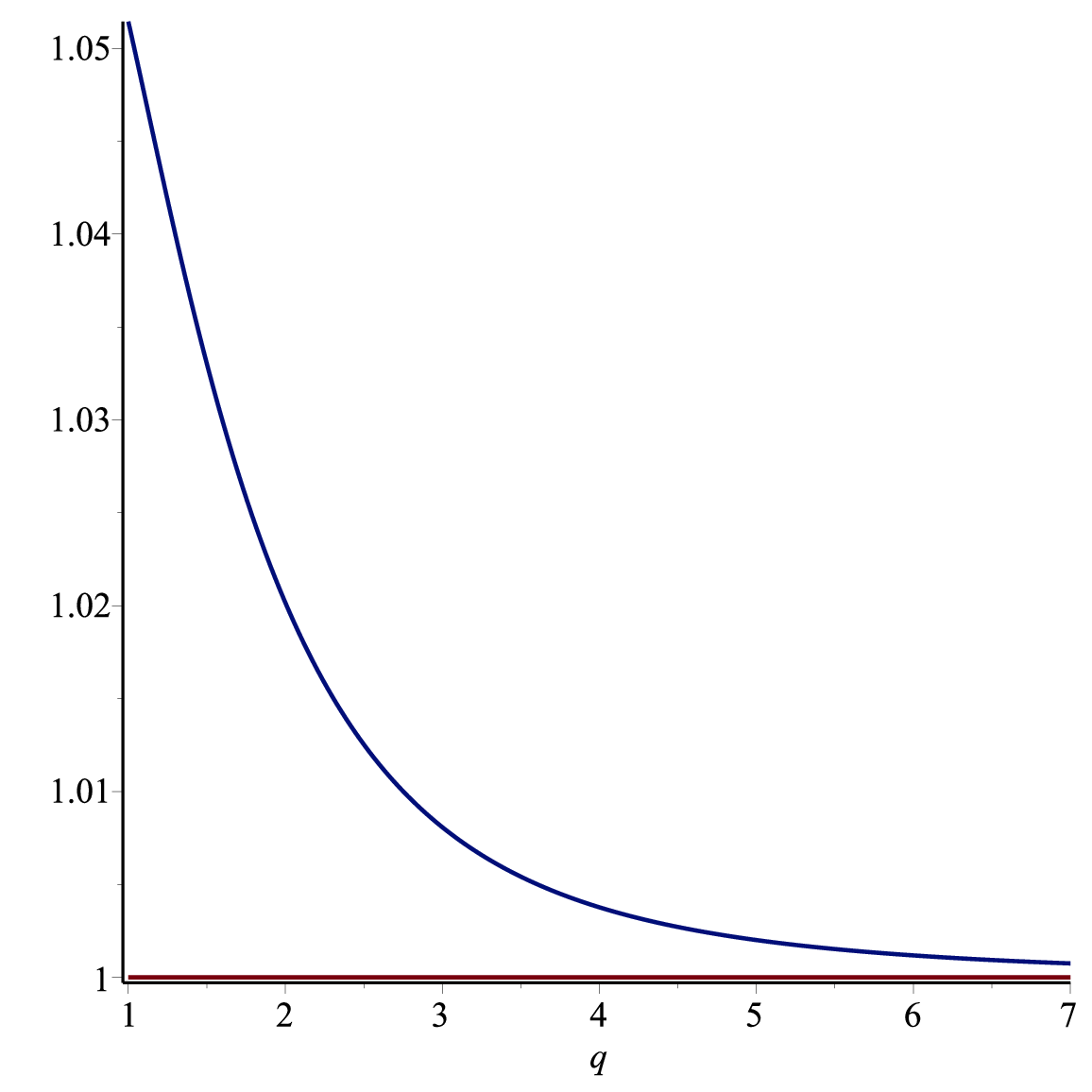} 
	\includegraphics[scale=0.28]{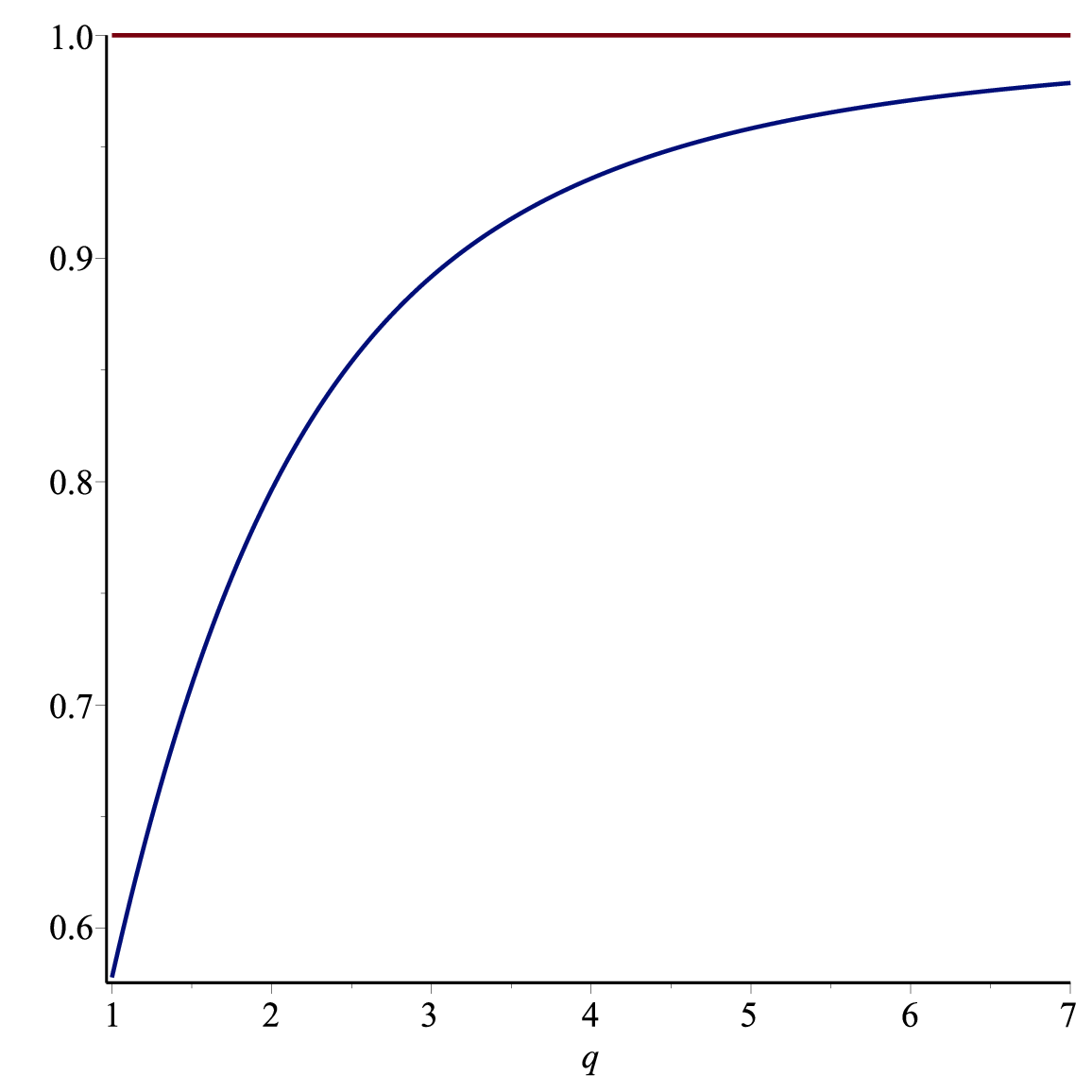}
	\caption{ The ratios  $\frac{\langle\mathsf{H}\rangle} {E(\bar{p}) }\to 1_+$ (left panel) and  $\frac{2\gamma\,{\rm disp}(\mathsf{H})} {\bar{p} }\to 1_-$ (right panel) as functions of $q=\gamma\bar{p}$ in the domain  $1<q\le 7$  for $\gamma m=1$. }
\end{figure}

\subsection{Isotropic wave-packet}

As a simple example, we consider now an  isotropic wave-packet  for which  it is convenient to use spherical coordinates in momentum space with ${\vec p}=p\, {\vec n}_p$ and 
\begin{equation}
	{\vec n}_p=(\sin\vartheta\cos\varphi, \sin\vartheta\sin\varphi,\cos \vartheta)\,.	
\end{equation}
We assume that at the initial time $t_0=0$  the observer $O$ prepares the wave-packet (\ref{packO}) whose wave spinor  (\ref{alphapm})  is equipped with the isotropic function  
\begin{equation}
	\phi({\vec p})\to\phi(p)=N
	p^{\gamma\bar{p}-\frac{3}{2}}e^{-\gamma p}\,,\quad \gamma,\,\bar{p}>0\,,
\end{equation}
depending on the real parameters $\gamma$ and $\bar{p}$ and the normalization factor
\begin{equation}
	N=\frac{(2\gamma)^{\gamma\bar{p}}}{2\sqrt{\pi\Gamma(2\gamma\bar{p})}}
\end{equation}
which guarantees that 
\begin{equation}
	\int d^3p \phi(p)^2=4\pi \int_0^{\infty}dp\, p^2\, \phi(p)^2=1\,.
\end{equation}
The parameter  $\bar{p}$ is just the expectation value of the {radial} momentum (\ref{Prad}) such that  
\begin{eqnarray}
	\langle \mathsf{P}\rangle&=&4\pi \int_0^{\infty}dp\, p^3\, \phi(p)^2=\bar{p}\,,\\
	{\rm disp}(\mathsf{P})&=&4\pi \int_0^{\infty}dk\, p^4\, \phi(k)^2-\bar{p}^2=\frac{\bar{p}}{2\gamma}\,.
\end{eqnarray}
In this isotropic case the Cartesian momentum and velocity components measured by $O$ have vanishing expectation values, $ \langle \mathsf{P}^i\rangle=0$ and $ \langle \mathsf{V}_+^i\rangle=0$, but relevant dispersions that read
\begin{eqnarray}
	{\rm disp}(\mathsf{P}^i)&=&\frac{4\pi}{3}\langle \mathsf{P}^2\rangle=\frac{4\pi}{3}\left( \bar{p}^2+\frac{\bar{p}}{2\gamma}\right)\,,\\
	{\rm disp}(\mathsf{
		V}_+^i)&=&\frac{4\pi}{3}\langle \mathsf{V}_+^2\rangle\,,\label{disVi}
\end{eqnarray}
as the angular integrals  give $\int (n^i_p)^2d\Omega=\frac{4\pi}{3}$.   Moreover, the observer $O$ measures the components of the initial position operator with expectation values (\ref{evXplus}) and dispersions (\ref{disXplus}) that now read
\begin{equation}
	{\rm disp}(\mathsf{X}_+^i)=\frac{1}{6}\frac{\gamma^2}{\gamma \bar{p}-1}~~\Rightarrow~~ \gamma \bar{p}>1\,,
\end{equation}
imposing a mandatory condition for our parameters.

\begin{figure}
	\centering
	\includegraphics[scale=0.28]{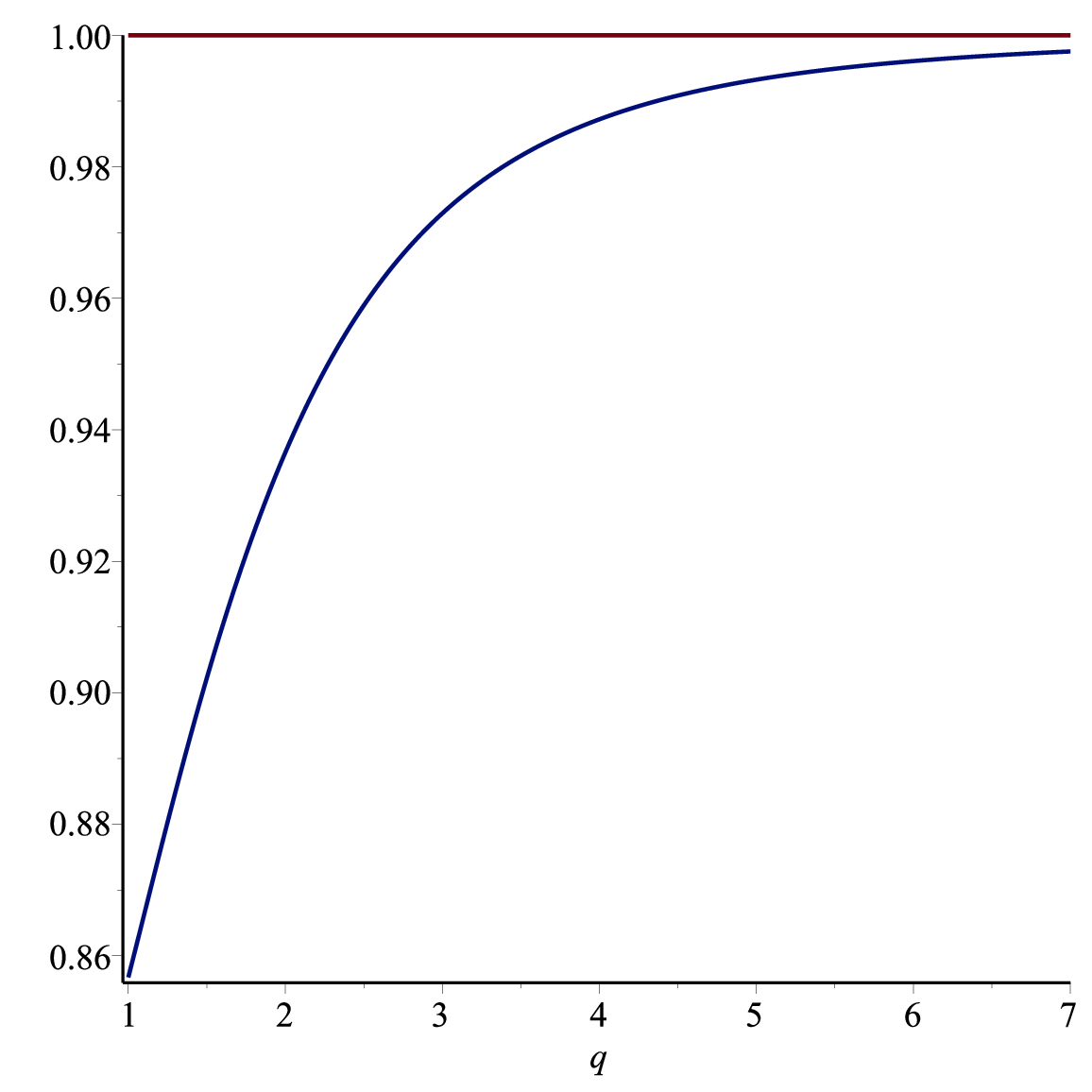}
	\includegraphics[scale=0.28]{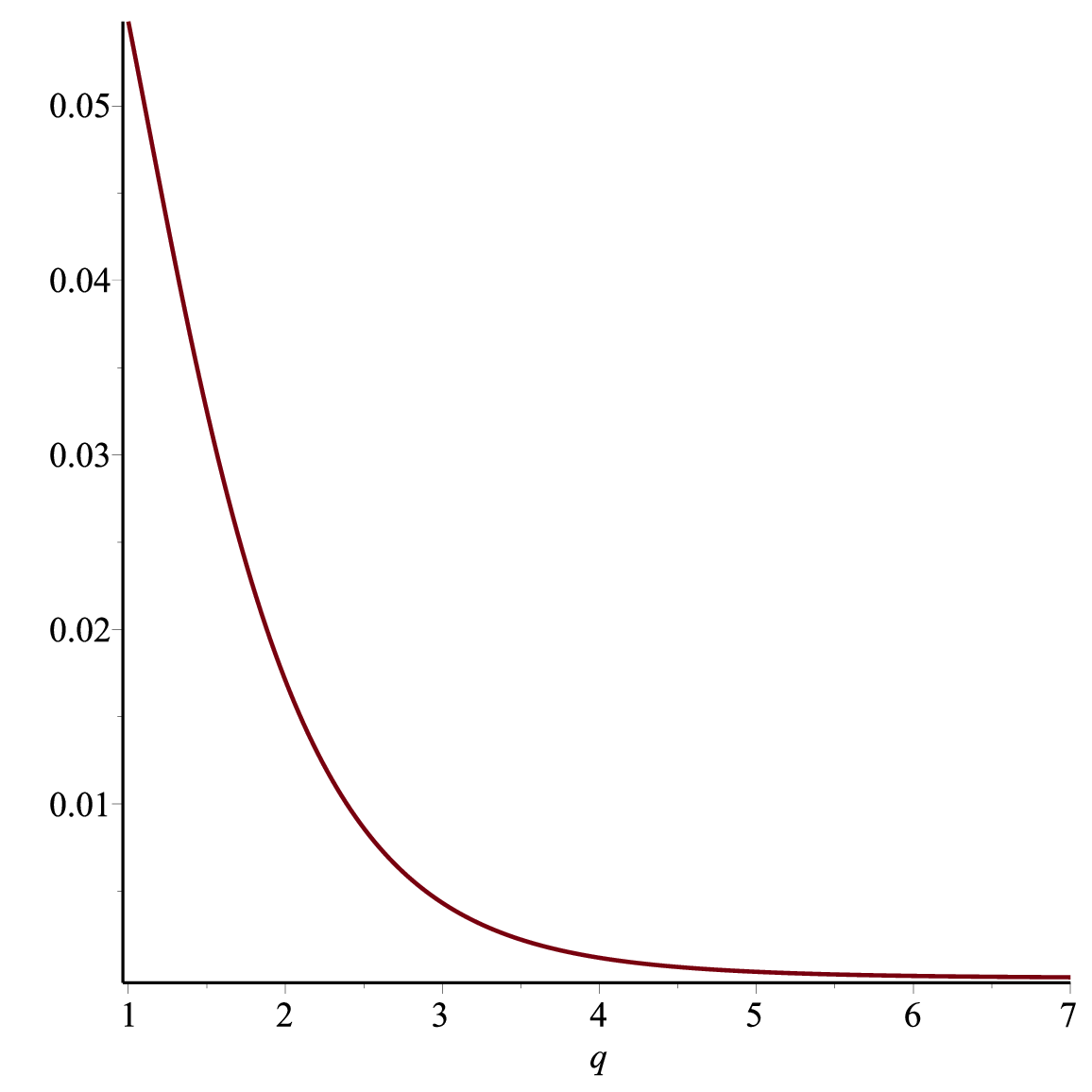}
	\caption{The ratio $\frac{\langle\mathsf{V}_+\rangle} {V(\bar{p}) }\to 1_-$ (left panel)
		and  the velocity dispersion ${\rm disp}(\mathsf{V}_+)$ (right panel) as functions of $q=\gamma\bar{p}$ in the domain  $1<q\le 7$  for $\gamma m=1$.}
\end{figure}

The observer $O'$ detects the one-dimensional wave-packet with 
\begin{equation}
	\kappa=\frac{1}{4\pi}	~~~\Rightarrow ~~~\phi'(p)=\sqrt{4\pi}\, p\, \phi(p)\,,
\end{equation}
which means the statistical quantities of the  operators (\ref{list1}-\ref{list2})  coincide with those given by Eqs. (\ref{bau1}-\ref{dispxt}) measured by the observer $O$. For writing down the expressions of these quantities  we  consider integrals of general form
\begin{eqnarray}
		G(\nu,\rho;\mu)&=&\int_0^{\infty}dp \,p ^{2\nu-1}\left( p^2+m^2\right)^{\rho-1}e^{-\mu p}\nonumber\\
	&=&\frac{m^{2\nu+2\rho-2}}{2\sqrt{\pi}\Gamma(1-\rho)}G^{31}_{13}\left(\left. \frac{m^2\mu^2}{4}\right|
	\begin{array}{l}
		1-\nu\\
		1-\rho-\nu, 0,\frac{1}{2}	
	\end{array}	
	\right)\,,~~~~~~
\end{eqnarray}  
which can be solved in terms of Meijer's $G$-functions, according to Eq. (3.389) of Ref. \cite{GR}.  With their help we may write
\begin{eqnarray}
	\langle \mathsf{H}\rangle'	= \langle \mathsf{H}\rangle&=& 4\pi N^2 G\left(\gamma \bar{p},\frac{3}{2}; 2\gamma \right)\,, \\
	\langle \mathsf{V}_+\rangle'=  \langle \mathsf{V}_+\rangle&=& 4\pi N^2 G\left(\gamma \bar{p}+\frac{1}{2},\frac{1}{2}; 2\gamma \right)\,, \\
	\langle \mathsf{V}_+^2\rangle'=   \langle \mathsf{V}_+^2\rangle&=& 4\pi N^2 G\left(\gamma \bar{p}+1,0; 2\gamma \right)\,, 
\end{eqnarray}
while for $\mathsf{H}^2$ we find the closed expression
\begin{equation}
	\langle \mathsf{H}^2\rangle'=\langle \mathsf{H}^2\rangle	=\bar{p}^2+m^2+\frac{\bar{p}}{2\gamma}=E(\bar{p})^2+\frac{\bar{p}}{2\gamma}\,. 
\end{equation}
We have now all  we need for writing down the dispersions (\ref{disVi}) and those   of  the radial operators $\mathsf{H}$ and $\mathsf{V}_+$.

The analytical results derived above are less intuitive because of the functions $G$ which are quite complicated. Therefore, for convincing ourselves that these results are plausible, we have to resort to a brief graphical analysis for comparing the above expectation values with the corresponding classical quantities $E(\bar{p}) $ and $V(\bar{p})=\frac{\bar{p}}{E(\bar{p})}$. In Fig. 1 we plot the ratios $\frac{\langle\mathsf{H}\rangle} {E(\bar{p}) }$ and $\frac{2\gamma\,{\rm disp}(\mathsf{H})} {\bar{p} }$ as functions of $q=\gamma\bar{p}>1$ observing that $\langle\mathsf{H}\rangle$ is very close to $E(\bar{p})$ while the dispersion ${\rm disp}(\mathsf{H})<\frac{\bar{p}}{2\gamma}$ tends asymptotically to its maximal value. In  Fig. 2  we plot the ratio $\frac{\langle\mathsf{V}_+\rangle} {V(\bar{p}) }$ and  ${\rm disp}(\mathsf{V}_+)$ observing again that $\langle\mathsf{V}_+\rangle$ is very close to the classical velocity having a small dispersion. We see thus that in the case of Dirac's massive fermions the quantum corrections to the classical  motion  are relatively small but not negligible. Note that these corrections diminish when $\bar{p}$ is increasing, vanishing in the ultra-relativistic limit when the velocity approaches to the speed of light. This behaviour convinces us that the above model works properly describing a plausible physical reality.

\section{Concluding remarks}

We improved  here the quantum theory of Dirac's free field focusing on the  spin and position operators of the version Pryce(e) as fundamental observables of QFT.   We succeeded to do this by using the method of associated operators allowing us to derive the principal operators of QFT.   The original results at the level of RQM,  presented in sections  3.3, 3.4 and 4.1-4.4,  prepare the quantization procedure leading to the new results reported in sections 5.2 and 5.3. The study of the wave-packet in section 6 is presented here for the first time.

In QFT we have the benefit of a correct physical interpretation that is not similar with the interpretations at the level of RQM or even classical theory. An example is the position operator of the version Pryce(e) which was proposed as a mass-center position operator satisfying the canonical coordinate-momentum commutation relations \cite{B}.  The quantization preserves this property but transforms the would-be mass center operator into the dipole one  interpreting correctly the antiparticle term.  For this reason we defined separately the position operators of particle and antiparticle centers, (\ref{Xplus}) and respectively (\ref{Xmin}),  whose linear combinations give both the dipole and mass-center operators. Apart from these operators  we show that  the one-particle operator (\ref{incoord}), interpreted as the initial coordinate operator, complies with Pryce's hypothesis (c) being related to the Pryce(c)-Czochor one-particle operator (\ref{SCzz}). Similarly,  the Frankel spin operator (\ref{SFrr}) corresponding to the hypothesis Pryce(d) is related to a specific position operator but which  does not have yet an obvious physical meaning.  In addition, we note that the orbital boost generators (\ref{Kt0s}) may be interpreted as components of a position operator with spin induced non-com\-mutativity \cite{D} but with  orbital angular momentum instead of  spin.  

Released on QFT  we do not abandon  the RQM but we reconsider each particular system we investigate  as a restriction of QFT,   keeping thus the correct physical interpretation. An example is the Dirac wave-packet we studied in Sec. 5 where finally all the statistical quantities are derived using  associated operators in MR and wave spinors.  It is worth noting that in the one-particle RQM derived from QFT  the associated spin, orbital angular momentum and position operators in MR have familiar forms such that  in momentum-spin basis they become just the corresponding operators, $\tilde S_i=\frac{1}{2}\sigma_i$,  $\tilde L_i =-i\epsilon_{ijk}p^j \partial_{p^k}$  and $\tilde X^i=i\partial_{p^i}$,  of the original non-relativistic Pauli's theory  but describing now relativistic systems as, for example, the spin-orbit interactions of photons and electrons \cite{SO}.

In momentum bases with peculiar polarization these operators become more complicated depending explicitly on polarization through the matrices  (\ref{Dxx}) and (\ref{Omega})   that can have non-trivial forms as in the case of the momentum-helicity basis  where these quantities are given by Eqs. (\ref{Sigp1}) and (\ref{Omega1}). The matrices (\ref{Dxx}) are the Pauli operators written in a new basis but the role of the matrices (\ref{Omega}) defining the covariant derivatives remains obscure for a while until we will study concrete examples of orbital operators in bases with peculiar polarization. Unfortunately, we do not have other momentum bases with peculiar polarization as the helicity is the only one used so far. We hope that our  approach  will offer one the opportunity of defining new types of peculiar polarization that could be observed in further experiments.

However, one may ask why the theory of quantum free field deserves this effort based on a relatively complicated mathematics.  This is because we cannot solve analytically the equations of interacting fields for obtaining closed forms of interacting quantum fields or other operators of QFT. Instead we may resort to perturbations in terms of  $in$ and $out$ fields which are just free fields for which we  constructed the approach presented here. For example, we can calculate using perturbations the expectation values of the new position operators (\ref{XVmc}) or the traditional one (\ref{coord}) in $out$ states if we know the incident wave-packets in  $in$ states.  In this manner we may better understand  the role of the radiative corrections in the fermion propagation affected by Zitterbewegung.     

We conclude that our approach may open the door to new directions of investigation improving in the same time  the traditional ones as, for example, the QED processes involving polarized fermions  studied for the first time in momentum-spin basis long time ago \cite{AS}. Moreover, we hope that solving the inherent new technical problems and working out various examples of systems of free or interacting polarized fermions we ay  improve the theory  filling out  the gap between the actual notorious successes of Dirac's theory, the Hydrogen atom and QED.

\appendix

	\section{Dirac representation }

		\renewcommand{\theequation}{A\arabic{equation}}
		\setcounter{equation}{0} \renewcommand{\theequation}
		{A.\arabic{equation}}
		
		The Dirac $\gamma$-matrices which satisfy $\{\gamma^{\mu},\gamma^{\nu}\}=2\eta^{\mu\nu}$  give rise to the  generators
		$s^{\mu\nu}=\frac{i}{4}\left[\gamma^{\mu},\gamma^{\nu}\right]=\overline{s^{\mu\nu}}$ of the Dirac reducible representation $\rho_D=(\frac{1}{2},0)\oplus(0,\frac{1}{2})$ of the $SL(2,\mathbb{C})$ group in the four-dimensional space ${\cal V}_D={\cal V}_P\oplus{\cal V}_P$ of  Dirac spinors.  Remarkably, this space hosts the fundamental representation of the group $SU(2,2)$ \cite{Yao} in which $SL(2,\mathbb{C})$ is a subgroup. A basis of the  Lie algebra $su(2,2)$ may be formed by those of the $sl(2,\mathbb{C})$ subalgebra, $\sigma_{\mu\nu}$, and the  matrices $\gamma^{\mu}$, $\gamma^5\gamma^{\mu}$ and $i\gamma^5$.

		All these matrices, including the $SL(2,\mathbb{C})$ generators, are Dirac self-adjoint such that the transformations
		\begin{equation}\label{tr}
			\lambda(\omega)=\exp\left(-\frac{i}{2}\omega^{\alpha\beta}s_{\alpha\beta}\right)\in \rho_D[SL(2,\mathbb{C})]\,, 
		\end{equation}
		having  real-valued parameters, $\omega^{\alpha \beta}=-\omega^{\beta\alpha}$,   leave invariant the Hermitian form $\overline{\psi}\psi$  as $\overline{\lambda(\omega)}=\lambda^{-1}(\omega)=\lambda(-\omega)$. The corresponding  Lorentz transformations,  
		$\Lambda^{\mu\,\cdot}_{\cdot\,\nu}(\omega)\equiv	\Lambda^{\mu\,\cdot}_{\cdot\,\nu}[\lambda(\omega)]=\delta^{\mu}_{\nu}
		+\omega^{\mu\,\cdot}_{\cdot\,\nu}+\frac{1}{2}\,\omega^{\mu\,\cdot}_{\cdot\,\alpha}\omega^{\alpha,\cdot}_{\cdot\,\nu}$ $+\cdots$
		satisfy the identities
		\begin{equation}\label{canh}
			\lambda^{-1}(\omega)\gamma^{\alpha}\lambda(\omega)	=\Lambda(\omega)^{\alpha\,\cdot}_{\cdot \,\beta}\gamma^{\beta}\,,
		\end{equation} 
		which encapsulate the canonical homomorphism \cite{WKT}.
		
		In the chiral representation we consider here, the Dirac matrices are expressed in terms of Pauli matrices, $\sigma_i$, and ${\bf 1}=1_{2\times 2}$ as
		\begin{equation}
			\gamma^0=\left(\begin{array}{cc}
				0&{\bf 1}\\
				{\bf 1}&0
			\end{array} \right)	\,,\quad	
			\gamma^i=\left(\begin{array}{cc}
				0&\sigma_i\\
				-\sigma_i&0
			\end{array} \right)	\,,\quad
			\gamma^5=\left(\begin{array}{cc}
				-{\bf 1}&0\\
				0&{\bf 1}
			\end{array} \right)\,,
		\end{equation}
		such that the transformations  $\lambda(\omega)$  generated by the matrices $s^{\mu\nu}$ are reducible to the  subspaces of Pauli spinors ${\cal V}_P$ carrying the irreducible representations $(\frac{1}{2},0)$ and $(0,\frac{1}{2})$ of ${\rho_D}$ \cite{WKT,Th}. 
		We denote  by 
		\begin{equation}\label{r0}
			r={\rm diag}(\hat r,\hat r)\in {\rho_D}\left[SU(2)\right]	
		\end{equation}
		the transformations we call here simply rotations, and by 
		\begin{equation}\label{l0}
			l={\rm diag}(\hat l,\hat l^{-1})\in {\rho_D}\left[ SL(2,\mathbb{C})/SU(2)\right]	
		\end{equation}
		the Lorentz boosts. For rotations we use the  generators  
		\begin{equation}\label{si}
			s_i= \frac{1}{2}\epsilon_{ijk}s^{jk}  ={\rm diag}(\hat s_i,\hat s_i)=-\frac{1}{2}\gamma^0\gamma^5\gamma^i\,, \quad \hat s_i=\frac{1}{2}\sigma_i\,,
		\end{equation}
		and Cayley-Klein parameters  $\theta^i=\frac{1}{2}\epsilon_{ijk}\omega^{jk}$ such that
		\begin{eqnarray}
			r(\theta)={\rm diag}(\hat r(\theta),\hat r(\theta))\,,\quad~~ \hat r(\theta)=e^{-i \theta^i \hat s_i}=e^{-\frac{i}{2} \theta^i \sigma_i}\,. \label{r}
		\end{eqnarray}
		Similarly, we chose the parameters  $\tau^i=\omega^{0i}$ and the generators 
		\begin{equation}\label{s0i}
			s_{i0}=s^{0i}={\rm diag}(-i\hat s_i, i\hat s_i) =\frac{i}{2}\gamma^0\gamma^i\,,
		\end{equation}
		for the Lorentz boosts that read 
		\begin{eqnarray}
			l(\tau)={\rm diag}(\hat l(\tau),\hat l^{-1}(\tau))\,,\quad \hat l(\tau)=e^{ \tau^i \hat s_i}=e^{\frac{1}{2} \tau^i \sigma_i} \,.\label{l}
		\end{eqnarray}
		The corresponding transformations of the group $L_+^{\uparrow}$ will be denoted as $R(r)\equiv R(\hat r)=\Lambda (r)$ and $L(l)\equiv L(\hat l)=\Lambda(l)$. We say that ${\vec s}$ is the Pauli-Dirac spin operator reducible to a pair of  Pauli spin operators,  ${\vec {\hat s}}$. Note that these operators satisfy the identities
		\begin{equation}\label{rsigr}
			\hat r^{-1}\sigma_i\hat r=R_{ij}(\hat r)\sigma_j ~~~~\Rightarrow ~~~~	r^{-1}\sigma_i r=R_{ij}(\hat r)\sigma_j\,,
		\end{equation}
		resulted from the canonical homomorphism.
		
		The boosts (\ref{l}) with parameters $\tau^i=-\frac{p^i}{p}{\rm tanh}^{-1} \frac{p}{E(p)}$ can be written as  \cite{Th}
		\begin{eqnarray}\label{Ap}
			l_{{\vec p}}=\frac{E(p)+m+\gamma^0{\vec\gamma}\cdot {\vec p}}{\sqrt{2m(E(p)+m)}}=	l_{{\vec p}}^+\,,\quad l^{-1}_{\vec p}=l_{-\vec p}=\bar{l}_{\vec p}\,.
		\end{eqnarray}
		They give rise to the Lorentz boosts $L_{\vec p}=\Lambda(l_{\vec p})$ with the matrix elements,  
		\begin{eqnarray}
			\left<L_{{\vec p}}\right>^{0\,\cdot}_{\cdot\, 0}&=&\frac{E(p)}{m}\,,\quad \left<L_{{\vec p}}\right>^{0\,\cdot}_{\cdot\, i}=\left<L_{{\vec p}}\right>^{i\,\cdot}_{\cdot\, 0}=\frac{p^i}{m}\,,\nonumber\\ 
			\left<L_{\vec p}\right>^{i\,\cdot}_{\cdot\, j}&=&\delta_{ij}+\frac{p^i p^j}{m(E(p)+m)}\,,~~~~~~\label{Lboost}
		\end{eqnarray}
		which transforms the representative momentum $\mathring{p}=(m,0,0,0)$  into the desired momentum ${\vec p}=L_{\vec p}\,\mathring{p}$.  Hereby it is convenient to separate the  three-dimensional tensor 
		\begin{equation}\label{tete}
			\Theta_{ij}({\vec p})\equiv \left<L_{\vec p}\right>^{i\,\cdot}_{\cdot\, j}~~\Rightarrow~~\Theta^{-1}_{ij} ({\vec p})=\delta_{ij}-\frac{p^i p^j}{E(p)(E(p)+m)}	\,,
		\end{equation}
		we need when we study space components. $\Theta^{-1}$ denotes the inverse of $\Theta$ on ${\mathbb R}^3$ which is different from the space part of $L^{-1}_{\vec p}=L_{-{\vec p}}$. 
		
		In Dirac's theory there are applications where we may use some properties as 
		\begin{equation}\label{Ap2}
			l_{\vec p}^2=\frac{E({p})+\gamma^0{\vec\gamma}\cdot {\vec p}}{m}\,,\qquad 	l_{-\vec p}^2=\frac{E(p)-\gamma^0{\vec\gamma}\cdot {\vec p}}{m}\,,
		\end{equation}
		giving rise to the following identities
		\begin{eqnarray}
			\frac{1\pm\gamma^0}{2}l_{\vec p}^2	\frac{1\pm\gamma^0}{2}= 	
			\frac{1\pm\gamma^0}{2}l_{-\vec p}^2	\frac{1\pm\gamma^0}{2}	
			&=&\frac{E(p)}{m}	\frac{1\pm\gamma^0}{2}\,,~~~~~~~~\label{idll}
		\end{eqnarray}
		which help us to recover the  operators (\ref{Pip}) and (\ref{Pim}) and to evaluate the quantities
		\begin{equation}\label{ullu}
			\mathring{u}^+_{\sigma}({\vec p})l_{\vec p}^2\mathring{u}_{\sigma'}({\vec p})=\mathring{v}^+_{\sigma}({\vec p})l_{\vec p}^2\mathring{v}_{\sigma'}({\vec p})=\frac{E(p)}{m}\delta_{\sigma\sigma'}\,,
		\end{equation}
		we need for normalizing the mode spinors. 
		
		Among the transformations of the set $SU(2,2)/SL(2,\mathbb{C})$  notorious ones are the Foldy-Wouthuysen unitary transformations \cite{FW}. The particular one,
		\begin{equation}\label{FW}
			U_{\rm FW}({\vec p})=U_{\rm FW}^+(-{\vec p})=\frac{E(p)+m+{\vec \gamma}\cdot {\vec p}}{\sqrt{2E(p)(E(p)+m)}} \,,
		\end{equation}
		brings the  Fourier transform of Dirac's Hamiltonian  in diagonal form 
		\begin{equation}\label{FW1}
			U_{\rm FW}({\vec p})	 \hat H_D({\vec p})U_{\rm FW}(-{\vec p})=\gamma^0 E(p)\,,
		\end{equation}
		and transforms the Fourier transform of the Pryce(e) spin operator into the Pauli-Dirac one \cite{FW}, 
		\begin{equation}\label{FW2}
			U_{\rm FW}({\vec p})	 \vec{\hat  S}({\vec p})U_{\rm FW}(-{\vec p})={\vec s}\,.
		\end{equation}
		Note that  Pryce proposed  previously a similar transformation which differs from  (\ref{FW})  only through a parity,   $U_{\rm Pryce}({\vec p})=\gamma^0 U_{\rm FW}({\vec p})$ \cite{B}.
		
		\section{Algebraic properties of associated operators}
		
		\setcounter{equation}{0} \renewcommand{\theequation}
		{B.\arabic{equation}}
		
		The generators $\{H,P^i,J_i,K_i\}$ form a basis of the algebra. Lie$({T})$ among them the $sl(2,\mathbb{C})$ ones satisfy,
		\begin{eqnarray}
			su(2)\sim so(3): \quad&& \left[J_i,J_j\right]=i\epsilon_{ijk}J_k\,,\label{su2}\\
			&&\left[J_i,K_j\right]=i\epsilon_{ijk}K_k\,, \nonumber\\ 
			&&\left[K_i,K_j\right]=-i\epsilon_{ijk}J_k\,,\label{sl2c}
		\end{eqnarray} 
		commuting with the Abelian generators as
		\begin{eqnarray}
			&\left[H,J_i\right]=0\,,~~~~~\qquad &\left[P^i,J_j\right]=i\epsilon_{ijk}J_k\,,\\
			&\left[H,K_i\right]=-iP^i\,,\qquad &\left[P^i,K_j\right]=-i\delta^{i}_{j}H\,.
		\end{eqnarray}
		In CR we cannot separate an orbital subalgebra as the operators $\underline{\vec x}\land{\vec P}$ and ${\vec s}$ are not conserved. For this reason it is convenient to analyze the algebraic properties in MR where the Abelian generators are diagonal as in Eq.  (\ref{tilHP}).

		In  MR the generators $\{ E(p), p^i, \tilde J_i,\tilde K_i\}$ of the  algebra Lie$(\tilde{T})$ associated to Lie$({T})$ satisfy similar commutation rules allowing the splittings (\ref{tilJ}) and (\ref{tilK})  which separate the orbital parts from the spin ones. In the case of rotations both the angular momentum and  spin operator are conserved separately their components forming two independent  $su(2)\sim so(3)$ algebras,
		\begin{equation}\label{ssuu2}
			\left[\tilde L_i,\tilde L_j\right]=i\epsilon_{ijk}\tilde L_k\,,\quad\left[\tilde S_i,\tilde S_j\right]=i\epsilon_{ijk}\tilde S_k\,,\quad \left[\tilde L_i,\tilde S_j\right]=0\,.
		\end{equation}
		In contrast, the operators $\tilde K^o$ and $\tilde K^s$ do not commute among themselves,
		\begin{eqnarray}\label{KiKjs}
			\left[\tilde K^o_i,\tilde K^s_j\right]=-\frac{i}{E(p)+m}\left[E(p)\epsilon_{ijk} \tilde S_k + p^i \tilde K_j^s\right]\,,
		\end{eqnarray}
		which means that the factorization (\ref{factor}) cannot be extended to the entire $sl(2,\mathbb{C})$ algebra. Nevertheless, the commutation relations
		\begin{eqnarray}
			&&~~~\left[\tilde L_i,\tilde K^o_j\right]=i\epsilon_{ijk}\tilde K^o_k\,,
			\quad\left[\tilde K^o_i,\tilde K^o_j\right]=-i\epsilon_{ijk}\tilde L_k\,,\label{sl2corb}\\
			&&~\left[\tilde L_i,E(p)\right]=0\,,\quad~~~~~~~~\quad \left[\tilde L_i, p^j\right]=i\epsilon_{ijk}p^k\,,\\
			&&\left[\tilde K^o_i, E(p)\right]=ip^i\,,\quad\qquad ~~\left[\tilde K^o_i, p^j\right]=i\delta^{i}_{j}E(p)\,.
		\end{eqnarray}
		convince us that the operators $\{E(p), p^i,\tilde L_i,\tilde K_i^o\}$ generate an orbital representation of the Poincar\'e algebra known as the  natural or scalar representation but now in MR instead of the configuration one. Note that $\tilde S_i$ commute with this entire algebra. Other useful relations in the spin sector,  
		\begin{eqnarray}
			\left[\tilde S_i,\tilde K^s_j\right]&=&\frac{i}{E(p)+m}\left[  p^i\tilde S_j-\delta_{ij} {\vec p}\cdot\tilde{\vec S}\right] \,, \label{KSSS} \\
			\left[\tilde K^s_i,\tilde K^s_j\right]&=&\frac{i}{(E(p)+m)^2}\epsilon_{ijk}p^k{\vec p}\cdot\tilde{\vec S} \,,	\label{KKSS}  
		\end{eqnarray}
		do not have an obvious physical meaning.
		
		The position operator in MR at the time $t$, $\vec{\tilde X}(t)=\tilde{\vec X}+t\tilde{\vec V}$, whose components are  given by Eqs. (\ref{tilX}) and (\ref{tilV}) do not have spin terms being genuine orbital operators satisfying,
		\begin{eqnarray}
			&&\left[ \tilde X^i(t), \tilde X^j(t) \right]=0\,, \qquad~~  \left[ \tilde X^i(t),  p^j\right]=i\delta_{ij}\,,\\
			&&\,\left[ \tilde X^i(t), E(p)\right]=i\tilde V^i\,, \qquad  \left[ \tilde V^i, E(p)\right]=0\,,\\
			&&\left[ \tilde K_i^o, \tilde X^j  \right]=\delta_{ij}\frac{1}{2E(p)} -i\frac{p^j}{E(p)}\tilde X^i-\frac{p^ip^j}{2E(p)^3}\,,\label{B14}\\
			&&\left[\tilde K^o_i, \tilde V^j \right]=E(p)\left[ \tilde X^i, \tilde V^j\right]=i\left[ \delta_{ij}-\frac{p^ip^j}{E(p)^2}    \right]\,.~~~~~~\label{comXV}
		\end{eqnarray}
		As was expected, $\vec{\tilde X}(t)$ behaves as a $SO(3)$ vector commuting as    
		\begin{equation}
			\left[ \tilde L_i, \tilde X^j(t) \right]=i\epsilon_{ijk}\tilde X^k(t)\,, 	\qquad \left[\tilde S_i, \tilde X^j(t)\right]=0 \,,
		\end{equation}
		with the components of the angular momentum and spin operators. In contrast, the commutators
		\begin{equation}
			\left[ \tilde K^s_i, \tilde X^j \right]=\frac{i}{E(p)+m}\left[- \epsilon_{ijk}\tilde S_k+\frac{p^j}{E(p)}\tilde K^s_i  \right]\,,\label{B17}
		\end{equation}
		do not have an intuitive interpretation. 
		
		The components (\ref{tilW0}) and (\ref{tilWi}) of the Pauli-Lubanski operator have well-known algebraic properties we complete here with the commutation relations with our new operators $\tilde S_i$ and $\tilde X^i$ that read
		\begin{eqnarray}
			&&\left[\tilde S_i, \tilde W^0 \right]=i (E(p)+m)\tilde K^s_i\,, \,,\\
			&&\left[\tilde S_i, \tilde W^j \right]=i\, m\, \epsilon_{ijk}\tilde S_k +ip^j \tilde K^s_i\,,\\
			&&	\left[\tilde X^i, \tilde W^0 \right]=i\tilde S_i\,,\\
			&&\left[\tilde X^i, \tilde W^j \right]=\frac{i}{E(p)+m}\left[ \delta_{ij} \tilde W^0 +p^i \tilde S_j^{(-)} \right]\,,
		\end{eqnarray}
		where $\tilde S^{(-)}$ are defined by Eq. (\ref{Sminus}). The operators $\tilde V^i$ are multiplicative commuting with all the components $\tilde W^{\mu}$.
		
		\section{Associated Pryce's (c) and (d) position operators}
		
		\setcounter{equation}{0} \renewcommand{\theequation}
		{C.\arabic{equation}}
		
		The  operators associated to the position operators (\ref{Prcd1}) can be derived taking into account that the Pryce(e) position operator is associated to the operators (\ref{tilX}) and using the Fourier transforms (\ref{Prcd2}) and (\ref{Prcd3}). We obtain thus the  the associated operators 
		\begin{eqnarray}
			{X}^i_{\rm Pr(c)}~~~\Rightarrow~~~	\tilde{X}^i_{\rm Pr(c)}&=&	\tilde{X}^{c\,i}_{\rm  Pr(c)}=i\tilde \partial_i+\frac{\epsilon_{ijk}p^j \tilde S_k}{E(p)(E(p)+m)}=\frac{1}{2}\left\{\tilde K_i, \frac{1}{E(p)}\right\}\,,\label{xili}\\
			{X}^i_{\rm Pr(d)}~~~\Rightarrow~~~	\tilde{X}^i_{\rm Pr(d)}&=&	\tilde{X}^{c\,i}_{\rm  Pr(d)}=i\tilde \partial_i-\frac{\epsilon_{ijk}p^j \tilde S_k}{m(E(p)+m)}\,.\label{xili1}
		\end{eqnarray}
		The  components of these operators do not commute among themselves such that the commutators
		\begin{eqnarray}
			&&	\left[\tilde{X}^i_{\rm Pr(c)}, 	\tilde{X}^j_{\rm Pr(c)}     \right]=-i\epsilon_{ijk}\tilde Y^k_{\rm Pr(c)}\,, \\
			&& 	\left[\tilde{X}^i_{\rm Pr(d)}, 	\tilde{X}^j_{\rm Pr(d)}     \right]=i\epsilon_{ijk}\tilde Y^k_{\rm Pr(d)}\,,
		\end{eqnarray}
		generate  new   associated components
		\begin{eqnarray}
			Y^i_{\rm Pr(c)}&\Rightarrow&\tilde Y^i_{\rm Pr(c)}	=-\tilde Y^{c\,i}_{\rm Pr(c)}=\frac{m}{E(p)^3}\tilde S^{(+)} =\frac{1}{E(p)^3}\tilde W^i\,,\\
			Y^i_{\rm Pr(d)}&\Rightarrow&\tilde Y^i_{\rm Pr(d)}	=-\tilde Y^{c\,i}_{\rm Pr(
				d)}=\frac{1}{m E(p)}\tilde S^{(+)}_i =\frac{1}{m^2 E(p)}\tilde W^i\,,\label{ili1}
		\end{eqnarray}
		proportional with those defined by Eqs. (\ref{Splus}) and (\ref{tilWi}),  giving rise to  new even one-particle operators.

		These operators have interesting algebraic properties but here we restrict ourselves to derive the commutation relations with the associated isometry generators, i. e. the translation generators,  $E(p)$ and $p^i$, and the $SL(2,\mathbb{C})$ ones,  $\tilde J_i$ and $\tilde K_i$,  defined by Eqs. (\ref{tilJ}) and (\ref{tilK}) whose terms are given in Eqs. (\ref{tilL}), (\ref{tilS}), (\ref{kaka0}) and (\ref{kakas}). We obtain the commutation rules with the components of the Pryce(c) position operator,
		\begin{eqnarray}
			\left[ E(p),\tilde X^j_{\rm Pr(c)} \right]&=&-i \tilde V^i\,, \nonumber\\
			\left[p_i,\tilde X^j_{\rm Pr(c)}     \right]&=&-i \delta_{ij} 1_{2\times 2}\,,\nonumber\\
			\left[\tilde J_i,\tilde X^j_{\rm Pr(c)} \right]&=&i\epsilon_{ijk} \tilde X^k_{\rm Pr(c)}\,,\nonumber\\
			\left[\tilde K_i,\tilde X^j_{\rm Pr(c)} \right]&=&\frac{1}{2E(p)}\left(\delta_{ij}-\frac{p^ip^j}{E(p)^2}\right)1_{2\times 2}
			- \frac{i}{E(p)^2}p^i \tilde X^j_{\rm Pr(c)}-\frac{i}{E(p)}\epsilon_{ijk} \tilde J_k\,,	
		\end{eqnarray}	
		and of those of the Pryce(d) ones 
		\begin{eqnarray}
			\left[ E(p),\tilde X^j_{\rm Pr(d)}     \right]&=&-i \tilde V^i\,,\nonumber\\
			\left[p_i,\tilde X^j_{\rm Pr(d)} \right]&=&-i \delta_{ij} 1_{2\times 2}\,,\nonumber\\
			\left[\tilde J_i,\tilde X^j_{\rm Pr(d)}     \right]&=&i\epsilon_{ijk} \tilde X^k_{\rm Pr(d)}\,,\nonumber\\
			\left[\tilde K_i,\tilde X^j_{\rm Pr(d)}     \right]&=&\frac{1}{2E(p)}\left(\delta_{ij}-\frac{p^ip^j}{E(p)^2}\right)1_{2\times 52}-\frac{i}{E(p)}p^j\tilde X^i_{\rm Pr(d)}\,,	
		\end{eqnarray}
		drawing the conclusion that the  components of these operators satisfy canonical momentum-coordinate commutation relations,  behaves as a $SO(3)$ vectors but having different commutation rules with the boost generators.

		The corresponding components of the one-particle operators,  $\mathsf{X}^i_{\rm Pr(c)}$,  $\mathsf{X}^i_{\rm Pr(d)}$, $\mathsf{Y}^i_{\rm Pr(c)}$ and $\mathsf{Y}^i_{\rm Pr(d)}$ have to be derived substituting in Eq. (\ref{Aq}) the associated operators (\ref{xili}-\ref{ili1}).   	
		
		\section{Spin and helicity momentum bases}
		
		\setcounter{equation}{0} \renewcommand{\theequation}
		{D.\arabic{equation}}
		
		In general,  the Pauli polarization spinors, $\xi_{\sigma}({\vec p})$,  and   $\eta_{\sigma}({\vec p})=i\sigma_2 \xi_{\sigma}^*({\vec p})$, which may depend on momentum,  form related  orthonormal,   
		\begin{equation}\label{xyort}
			\xi_{\sigma}^+({\vec p})\xi_{\sigma'}({\vec p})=\eta_{\sigma}^+({\vec p})\eta_{\sigma'}({\vec p})=\delta_{\sigma\sigma'}\,,	
		\end{equation}
		and complete systems, 
		\begin{equation}\label{xycom}
			\sum_{\sigma}\xi_{\sigma}({\vec p})\xi_{\sigma}^+({\vec p})=\sum_{\sigma}\eta_{\sigma}({\vec p})\eta_{\sigma}^+({\vec p})=1_{2\times2}\,,	
		\end{equation}
		representing bases in the subspaces of Pauli spinors,  ${\cal V}_P$, of the space of Dirac spinors, ${\cal V}_D={\cal V}_P\oplus{\cal V}_P$. 
		
		In the case of arbitrary common  polarization the spin projection is measured along to an  unit vector ${\vec n}$. In this case the Pauli polarization spinors $\xi_{\sigma}(\vec{n})$ and $\eta_{\sigma}(\vec{n})=i\sigma_2 \xi_{\sigma}(\vec{n})^*$  satisfy the eigenvalues problems
		\begin{equation}
			(\vec{n}\cdot \hat{\vec s})\,\xi_{\sigma}(\vec{n})=\sigma\, \xi_{\sigma}(\vec{n})~~\Rightarrow~~
			(\vec{n}\cdot \hat{\vec s})\,\eta_{\sigma}(\vec{n})=-\sigma\, \eta_{\sigma}(\vec{n})\,,
		\end{equation}
		where the matrices $\hat s_i$ are defined in Eq. (\ref{si}).  These spinors have the form 
		\begin{eqnarray}
			\xi_{\frac{1}{2}}(\vec{n})&=&\sqrt{\frac{1+n^3}{2}}\left(
			\begin{array}{c}
				1\\
				\frac{n^1+i n^2}{1+n^3}
			\end{array}\right)\,, \nonumber\\
			\xi_{-\frac{1}{2}}(\vec{n})&=&\sqrt{\frac{1+n^3}{2}}\left(
			\begin{array}{c}
				\frac{-n^1+i n^2}{1+n^3}\\
				1
			\end{array}\right)\,,\label{xi1}
		\end{eqnarray}
		satisfy the normalization and completeness conditions and, in addition, have the property
		\begin{equation}\label{xycom1}
			\sum_{\sigma}\sigma \xi_{\sigma}({\vec n})\xi_{\sigma}^+({\vec n})=\sum_{\sigma}\sigma\eta_{\sigma}({\vec n})\eta_{\sigma}^+({\vec n})= n^i\sigma_i\,,	
		\end{equation}
		we may use in concrete calculations. 
		
		The well-known example is the momentum-spin basis \cite{BDR} with ${\vec n} ={\vec e}_3$ and   
		\begin{equation}\label{etapm}
			\xi_{\frac{1}{2}}=\left(\begin{array}{c}
				1\\
				0
			\end{array}\right)\,,\quad
			\xi_{-\frac{1}{2}}=\left(\begin{array}{c}
				0\\
				1
			\end{array}\right)  \,,
		\end{equation} 
		largely used in applications.

		The only peculiar polarization used so far is the helicity giving rise to the momentum-helicity basis in which the spinors $\xi_{\sigma}({\vec n}_p)$ have the forms (\ref{xi1})  with ${\vec n}={\vec  n}_{p}=\frac{\vec p}{p}$. For writing down the spin components (\ref{Spin}) in this basis we derive the matrices (\ref{Dxx})  that read \cite{Cot}, 
		\begin{eqnarray} 
			\Sigma_{1}({\vec p})&=&\frac{p^1}{p}\,\sigma_3 -p^1\frac{p^1 \sigma_1+p^2\sigma_2}{p(p+p^3)}+\sigma_1\,,\nonumber\\
			\Sigma_{2}({\vec p})&=&\frac{p^2}{p}\,\sigma_3 -p^2\frac{p^1 \sigma_1+p^2\sigma_2}{p(p+p^3)}+\sigma_2\,,\nonumber \\
			\Sigma_{3}({\vec p})	&=&\frac{p^3}{p}\sigma_3-\frac{p^1 \sigma_1+p^2\sigma_2}{p}\,,\label{Sigp1}
		\end{eqnarray}
		verifying that these satisfy 
		\begin{equation}\label{pSig}
			p^i\Sigma_i({\vec p})=p\sigma_3	\,.
		\end{equation}
		The form of the covariant derivatives $\tilde\partial_i=\partial_{p^i} 1_{2\times2}+\Omega_i({\vec p})$ is determined by  the matrices (\ref{Omega})  \cite{Cot},
		\begin{eqnarray}
			\Omega_1({\vec p})&=&\frac{-i}{2p^2(p+p^3)}\left[ p^1p^2\sigma_1 +pp^2\sigma_3+(pp^3+{p^2}^2+{p^3}^2)\sigma_2\right]\,, \nonumber\\
			\Omega_2({\vec p})&=&\frac{i}{2p^2(p+p^3)}\left[ p^1p^2\sigma_2 +pp^1\sigma_3+(pp^3+{p^1}^2+{p^3}^2)\sigma_1\right]\,,\nonumber\\
			\Omega_3({\vec p})&=&\frac{i}{2p^2}\left(	p^1\sigma_2-p^2\sigma_1\right)\,,\label{Omega1} 
		\end{eqnarray}
		satisfying $p^i\Omega_i({\vec p})=0$. We obtain thus  apparently complicated  matrices $\Sigma_i$ and $\Omega_i$ but whose algebra is the same as in momentum-spin basis where $\Omega_i=0$ and $\Sigma_i=\sigma_i$.

\end{document}